\documentclass[showpacs,preprintnumbers,amsmath,amssymb,nofootinbib]{revtex4}

\usepackage{color}   

\usepackage{graphicx}
\usepackage{dcolumn}
\usepackage{bm}

\usepackage[utf8]{inputenc}
\usepackage[T1]{fontenc}
\usepackage{ulem}
\usepackage{amsmath}
\usepackage{amssymb}
\usepackage{float}

\begin{document}

\def\bb    #1{\hbox{\boldmath${#1}$}}
\def\bb    #1{\hbox{\boldmath${#1}$}}

\def\blambda{{\hbox{\boldmath $\lambda$}}}
\def\eeta{{\hbox{\boldmath $\eta$}}}
\def\bxi{{\hbox{\boldmath $\xi$}}}
\def\bzeta{{\hbox{\boldmath $\zeta$}}}
\def\sD{D \!\!\!\!/}
\def\sd{\partial \!\!\!\!/}
\def\EQ{{\hbox{\boldmath $Eq.(\ref$}}}

\def\qcdu{{{}_{ \rm QCD}}}   
\def\qedu{{{}_{\rm QED}}}   
\def\qcdd{{{}^{ \rm QCD}}}   
\def\qedd{{{}^{\rm QED}}}   
\def\qcd{{{\rm QCD}}}   
\def\qed{{{\rm QED}}}   
\def\2d{{{}_{\rm 2D}}}         
\def\4d{{{}_{\rm 4D}}}         
\def\sg#1{ {\rm \,sign}(#1)\, } 

\large

\title{On the question of  quark confinement in the Abelian
  U(1) QED gauge interaction}

\author{Cheuk-Yin Wong}

\affiliation{ Physics Division, Oak Ridge National
  Laboratory\footnote{This research was supported in part by the
  Division of Nuclear Physics, US Department of Energy under contract
  DE-AC05-00OR22725 with
  UT-Battelle, LLC.  The publisher, by
  accepting the article for publication, acknowledges that the US
  government retains a nonexclusive, paid-up, irrevocable, worldwide
  license to publish or reproduce the published form of this
  manuscript, or allow others to do so, for US government
  purposes. DOE will provide public access to these results of
  federally sponsored research in accordance with the DOE Public
  Access Plan (http://energy.gov/downloads/doe-public-access-plan),
  Oak Ridge, Tennessee 37831, USA }\!, Oak Ridge, Tennessee
37831, U.S.A.\\
~~wongc@ornl.gov}

\begin{abstract}

If we approximate light quarks as massless and apply the Schwinger
confinement mechanism to light quarks, we will reach the conclusion
that a light quark $q$ and its antiquark $\bar q$ will be confined as
a $q\bar q$ boson in the Abelian U(1) QED gauge interaction in (1+1)D,
as in an open string.  From the work of Coleman, Jackiw, and Susskind,
we can infer further that the Schwinger confinement mechanism persists
 even for massive quarks in (1+1)D.  Could such a QED-confined $q\bar
q$ one-dimensional open string in (1+1)D be the idealization of a flux
tube in the physical world in (3+1)D, similar to the case of
QCD-confined $q\bar q$ open string?  If so, the QED-confined $q\bar q$
bosons may show up as neutral QED mesons in the mass region of many
tens of MeV (PRC81(2010)064903 \& JHEP2020(8)165).  Is it ever
possible that a quark and an antiquark be produced and interact in QED
alone to form a confined QED meson?  Is there any experimental
evidence for the existence of a QED meson (or QED mesons)?  The
observations of the anomalous soft photons, the X17 particle, and the
E38 particle suggest that they may bear the experimental evidence for
the existence of such QED mesons.  Further confirmation and
investigations on the X17 and E38 particles will shed definitive light
on the question of quark confinement in QED in (3+1)D. Implications of
quark confinement in the QED interaction are discussed.

\end{abstract}


\maketitle

\section{Introduction}

As is well-known, quarks\footnote{ We use the term ``quarks'' to
include also the antiquarks implicitly, if no ambiguity arises.  The
term ``antiquark(s)'' will be explicitly used, if ambiguities may
arise.  } interact in the QCD (quantum chromodynamical) interaction
and the QED (quantum electrodynamical) interaction.  The general
understanding is that the confinement of quarks arises from the
non-Abelian nature of the QCD interaction in which the gluons mediate
the QCD interaction between quarks and the gluons also interact among
themselves.  The self-interactions of gluons build the bridges
connecting the quarks and confining the quarks.

The confinement of quarks is a peculiar phenomenon because quarks
cannot be isolated.  We can get an idea about such a peculiar property
by asking whether a quark and its antiquark are confined in the QCD
interaction at a certain energy.  The answer is that 
only 
specifically at the eigenenergies in the QCD
eigenstates, whose spatial extension spans a confined region,
do a quark and an antiquark exist and are confined.
However, at all other energies different from those of the QCD
quark-antiquark eigenenergies, states of a quark and an antiquark  do not exists ----  they do not exists as bound states, nor do they exist as continuum states of an
isolated quark and antiquark.  In contradistinction, for isolatable
particles such as an electron and a positron interacting in QED,
states of an electron and a positron exist at all energies above the positronium ground state energy, either as bound $e^+e^-$ states or as
continuum states of an isolated electron and positron.  Therefore, a
quark and an antiquark can be described as being confined in a certain
interaction, if there exist confined $q \bar q$ eigenstates at the
eigenenergies when a quark and an antiquark interact in that
interaction, in conjunction with the absence of continuum states of an
isolated quark and antiquark at other energies.

We would like to study quark confinement in the lowest energy states
of a system of quarks interacting in the QED interaction.  Such a
study will benefit from the study of quarks interacting in the QCD
interaction and vice versa.  For this reason, we include also the QCD
interaction in our consideration when it is appropriate.  In order for
quark-antiquark states to be observable, they must be among the bound
and confined eigenstates arising from the quark and the antiquark
interacting non-pertubatively at the appropriate eigenenrgies.  As
bound state are involved, we shall therefore consider the QCD and QED
interactions between the quark and the antiquark to be implicitly
non-perturbative in nature, and limit our attention to those systems
that can potentially be bound and confined at possible eigenenergies.

For the non-Abelian QCD interaction, the question of quark confinement
in QCD can be inferred from the QCD potential between a static quark
and a static antiquark.  For example, the quark and the antiquark
appear as static external probes represented by time-like world lines
at fixed spatial separations in a Wilson loop, given in terms of the
product of link variables.  The area law of the Wilson loop gives a
linear QCD interaction potential between the quark and the antiquark,
which leads to QCD meson eigenstates at eigenenergies in the confining
quark-antiquark potential and the absence of continuum states of an
isolated quark and antiquark at other energies.

However, for quarks interacting in the QED interaction, the question
of quark confinement in (3+1)D cannot be answered by just studying the
static QED potential between a static quark and a antiquark as
inferred from static lattice gauge calculations only, because there is
an important Schwinger confinement mechanism\footnote{
The Schwinger confinement mechanism, also known as
the Schwinger model or the Schwinger QED2 model, refers to the
mechanism in which a massless charged fermion and its antifermion
interacting in QED in (1+1)D are confined and bound into a neutral
boson with a mass $m=g_\2d/\sqrt{\pi}$, where $g_\2d$ is the magnitude
of the dimensional coupling constant in (1+1)D \cite{Sch62,Sch63}.  It
differs from the Schwinger pair production mechanism which refers to
the mechanism for the production of charge pairs in a strong
electric field examined in \cite{Sch51}.  For a pedagogical review of
the Schwinger pair-production mechanism and the Schwinger confinement
mechanism, see Chapter 5 and Chapter 6 of \cite{Won94} respectively.
Recent generalizations and extensions of the Schwinger model have been
presented in \cite{Geo19,Geo19a,Geo20,Geo20a,Geo22}  Recent 
lattice gauge solution  of the massive Schwinger model
 has been obtained in \cite{Dem22}. }
associated with the
interplay between gauge fields $A^\mu$ and dynamical quark currents
$j^\mu$ \cite{Sch62,Sch63,Col75,Col76} that may play an important role
in the question of confinement as we shall examine in Section 3.  In
particular, if we consider quark confinement just from the viewpoint
of the static QED potential between a static quark and an antiquark as
inferred from static lattice gauge calculations, we will reach the
conclusion that a static quark and a antiquark will be deconfined in
compact QED in (3+1)D because the compact QED interaction\footnote{ To study
quark confinement in QED in lattice gauge calculations in (3+1)D, we
need to consider only the compact QED because in non-compact QED a
quark and an antiquark are always deconfined in lattice gauge
calculations in (3+1)D \cite{Pol87}.  For the Schwinger confinement
mechanism in (1+1)D, the confinement mechanism occurs in the continuum
limit, which is the same for both the compact QED or the non-compact
QED.  There is no need to specify the compact or non-compact nature of
the QED interaction in the Schwinger confinement mechanism.}  
belongs
to the weak-coupling regime in lattice gauge calculations in (3+1)D
\cite{Wil74,Kog75,Man75,Pol77,Pol87,Ban77,Gli77,Pes78,Dre79,Gut80,Kon98,Mag20,Arn03,Lov21},
as we shall discuss in more detail in Section 6.  However, a serious
question arises because if a static quark and a static antiquark are
deconfined in compact QED in (3+1)D, the isolated quark and antiquark will
appear as fractional charges, because there exists no physical laws to
forbid quarks and antiquarks to interact in QED alone below the pion mass gap $m_\pi$.  In a
contradicting manner, no such deconfined quarks and antiquarks in the
form of isolated fractional charges have ever been observed in (3+1)D.
The absence of fractional charges suggests that previous conclusion of
deconfined static quark and antiquark in compact QED in (3+1)D may not be as
definitive as it may appear to be.  The Schwinger confinement
mechanism may need to be included in future lattice gauge
calculations.  We may need to return to the basic description of quark
confinement in terms of the existence of confining eigenstates at
quark-antiquark eigenenergies when quarks are treated as dynamical
fields in an Abelian QED  gauge interaction, in conjunction with the absence of
continuum states of an isolated quark and antiquark at other energies.
We shall return to this question and discuss the relevant salient
points in Section 6. 

Out of scientific curiosity with encouraging suggestions from theories
and experiments, we study whether quarks are confined when they
interact in QED alone, without the QCD interaction.  If we approximate
light quarks as massless and we apply the Schwinger confinement
mechanism
\cite{Sch62,Sch63} to light quarks, we will reach the conclusion that a
light quark and a light antiquark will be confined in QED in (1+1)D as
in an open string
\cite{Won10,Won11,Won14,Won20,Won22,Won22a,Won22b,Won22c}.  From the
work of Coleman, Jackiw, and Susskind on massive Schwinger model
\cite{Col75,Col76}, we can infer further that the Schwinger
confinement mechanism persists in (1+1)D even for massive quarks.
Could such a QED-confined one-dimensional $q\bar q$ open string in
(1+1)D space-time be the idealization of a flux tube in the physical
(3+1)D, with the quark and the antiquark at the two ends of the flux
tube?  If so, the confined $q\bar q$ states in (1+1)D will show up as
neutral QED mesons in (3+1)D.  Is it ever possible for a quark and
an antiquark to be produced and to interact in QED alone so as to form
a confined QED meson?  Is there any experimental evidence to indicate
the possible existence of the QED-confined $q\bar q$ meson or mesons?

Such questions have not been brought up until recently for obvious
reasons.  It is generally perceived that a quark and an antiquark
interact with the QCD and the QED interactions simultaneously, with
the QED interaction as a perturbation, and the occurrence of a stable
and confined state of the quark and the antiquark interacting in the
QED interaction alone, without the QCD interaction, may appear
impossible.  Furthermore, the common perception is that only the QCD
interaction with its non-Abelian properties can confine a quark and an
antiquark, whereas the QED interaction is Abelian.  It has been argued
that even if a quark and an antiquark can interact in the QED
interaction alone, the QED interaction by itself cannot confine the
quark and the antiquark, as in the case of an electron and a positron,
so the quark and the antiquark cannot be confined even if they can
interact with the QED interaction alone.

Experimentally, the occurrence of anomalous neutral bosons with masses
in the region of many tens of MeV suggests a need to re-examine the
above common perceptions with regard to the question of quark
confinement in the QED interaction.  Specifically, (i) the observation
of the anomalous soft photons in high-energy hadron-hadron collisions
\cite{Chl84,Bot91,Ban93,Bel97,Bel02pi,Bel02,Per09} and $e^+e^-$
annihilation collisions \cite{Per09,DEL06,DEL08,DEL10}, (ii) the
observation of the X17 particle at about 17 MeV
\cite{Kra16,Kra19,Kra21,Kra21a,Sas22,Kra22}, and (iii) the observation of
the E38 particle at about 38 MeV \cite{Abr12,Abr19} point to the
possible existence of anomalous neutral particles at energies of many
tens of MeV 
\cite{Won10,Won11,Won14,Won20,Won22,Won22a,Won22b,Won22c}.  
These anomalous particles may
apparently place them outside the known families of the Standard
Model.  However, if we wish to include only particles and interactions
within the Standard Model, a consistent and viable picture of these
anomalous particles emerges to describe them as composite particles of
a quark and an antiquark interacting non-perturbatively in the QED
interaction \cite{Won10,Won11,Won14,Won20,Won22,Won22a,Won22b,Won22c}.
We would like to explain here how such a description of quark
confinement in QED in (3+1)D emerges as a reasonable theoretical
concept consistent with the experimental observations.  The present
review is also timely since it predicts an isoscalar $q\bar q$
composite particle with a mass of about 17 MeV that is a good
candidate for the observed X17 particle, and the confirmation of the
X17 particle is actively pursued by many laboratories as summarized in
\cite{X1722}, including ATOMKI \cite{x17Kra}, Dubna \cite{x17Abr},
STAR \cite{x17STAR}, MEGII \cite{x17MEG}, TU Prague \cite{x17Prague},
NTOF \cite{x17NTOF}, NA64 \cite{x17NA64}, INFN-Rome
\cite{x17INFNRome}, NA48 \cite{x17NA48}, Mu3e \cite{x17Mu3e},
MAGIX/DarkMESA \cite{x17MAGIX}, JLAB PAC50 \cite{x17JLAB,x17JLAB1},
PADME \cite{x17PADME},  DarkLight \cite{ARIEL,Tre22}, LUXE \cite{Hua22}, and Montreal Tandem\cite{Mon22}.

This paper is organized as follows.  In Section 2, we present examples
of reactions in which a $q\bar q$ pair may be produced and may
interact in QED alone.  In Section 3, we apply the Schwinger
confinement mechanism to quarks interacting in QED and show that a
quark and an antiquark approximated as massless are confined in QED in
(1+1)D. We discuss the effects of quark masses on the Schwinger
confinement mechanism.  We infer from the works of Coleman, Jackiw, 
and Susskind that the Schwinger confinement mechanism persists 
even for massive quarks in (1+1)D.  In Section 4, we make the quasi-Abelian
approximation of the non-Abelian QCD dynamics to search for stable
collective excitations in QCD.  The quasi-Abelian approximation allows
the generalization of the Schwinger mechanism from QED in (1+1)D to
(QCD+QED) in (1+1)D.  We obtain the open string model of QCD and QED
mesons.  We use the open string model of QCD and QED mesons in (1+1)D
as a phenomenological model to study the masses of $\pi^0$, $\eta$,
and $\eta'$ in order to determine the QCD coupling constant and the
flux tube radius.  Extrapolation of the QCD open string model to the
QED open string model with the QED fine-structure coupling constant,
we predict the masses of the QED mesons.  In Section 5, we discuss the
decay modes of the QED mesons and examine recent experimental
observations of anomalous particles in the mass region of many tens of
MeV produced in low-energy $pA$, and high-energy $e^+$-$e^-$,
hadron-hadron, and nucleus-nucleus collisions.  We compare the masses
of the experimental anomalous particles with the predicted QED meson
masses.  There is a reasonable agreement of the predicted QED meson
masses with the observed experimental masses of the X17 and the E38
particles, placing the QED mesons as good candidates to  describe these
anomalous particles.  In Section 6, we examine the question of quark
confinement in QED from the viewpoint of lattice gauge calculations
which indicate that quarks in compact QED in (3+1)D are not confined.
We discuss the lattice gauge calculation results of deconfined quarks
in compact QED in (3+1)D, which contradicts the experimental absence
of fractional charges.  It is therefore suggested that the Schwinger
confinement mechanism may need to be included in future lattice gauge
calculations for quarks in compact QED in (3+1)D.  
We propose a ``stretch (2+1)D'' model to study the
importance of the Schwinger confinement mechanism by combining the
Schwinger confinement mechanism of QED in (1+1)D with Polyakov's
transverse confinement in (2+1)D.  In Section 7, We discuss the
implications of quark confinement in the QED interaction. Many new
phenomena and molecular states may emerge if quarks are confined in
the QED interaction.  We present our conclusions and discussions in
Section 8.
 
\section{Could a $q\bar q$  pair be produced and interact
  non-perturbatively in QED alone?}

The proposal of quark confinement in QED
\cite{Won10,Won11,Won14,Won20,Won22,Won22a,Won22b,Won22c} involves the
hypothesis that a quark and an antiquark could be produced below the
QCD pion mass gap $m_\pi$ and could interact non-perturbatively in QED
to lead to QED collective excitation, whereas the collective
excitations of the QCD interaction will not be excited as they require
an excitation energy higher then the pion mass gap and the QCD
interaction appears as a spectator interaction.  From the static quark
and antiquark viewpoint, the common perception is that a quark and an
antiquark interact simultaneously in QCD and QED.  However, from the
dynamical viewpoint of the quantum field theory of quarks interacting
in the QED and the QCD interaction, we envisage that quarks can
exchange a virtual gluon to interact non-perturbatively in the QCD
interaction.  They can also exchange a virtual photon to interact
non-pertuabatively in the QED interaction.  There is no theorem nor
basic physical principle that forbids a quark and an antiquark to
exchange a virtual photon and interact non-perturbative in the QED
interaction alone.  What is not forbidden is allowed, in accordance
with Gell-Mann's Totalitarian Principle \cite{Gel56}.  Therefore, it
is theoretically permitted to explore the hypothesis that a quark and
an antiquark could interact in QED alone.
\begin{figure} [h]
\centering
\vspace*{-0.cm}\hspace*{-0.6cm}
\includegraphics[scale=0.65]{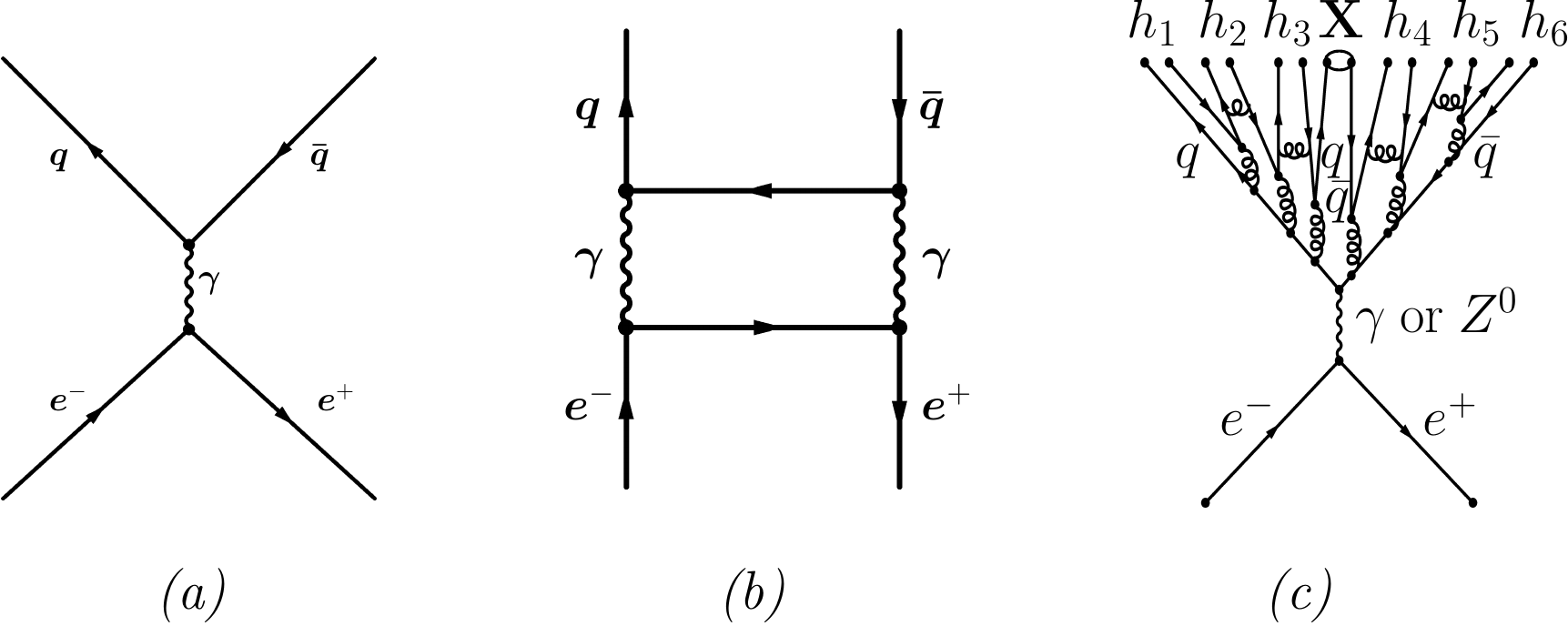}
\caption{ ($a$) The production of a $q\bar q$ pair by the $e^+$+$e^-$
  collision with a single intermediary virtual photon at low energies,
  ($b$) the production of a $q\bar q$ pair by the $e^+$+$e^-$
  collision with two intermediary virtual photons at low energies, and
  ($c$) the production of many $q\bar q$ pairs by the $e^+$+$e^-$
  collision at high energies. }
\label{fig1}
\end{figure}

Experimentally, we can consider the production of $q\bar q$ pairs in
$e^+ $ and $ e^- $ collisions in Fig. 1 as examples.  The energy
threshold for such a reaction is $m_q+m_{\bar q}$ which is about a few
MeV \cite{PDG19}.  At low collision energies, a single $q\bar q$ pair
may be produced as shown in the diagrams in Figs.\ 1($a$) and 1($b$).
As the collision energy increases, many $q\bar q$ pairs may be
produced in high-energy collisions, as shown in Fig.\ 1($c$):
\begin{subequations}
\begin{eqnarray}
 ~~&&e^+ + e^- \to \gamma \to q +\bar q,
\\
~~&& e^+ + e^-
\to \gamma +\gamma \to q + \bar q,
\\
~~&&  e^+ + e^- \to \gamma {\rm ~or~ }\, Z^0 \to  (q  \bar q)^n.~~~~~~
\end{eqnarray}
\end{subequations}
The incident $e^+$ and $ e^-$ pair is in a colorless color-singlet
state in reactions 1($a$) and 1($b$).  The produced $q$ + $\bar q$
pair must combine with their interacting virtual gauge boson $\gamma$
or $g$ to result in a colorless color-singlet final state.  The
produced $q$ resides in the color-triplet $\bb 3$ representation, and
the produced $\bar q$ in the anti-triplet $\bb 3^*$ representation.
They can combine to form the color-singlet $\bb 1$ and the color-octet
$\bb 8$ configurations,
\begin{eqnarray}
{\bb 3} \otimes {\bb 3}^* = {\bb 1} \oplus {\bb 8}.
\end{eqnarray}
The produced $q$ and $\bar q$ in their coupled color-singlet
configuration can interact non-perturbatively in the QED interaction
and combine with a virtual photon $\gamma$ to form a color-singlet
$[(q\bar q)^1 \gamma ^1 ]^1$ final state, where the superscripts
denote color multiplet indices.  Similarly, the produced $q$ and $\bar
q$ in their color-octet configuration can interact non-perturbatively
in the QCD interaction and combine with a virtual gluon $g$ to form a
color-singlet $[(q\bar q)^8 g^8 ]^1$ final state.  A $q\bar q$ pair
will be produced at the eigenenergy of a QCD-confined $[(q\bar q)^8
  g^8 ]^1$ eigenstate, as a QCD meson.  In a similar way, a $q\bar q$
pair will be produced at the eigenenergy of a QED-confined $[(q\bar
  q)^1 \gamma^1 ]^1$ eigenstate as a QED meson, if there is such an
eigenstate at that eigenenergy.  At all other energies, no $q\bar q$
pair will be produced because a confined $q\bar q$ or a continuum
state of a quark $q$ and an antiquark $\bar q$ do not exist at these
energies.  Reactions involving the production of a $q \bar q$ pair
contain the density of final-states factor, which is a delta-function
at the eigenenergies of the confined $q\bar q$ eigenstates.

We can examine the $e^+$ + $e^-$ $\to$ $q$ + $\bar q$ reactions in
Figs.\ 1($a$) and 1($b$) with a center-of-mass energy $\sqrt{s}(q\bar
q)$ in the range $(m_q + m_{\bar q}) <\sqrt{s}(q\bar q) < m_\pi$,
where the sum of the rest masses of the light quark and light
antiquark is of order a few MeV and $m_\pi\sim 135$ MeV \cite{PDG19}.
If there is a confined $[(q\bar q)^1 \gamma ^1 ]^1$ QED eigenstate in
this energy range, then a confined $q\bar q$ pair can be produced as
a QED meson.  Such a QED meson can come only from the non-perturbative
QED interaction but not from the non-perturbative QCD interaction,
because the non-perturbative QCD interaction with a virtual gluon
exchange would endow the $q$$\bar q$ pair as a composite $[(q\bar q)^8
  g^8 ]^1$ QCD meson with a center-of-mass energy $\sqrt{s}(q\bar q)$
beyond this energy range, in a contradictory manner.  It is therefore
possible for a quark and an antiquark to be produced and interact
non-pertubatively in the QED interaction to lead to a QED meson, if
there exists a QED meson eigenstate in this energy range.  At energies
other than the QED meson eigenenergies, in this energy range below
$m_\pi$, the $e^+$ + $e^-$ collisions will probe the dynamics of a
quark and antiquark interacting in QED alone, without the QCD
interaction.  In this energy range, the absence of fractional charges
in $e^+$ + $e^-$ collisions will indicate the absence of the continuum
isolated quark and antiquark states when a quark and an antiquark
interact in the QED interaction alone.

As the $e^+$ + $e^-$ collision energy increases, many $q\bar q$ pairs
will be produced as shown in Fig. 1($c$).  At the $Z^0$ resonance
energy in the DELPHI experiments \cite{DEL06,DEL08,Per09,DEL10}, most
of the produced $q\bar q$ pairs will materialize as $q \bar q$ QCD
mesons labeled schematically as $h_i$ in Fig.\ 1($c$).  However, there
may be a small fraction of the $q\bar q$ pairs which will have
invariant masses below the pion mass.  If there is a confined $q\bar
q$ QED meson state at the appropriate eigenenergy below $m_\pi$, then
the $q\bar q$ pair will be produced as a QED meson, shown
schematically as the $X$ particle in Fig.\ 1($c$).  The decay of the
QED mesons into $e^+$ and $e^-$ pairs may be the source of the
anomalous soft photons observed at DELPHI
\cite{DEL06,DEL08,Per09,DEL10}.

 \begin{figure} [h]
\centering
\vspace*{-0.cm}\hspace*{-0.6cm}
\includegraphics[scale=0.65]{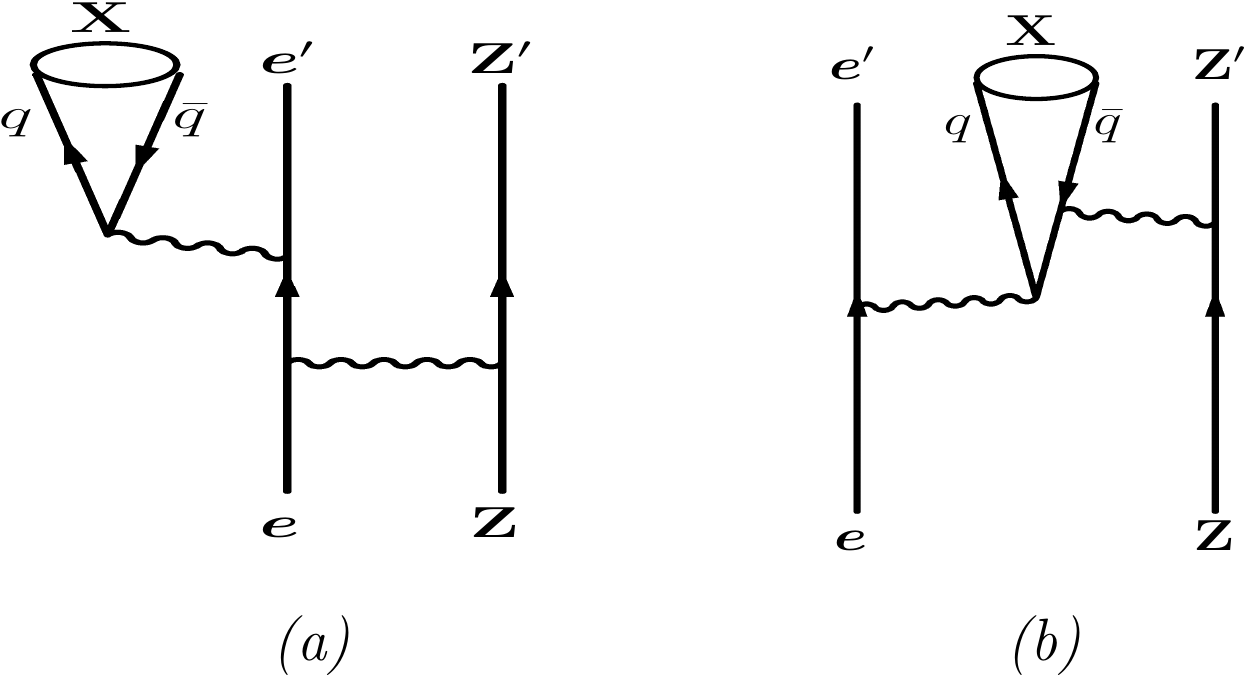}
\caption{ In high-energy electron scattering off an atomic nuclei with
  charge $Z$, the bremsstrahlung-like reaction may lead to the
  emission of a photon, with the creation of a $q\bar q$ pair \cite{x17JLAB,x17JLAB1}.  A QED
  meson may be produced if the $\sqrt{s}$ of the emitted $q\bar q$
  pair coincides with the eigenenergies of the QED meson, $X$, by a
  single photon emission as in ($a$), or by the fusion of two photons,
  as in ($b$).  }
\label{fig2}
\end{figure}

As an illustrative case for a $q\bar q$ pair to be produced and to interact in the QED
interaction alone, we have examined the above reaction,
$e^+$+$e^-$$\to$ $ q$+$\bar q$ below the pion mass energy --- as good
examples.  There are actually other circumstances in which a $q\bar q$
pair can also be created with a total pair energy lower than the QCD
pion mass gap in other reactions, and the produced $q\bar q$ pair can
interact in the QED interaction alone without the QCD interaction.
For example, in the proposed experiment at JLAB with the collision of
an electron on a nuclear charge $Z$ \cite{x17JLAB,x17JLAB1}, a single
or many $q\bar q$ pairs may be produced by the bremsstrahlung-type
reaction as shown in Fig. 2($a$) or in the fusion of two virtual
photons as shown in Fig.\ 2($b$):
\begin{eqnarray}
e+ Z \to e' + Z' + q + \bar q.
\end{eqnarray}
In such a reaction, in addition to the production of QCD mesons which
are composite QCD-confined $q\bar q$ bound states, a QED meson may
also be produced (shown schematically as the $X$ particle in Fig.\ 2),
if there is a QED-confined $q\bar q$ meson state at the appropriate
eigenenergy below $m_\pi$.  The produced QED may be detected by its
decay products of an $e^+ e^-$ pair, a pair of real photons, or a pair
of virtual photons as two pairs of dileptons.

  \begin{figure} [h]
\centering
\vspace*{-0.cm}\hspace*{-0.6cm}
\includegraphics[scale=0.65]{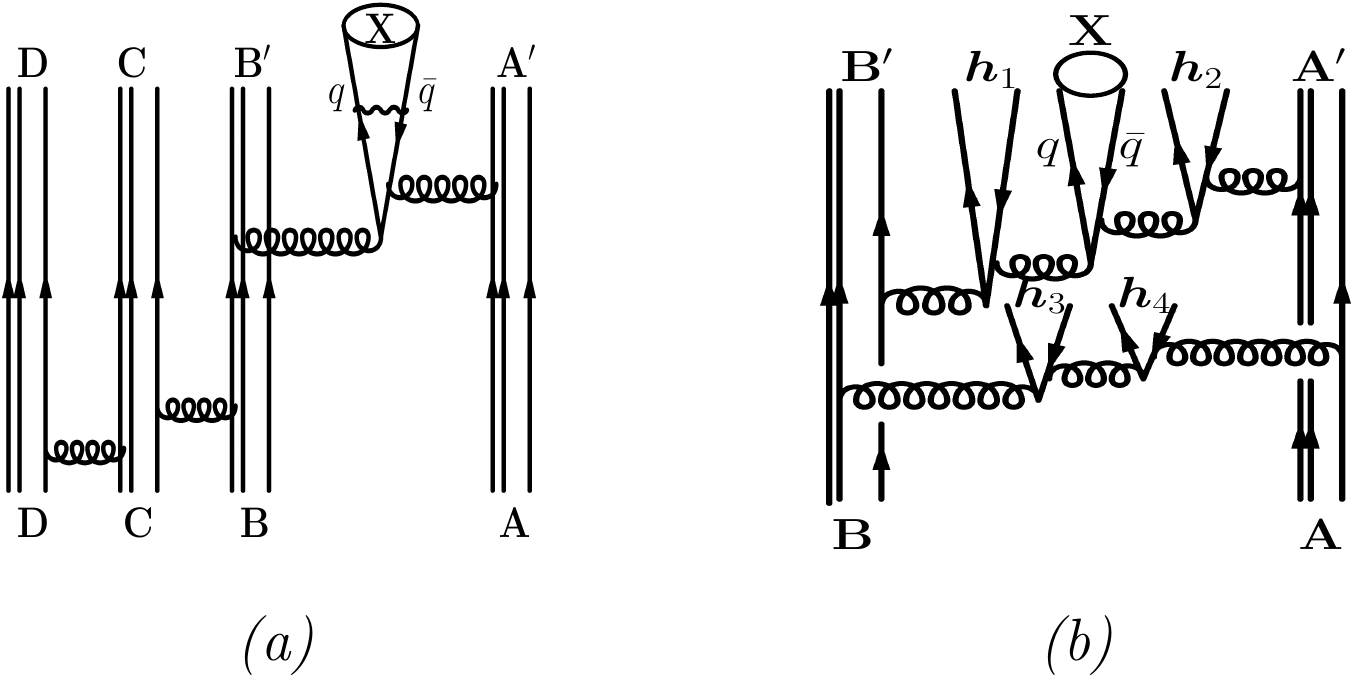}
\caption{ ($a)$ $q\bar q(X)$ production by the fusion of two virtual
  gluons in the de-excitation of a highly-excited $^4$He$^*({\rm
    ABCD}) \to  ^4$He(ground state)({\rm A'B'CD}) + $q$$\bar q$(X)
  with the 
  fusion of two virtual gluons between B and A.  $(b)$
  $q\bar q(X)$ production in hadron $q\bar q$-hadron or a
  nucleus-nucleus collision by A + B $\to$ A' + B' + $(q\bar q)^n \to$
  A' + B' +$ \sum\limits_i h_i + q \bar q(X)$. }
\label{fig3}
\end{figure}

In another example as shown in Fig.\ \ref{fig3}($a$), we show an
excited state of $^4$He nucleus, that has been prepared in a
low-energy $p$+$^3$H proton fusion reaction.  For example, the $J^\pi
I=$ 0$^-0$ excited state at 20.02 MeV of $^4$He can be formed by
placing a proton in the stretched-out $p$ state interacting with the
$^3$H core in Fig.\ \ref{fig3}($a$) \cite{Kra16,Kra19,Kra21}.  The
de-excitation of the 20.02 MeV $J^\pi I=0^-0$ $^4$He$^*$ excited state
to the $^4$He$_{\rm g.s.}$ ground state can occur by the proton
emitting a virtual gluon which fuses with the virtual gluon from the
$^3$H core, to lead to the production of a $q\bar q$ pair as shown in
Fig.\ 3($a$).  Note that in Fig.\ \ref{fig3}($a$), the projectile
A and target $^3$H triton (BCD) are initially colorless prior to the
collision. After each emitting a gluon such that the fusion of the two
gluons lead to the production of the colorless QED meson, the
scattered projectil nucleon A$'$ and  the triton target (BCD$'$) are colored objects.  The colored projectil nucleon A$'$
and the colored triton target (BCD$'$) can become colorless by exchanging a
gluon and fusing into the colorless ground state of $^4$He. Only by
the exchanging a gluon and fusing into the $^4$He ground state will
the scattered proton A$'$ and $^3$H$'$ lead to a colorless observable
final state.  Therefore, the emission of the the QED meson is
necessarily accompanied by the fusion of the proton and the triton
core.  It is interesting to note that there appears to be a similar
requirement that the emission of the ABC resonance at $\sim$310 MeV is
accompanied by the fusion of the reacting nucleons
\cite{Aba60,Boo61,Ban73,Adl11,Bas17,Kom18}, as will be discussed in
Section 7F.

Gluons reside in the color-octet ${\bb 8}$ representation.  In the
fusion of gluons in the reaction $ g + g \to q+ \bar q $, the fusion
gives rise to color multiplets as
\begin{eqnarray}
\bb 8 \otimes \bb 8 = {\bb 1} \oplus {\bb 8} \oplus {\bb 8} \oplus
    {\bb {10}} \oplus {\bb {10}} \oplus {\bb {27}},
\end{eqnarray}
which contains the color-singlet component, $\bb 1$, among other color
multiplets.  There is thus a finite probability in which a
color-singlet $q\bar q$ pair can be produced by gluon fusion.  At very
low energies in the de-excitation of the $^4$He$^*$ nucleus, this
gluon fusion process may lead to the production of a QED meson, if a
QED meson eigenstate $X$ exists at the appropriate energy, as labeled
schematically in Fig.\ 3($a$).  

In high-energy hadron-hadron collisions at CERN
\cite{Chl84,Bot91,Ban93,Bel97,Bel02pi,Bel02,Per09} and nucleus-nucleus
collisions at Dubna \cite{Abr12,Abr19}, many $q\bar q$ pairs may be
produced as depicted schematically in Fig.\ 3($b$),
\begin{eqnarray}
A + B \to A' + B' + (q\bar q)^n .
\end{eqnarray}
The invariant masses of most of the produced $q\bar q$ pairs will
exceed the pion mass, and they will materialize as QCD mesons and
labeled as $h_i$ in Fig.\ 2($b$).  However, there may remain a small
fraction of the color-singlet $[q\bar q]^1$ pairs with an invariant
masses below $m_\pi$.  The $q \bar q$ pairs in this energy range below
mass $m_\pi$ allow the quark and the antiquark to interact in QED
alone, without the QCD interaction, to lead to possible QED meson
eigenstates labeled schematically as $q\bar q(X)$ in Fig.\ \ref
{fig3}($b$).

 In other circumstances in the deconfinement-to-confinement phase
 transition of the quark-gluon plasma in high-energy heavy-ion
 collisions, a deconfined quarks and a deconfined antiquark in close
 spatial proximity can coalesce to become a $q\bar q$ pair with a pair
 energy below the pion mass, and they can interact in QED alone to
 lead to a possible QED meson, if there is QED-confined eigenstate in
 this energy range.
  
\section{Quarks are confined in QED in (1+1)D}

Having settled on the possibility for the production of a quark and an
antiquark interacting in QED alone, we proceed to explore whether
there can be confined $q\bar q$ eigenstates for quarks interacting in
QED in (3+1)D.  From the discussions in the Introduction, we are well
advised to limit our attention only to those $q\bar q$ systems which
potentially will be bound and confined.  For this reason, we shall
first examine the case of (1+1)D space-time and then proceed
subsequently to the case of (3+1)D space-time.

Schwinger showed previously that a massless fermion and its
antifermion interacting in an Abelian QED U(1) gauge interaction with
a coupling constant $g_\2d$ in (1+1)D are bound and confined as a
neutral boson with a mass \cite{Sch62,Sch63},
\begin{eqnarray}
m=\frac{g_\2d}{\sqrt{\pi}},
\label{eq6}
\end{eqnarray}
where the coupling constant $g_\2d$ has the dimension of a mass.  We
take the convention that the coupling constant $g_{\2d}$ to describe
the interaction between a quark and an antiquark is positive for the
QED and QCD interactions.  We call this mechanism the Schwinger
confinement mechanism$^{2}$.  Such a  Schwinger confinement mechanism
occurs for massless fermions interacting in Abelian gauge interactions
of all strengths of the coupling constant, including the interactions
with a weak coupling strength (e.g. QED), as well as the interactions
with a strong coupling strength (e.g. QCD).  From the works of
Coleman, Jackiw, and Susskind \cite{Col75,Col76}, we can infer  further that the Schwinger confinement mechanism persists even for massive quarks  in (1+1)D.   It is in fact a
general property for fermions and (quarks) of all masses  interacting in the Abelian
and quasi-Abelian gauge interaction with  all coupling strengths   in (1+1)D.

 The masses of light quarks   are about  a
few MeV \cite{PDG19}, and they can be first approximated as massless.  
A light quark and its antiquark can be produced and can
interact in a U(1) Abelian QED gauge interaction alone, as discussed in
the last section.   Because they cannot be isolated, they reside predominantly in (1+1)D.
The conditions for which the Schwinger confinement mechanism can be applicable are met by the light quark and its antiquark interacting in the QED interaction. 
 We can therefore apply the Schwinger confinement
mechanism to light quarks interacting in the QED interaction to infer
that a light quark and its antiquark  are bound and confined in QED in (1+1)D.

It is instructive to review here the Schwinger confinement mechanism
how light quarks approximated as massless are confined in QED in
(1+1)D.  We shall subsequently discuss how the confinement property
persists when quarks are massive.  We envisages the vacuum of the
interacting system as the lowest-energy state consisting of quarks
filling up the (hidden) negative-energy Dirac sea and interacting with
the QED interaction in (1+1)D space-time with coordinates
$x=(x^0,x^1)$.  The vacuum is defined as the state that contains  
no valence
quarks as particles above the Dirac sea and no valence antiquarks as
holes below the Dirac sea.  We wish to describe in detail how an
applied local QED disturbance in the form of $A^\mu(x)$ will generate
self-consistently stable collective particle-hole excitations of the
quark-QED system as in a quantum fluid.  Subject to the applied
disturbing gauge field $A^\mu(x)$ with a coupling constant $g_\2d$ in
(1+1)D, the massless quark field $\psi(x)$ satisfies the Dirac
equation,
\begin{eqnarray}
\gamma_\mu ( p^\mu - g_{\2d} A^\mu) \psi =0.
\label{eq7}
\end{eqnarray}
The applied gauge field $A^\mu(x)$ instructs the quark field $\psi(x)$
how to move.  From the motion of the quark field $\psi(x)$, we can
obtain the induced quark current $j^\mu(x)=\bar \psi (x) \gamma^\mu
\psi(x)$.  If we consider only the sets of states and quark currents
that obey the gauge invariance by imposing the Schwinger modification
factor to ensure the gauge invariance of the quark Green's function,
the quark current $j^\mu(x) $ at the space-time point $x$ induced by
the applied $A^\mu(x)$ can be evaluated.  After the singularities from
the left and from the right cancel each other, the gauge-invariant
induced quark current $ j^\mu(x)$ is found to relate explicitly to the
applied QED gauge field $A^\mu(x)$ by \cite{Sch62,Sch63,Won94}
\begin{eqnarray}
j^\mu (x)  = -\frac{g_\2d}{\pi }\left ( A^\mu  (x) - \partial ^\mu \frac{1}{\partial _\eta \partial ^\eta} \partial _\nu A ^\nu (x) \right ).
\label{eq8}
\end{eqnarray}
The derivation of the above important equality can be found in
\cite{Sch63} and Chapter 6 of \cite{Won94}.  We can provide an
intuitive understanding of the above relation between applied
$A^\mu(x)$ and the induced $j^\mu(x) $ in the following way.  The induced
current $j^\mu(x)$ is a function of the applied gauge field
$A^\mu(x)$, and $j^\mu(x)$ gains in strength as the applied strength
of the gauge field $A^\mu(x)$ increases.  So, the linear relation in Eq.\ (\ref{eq8}) 
between $j^\mu(x)$ and $A^\mu(x)$ and its coupling constant $g_\2d$,
is a reasonable concept.  The proportional property involving the
coupling constant $g_\2d$ in Eq.\ (\ref{eq8}) can also inferred by
dimensional analysis and the negative sign from simple intuitive
physics arguments \footnote{ From the Dirac equation (\ref{eq7}), the
quantities $g_\2d A^0$ and $p^0$ have the same dimension.  The
quantity $p^0$ has dimension 1/length (in high energy units and conventions with
$\hbar=1$, $c$=1) which is also the dimension of the particle density
$j^0$ in (1+1)D. Hence, from dimensional analysis, $g_\2d A^\mu$ has
the same dimension as $j^\mu$ to establish the proportional relation
of Eq. (\ref{eq8}).  We can infer the sign of their relation by
introducing a positive Coulomb field $A^0(x^1)$ from a positive charge
source at $x^1$, which will induce locally a negative charge density
$j^0(x^1)$.  Hence, the sign of their proportional relation has a
negative sign on the right hand side of Eq.\ (\ref{eq8}).  }.
However, $j^\mu(x)$ is a gauge-independent quantity, whereas
$A^\mu(x)$ is a a gauge-dependent quantity.  The right hand side must
be made gauge-independent and gauge invariant.  The additional term on
the right-hand side of the above relation consists of the linear
function of $A^\mu(x)$ and a unique functional combination of partial
derivatives that ensures the gauge-independence and gauge-invariance
between $j^\mu(x)$ and $A^\mu(x)$.  The gauge invariance of the
relationship between $j^\mu(x)$ and $A^\mu(x)$ can be easily
demonstrated directly upon a change of the gauge in $(A^\mu)' (x) \to
A^\mu(x) - \partial ^\mu \Lambda(x)$, for any local function of
$\Lambda(x) $ in Eq.\ (\ref{eq8}).

The induced gauge-invariant quark current $j^\mu(x)$ in turn
generates a new gauge fields $\tilde A^\mu(x)$ through the Maxwell
equation,
\begin{eqnarray}
\hspace*{-0.3cm}\partial_\mu F^{\mu \nu}(x) =\partial_\mu  \{ \partial^\mu \tilde A^\nu (x)  -  \partial^\nu \tilde A^\mu (x) \}=g_\2d j^\nu (x) = g_\2d \bar \psi (x) \gamma^\nu \psi (x) .
\label{eq9}
\end{eqnarray}
A stable collective particle-hole excitation of the quark system
occurs, when the initial applied $A^\mu(x)$ gives rise to the induced
quark current $j^\mu$ which in turn leads to the new gauge field
$\tilde A^\mu(x)$ self-consistently.  We impose this self-consistency
condition of the gauge field, $A^\mu(x)=\tilde A^\mu(x) $, by
substituting the relation (\ref{eq8}) into the Maxwell equation
(\ref{eq9}).  We get both $j^\mu(x)$ and $A^\mu(x)$ satisfy the
Klein-Gordon equation
\begin{eqnarray}
\hspace*{-0.8cm}\partial _\nu \partial ^\nu A^\mu (x) + \frac{g_\2d^2}{\pi }A^\mu(x) =0,  ~~{\rm and~} ~~
\partial _\nu \partial ^\nu j^\mu (x) +\frac{g_\2d^2}{\pi }j^\mu (x) =0,\!
\end{eqnarray}
 for a bound and confined boson with a mass $m=g_\2d/\sqrt{\pi}$ as
 given by Eq.\ (\ref{eq6}).  Hence, massless quarks in an Abelian QED
 U(1) gauge in (1+1)D are bound and confined as a neutral boson with a
 mass \cite{Sch62,Sch63}.  In reaching the above Klein-Gordon
 equations for $A^\mu$ and $j^\mu$ for a bound boson, the second term
 on the right hand side of Eq.\ (\ref{eq8}) plays a crucial role, and
 such a term arises from the gauge invariance of the current.  From
 this viewpoint, the confinement of the fermion-antifermion pair or
 the quark-antiquark into a boson owes crucially to the dynamics
 originating from the gauge invariance of the current, as emphasized
 by Schwinger \cite{Sch62,Sch63}.
 
Light quarks have small but non-zero rest masses.  From the Particle
Group Tables, we have $m_u$=2.16$^{+ 0.49}_{-0.26}$ MeV, and
$m_d$=4.67$^{+0.48}_{-0.17}$ MeV \cite{PDG19}.  It is necessary to
examine the effects of quark rest masses on the property of quark
confinement because quarks are not exactly massless.  How good is the
massless approximation for light quarks in QCD and in QED in (1+1)D?
Do the quark masses affect the quark confinement property of the
Schwinger mechanism in (1+1)D?
   
Coleman, Jackiw, and Susskind \cite{Col75,Col76} studied how the
fermion masses affect the Schwinger confinement mechanism in (1+1)D.
They showed that the Schwinger confinement mechanism persists even for
massive fermions.  Therefore, when we apply the results of Coleman,
Jackiw, and Susskind for fermions to quarks, we reach the conclusion
that a quark and its antiquark of the same flavor are always confined
in QED in (1+1)D, no matter whether they have a small or large quark
mass.  As the Schwinger confinement mechanism occurs for all strength
of the coupling constant, we can infer that a quark and an antiquark
are confined as a neutral $q \bar q$ boson in QED in (1+1)D for all
flavors of quarks, whatever their masses or the strengths of the
coupling constant.

While the confinement of a quark-antiquark pair is quite a general
result in (1+1)D for quarks with different flavors and coupling
strengths of the gauge interaction, the effect of the quark masses and
coupling constants manifest themselves in the spectral properties,
boson properties, and the number of bound states of the confined
bosons, depending on the comparison between the strength of the
coupling constant $g_\2d$ and the quark mass $m_q$.

In the Schwinger confinement mechanism in (1+1)D, the coupling
constant $g_\2d$ has the dimension of a mass and the coupling constant
$g_\2d$ can be compared directly with the quark mass $m_q$.  A quark
of mass $m_q$ interacting in a gauge interaction with a gauge coupling
constant $g_\2d$ in (1+1)D belongs to the strong 2D-coupling regime if
$ g_\2d >> m_q $, and it belongs to the weak 2D-coupling regime if $
g_\2d<< m_q $ \cite{Col76}.  We have added the label ``2D'' to the
coupling regime to refer to the coupling regime in the Schwinger
confinement mechanism in (1+1)D.

Coleman calculated the spectra of the quark systems in both the strong
and the weak 2D-coupling regimes in \cite{Col76}.  In the strong
2D-coupling regime, quark systems always contain at least a stable
meson, $m=g_\2d/\sqrt{\pi}=m_{\rm meson}$.  It display the properties
of a meson with a weak-strength interaction between mesons.  There
will be a weakly bound collection of $n$-meson states of mass $\sim nm
$ plus perturbation contributions that depend on the quark mass $m_q$.
The spectra can be obtained by non-relativistic reasoning.  In the
weak 2D-coupling regime, quarks display the theory of a weak Coulomb
interaction and the spectrum can be calculated by the
mass-perturbation theory. Coleman also calculated the number of stable
particles for the weak 2D-coupling regime and generalized the quark
system to two flavors \cite{Col76}.  Coleman's results are useful for
future studies of the spectral and boson properties of the
QED-confined and QCD-confined $q\bar q$ bosons in (1+1)D which may
find applications in the physical (3+1)D.

It is important not to confuse the strong and weak 2D-coupling regimes
in the Schwinger confinement mechanism in (1+1)D with the strong and
weak coupling regimes in quark confinement in static quark lattice
gauge calculations in (3+1)D.  In lattice gauge calculations in
(3+1)D, a static quark and a static antiquark belong to the strong
coupling regime with confined quarks for $g_\4d^2/4\pi >> \alpha_{\rm
  crit}$, and they belong to the weak coupling regime with deconfined
quarks if $g_\4d^2/4\pi << \alpha_{\rm crit}$
\cite{Wil74,Kog75,Man75,Pol77,Pol87,Ban77,Gli77,Pes78,Dre79,Gut80,Kon98,Mag20,Arn03,Lov21},
where $\alpha_{\rm crit}$=0.988989481 \cite{Arn03,Lov21}.  For quarks
interacting in the QED interaction for which
$\alpha^{\qed}=(g_\4d^{\qed})^2/4\pi$= 1/137, and $\alpha^\qed <
\alpha_{\rm crit}$, we can conclude that a static quark and a static
antiquark in QED interactions belongs to the deconfined weak-coupling
regime in (3+1)D static quarks lattice gauge calculations.  So, in
(3+1)D a static quark and a static quark are not confined in compact
QED lattice gauge calculations.  In contrast, in the Schwinger
confinement mechanism for QED in (1+1)D, a quark and an antiquark are
always confined, whatever the magnitude of the coupling constant.

In the Schwinger confinement mechanism, it is instructive to compare
the strength of coupling constant relative to the quark masses for
quarks interacting with its antiquark in the QED interaction in the
physical (3+1)D space-time.  We can carry out similar comparison when
we approximate non-Abelian QCD interaction as a quasi-Abelian
interaction, as described in the next Section.  The QED and QCD
coupling constants, $g_\4d$, are known in the physical (3+1)D world,
and we need to know their corresponding (1+1)D coupling constant,
$g_\2d$, when we approximate a flux tube with a radius $R_T$ as a
string without a structure.  We showed previously
\cite{Won09,Won10,Kos12,Won20,Won22} that in such an idealization,
$g_\2d^\lambda $ in (1+1)D is related to $g_\4d^\lambda$ in (3+1)D by
\begin{eqnarray}
g_\2d^\lambda=\left (\frac{1}{\sqrt{\pi}  R_T} \right )g_\4d^\lambda= \frac{\sqrt{4\alpha_\lambda}}{R_T},
~~~ ~~ \lambda=~~\begin{cases} 0 & \text{QED} \cr 1  & \text{QCD} \cr \end{cases},
\label{eq11}
\end{eqnarray}
where $R_T$ is the radius of the flux tube,
$\alpha_{\lambda}=(g_\4d^\lambda)^2/4\pi$, and $\lambda$ is the label
for the interaction.  
The qualitative consistency of such a relationship can be checked by dimensional analysis. 
By such a ``dimensional transmutation''
relation, the information on the structure of the flux tube radius
$R_T$ in (3+1)D is stored in the coupling constant $g_\2d$ for its
subsequent dynamics in (1+1)D.

 For the QCD interaction, the value of $R_T$ for a QCD open string has
 been found to be $R_T$=0.4 fm, and $\alpha_\qcd=0.67$ from the masses
 of the $\pi^0$, $\eta$, and $\eta'$ \cite{Won20}, and $R_T$=0.35 fm
 from the average transverse momentum of produced hadrons in $pp$
 collisions \cite{Won10} in the open-string description.  We have
 therefore
\begin{eqnarray}
g_\2d^\qcd = 807.5 ~{\rm to}~ 922.84 ~{\rm MeV}. 
\label{eq12}
\end{eqnarray}
Thus, as far as Schwinger confinement mechanism in (1+1)D is
concerned, $g_\2d^\qcd >> m_u, m_d, m_s$ and $u,d,$ and $s$ quarks
interacting in the QCD interaction belongs to the strong 2D-coupling
regime, whereas $g_\2d^\qcd << m_c, m_b, m_t$ and $c, b,$ and $t$
quarks interacting in QCD belong to the weak 2D-coupling regime.

We do not know the flux tube radius $R_T$ for the QED interaction.
The meager knowledge from the possible QED meson spectrum and the
anomalous soft photons suggests the flux tube may be an intrinsic
property of a quark that may be independent or weakly dependent on the
interactions, as such an assumption gives a reasonable description of
the X17, E38, and anomalous soft photon masses \cite{Won20}, subject
to future amendments as more experimental data become available.  So,
for the QED interaction with $\alpha_\qed$=1/137 and we get for $R_T$
$\sim$ 0.35 to 0.4 fm,
\begin{eqnarray}
g_\2d^\qed  \sim 84.28~~\text{to}~~ 96.32 ~{\rm MeV},
\end{eqnarray}
indicating $g_\2d^\qed >> m_u, m_d$.  Light $u$ and $d$ quarks
interacting in QED belong to the strong 2D-coupling regime whereas
$g_\2d^\qed \sim m_s$ and $g_\2d^\qed << m_c, m_b, $ and $ m_t$, the
$s$ quark has a mass comparable to the $g_\2d^\qed$.  The $c, b,$ and $t$
quarks interacting in QED belongs to the weak 2D-coupling regime.
Consequently, with light quarks interacting in QED belonging to the
strong 2D-coupling regime, $g_\2d^\qed>> m_q$, the quarks mass will
only be a perturbation in the light-quark QED meson spectrum.

 The QED-confined meson states in (1+1)D can be viewed in two
 equivalent ways \cite{Won10,Won20,Won22}.  It can be depicted
 effectively as a QED-confined one-dimensional open string, with a
 quark and an antiquark confined at the two ends of the open string
 subject to an effective linear two-body confining interaction.  A
 more basic and physically correct picture depicts the boson as the
 manifestation of a collective particle-hole excitation from the Dirac
 sea involving the coupled self-consistent responses of quark current
 $j^\mu$ and the gauge field $A^\mu$.  The quark confinement arises
 because the quark current $j^\mu$ and the gauge field $A^\mu$ depend
 on each other self-consistently and such a self-consistency leads to
 a stable and self-sustainable space-time variations of both the
 current $j^\mu$ and the gauge field $A^\mu$.  Specifically, through
 the Dirac equation for quarks, a space-time variation of the gauge
 field $A^\mu$ leads to a space-time variation in the quark current
 $j^\mu$, which in turn determines the space-time variation of the
 gauge field $A^\mu$ through the Maxwell equation
 \cite{Sch62,Sch63,Won94}.  As a consequence of such self-consistent
 dependencies, a quantized and locally-confined space-time collective
 variations of the quark current $j^\mu$ and the QED gauge field
 $A^\mu$ can sustain themselves indefinitely at the lowest eigenenergy
 of the QED-confined $q\bar q$ state in a collective motion as a
 one-dimensional open string with a mass, when the decay channels for
 the confined collective state are turned off for such an examination
 \cite{Won10,Won20}.  From such a viewpoint, the Schwinger confinement
 mechanism is a many-body phenomenon containing dynamical quark
 effects beyond the potential interaction between a static quark and a
 static antiquark alone.

\section{Generalizing the Schwinger confinement mechanism  in (1+1)D  \\
 from quarks in QED to  quarks in (QCD+QED) }
 
\subsection{Question of the compactification of QCD and QED from (3+1)D
  to (1+1)D} 

 From the works of Schwinger, Coleman, Jackiw, and Susskind
 \cite{Sch62,Sch63,Col75,Col76}, it can be inferred that the Schwinger
 confinement mechanism applies generally to all Abelian gauge theories
 in (1+1)D with fermions and quarks of all masses and coupling
 strengths.  Even though QCD is a non-Abelian gauge theory, if the
 non-Abelian QCD can be approximated as an Abelian or quasi-Abelian
 gauge theory in (1+1)D, then the Schwinger confinement mechanism will
 apply also to quarks in QCD dynamics.  Indeed, many features of the
 QCD mesons (such as quark confinement, meson states, and meson
 production), mimick those of the Schwinger model for the Abelian
 gauge theory in (1+1)D, as noted early on by Bjorken, Casher, Kogut,
 and Susskind \cite{Bjo73,Cas74}.  Such generic string feature in
 hadrons was first recognized even earlier by Nambu \cite{Nam70,Nam74}
 and Goto \cite{Got71}.  They indicate that in matters of confinement,
 quark-antiquark bound states and hadron production, an Abelian
 approximation of the non-Abelian QCD theory is a reasonable concept.
 Various nonlocal maximally Abelian projection methods to approximate
 the non-Abelian QCD by an approximate Abelian gauge theory have been
 suggested by t'Hooft \cite{tHo80}, Belvedere $et~al.$ \cite{Bel79},
 Sekeido $et ~al.$ \cite{Sei07}, and Suzuki $et~ al.$ \cite{Suz08}.
 Suganuma and Ohata \cite{Sug21} investigate Abelian projected QCD in
 the maximally Abelian gauge, and find a strong correlation between
 the local chiral condensate and magnetic fields in both idealized
 Abelian gauge systems and Abelian projected QCD.
 
Before we embark on the study of the dynamics of QCD and QED in the
lower-dimension (1+1)D space-time, it is necessary to know how the
dynamics of the lowest states in QCD and QED can be compactified from
(3+1)D to (1+1)D.  For the QCD dynamics in (3+1)D, t'Hooft showed that
in a gauge theory with an SU($N_{\rm color}$) gauge group approximated
as a U($N_{\rm color}$) gauge group in the large $N_{\rm color}$
limit, planar Feynman diagrams with quarks at the edges dominate, and
the QCD dynamics in (3+1)D can be well approximated as QCD dynamics in
(1+1)D \cite{tHo74a,tHo74b}.  Numerical lattice calculations for a
quark and antiquark system supports such concepts as they exhibits a
flux tube structure \cite{Hua88,Bal05,Cos17,Car13,Bic18,Bic19}.  Thus,
the compactification of QCD in (3+1)D to QCD in (1+1)D is a reasonable
concept.

For a quark and an antiquark interacting in the QED interaction in
(3+1)D, however, the compactification from (3+1)D to (1+1)D is much
more complicated and must be examined carefully.  It has been known
for a long time since the advent of Wilson's lattice gauge theory that
a static fermion and a static antifermion in (3+1)D in compact QED
interaction has a strong-coupling confined phase and a weak-coupling
deconfined phase \cite{Wil74}.  The same conclusion was reached
subsequently by Kogut, Susskind, Mandelstam, Polyakov, Banks, Jaffe,
Drell, Peskin, Guth, Kondo and many others
\cite{Wil74,Kog75,Man75,Pol77,Pol87,Ban77,Gli77,Pes78,Dre79,Gut80,Kon98,Mag20}.
The transition from the confined phase to the deconfined phases occurs
at the coupling constant $\alpha_{\rm crit}=g_{\rm
  crit}^2/4\pi$=0.988989481 \cite{Arn03,Lov21}.  The magnitude of the
QED coupling constant, the fine-structure constant $\alpha_c$=1/137,
places the QED interaction between a quark and an antiquark as
belonging to the weak-coupling deconfined regime.  Therefore, a static
quark and a static antiquark are deconfined in lattice gauge
calculations in compact QED in (3+1)D.  However, no such fractional
charges have ever been observed, even though there exists no physical
law to forbid a quark and an antiquark to interact in the QED
interaction alone.  On the other hand, according to Schwinger,
Coleman, Jackiw, and Susskind \cite{Sch62,Sch63,Col75,Col76}, there is
a confined QED regime in (1+1)D, and from the work of Polyakov, there
is in addition a confined regime for quarks in (2+1)D in compact QED
\cite{Pol77}, for all gauge coupling interaction strengths.  To study
the importance of the Schwinger confinement mechanism on quark
confinement in QED in (3+1)D, we can combine Polyakov's transverse
confinement of quarks in compact QED in (2+1)D with the the
longitudinal confinement of Schwinger's dynamical quarks in (1+1)D QED
in our investigation on the compactification in (3+1)D, as is carried
out in \cite{Won22c}.  We construct a stretch (2+1)D model to study
the importance of the Schwinger confinement mechanism in (3+1)D
\cite{Won22b,Won22c}.  We find there that the Polyakov's transverse
confinement can effectively maintain a flux tube structure and the
Schwinger longitudinal confinement can lead the compactification from
$q\bar q$ in (3+1)D in QED to $q\bar q$ in (1+1)D as reviewed in
Section 6.  In view of the observation that quarks cannot be isolated
and can be considered to reside predominantly  in (1+1)D space-time, we can proceed
to consider the compactification as a working hypothesis and examine
its consequences as in \cite{Won20}, while we await the theoretical
resolution on the question of QED compactification from (3+1)D to
(1+1)D.

\subsection{QCD and QED dynamics and bosonization  in (1+1)D}

Because of the three-color nature of the quarks, the quark current
$j^\mu(x)$ and the gauge field $A^\mu(x)$ are 3$\times$3 color
matrices with 9 matrix elements at space-time point
$x^\mu$=$(x^0,x^1)$ in (1+1)D space-time.  The 9 matrix elements in
the color space can be separated naturally into color-singlet and
color-octet subgroups of generators.  Specifically, quarks reside in
the $\bb 3$ representation and antiquarks reside in the $\bb 3^*$
representation, and they form a direct product of $\bb 3 \otimes {\bb
  3^*}$=$\bb 1 \oplus \bb 8$, with a color-singlet $\bb 1$ subgroup
and a color-octet $\bb 8$ subgroup.  The quark current $j^\mu$ and the
QED and QCD gauge field $A^\mu$ are 3$\times$3 color matrices which
can be expanded in terms of the nine generators of the U(3) group,
\begin{eqnarray}
\hspace*{-0.6cm}j^\mu = \sum_{i=0}^8 j^\mu_i t^i,~~A^\mu = \sum_{i=0}^8 A^\mu_i t^i,~~~
t^0=\frac{1}{\sqrt{6} }
\begin{pmatrix}
1 & 0 & 0 \cr
0 & 1 & 0 \cr
0 & 0 & 1 \cr
\end{pmatrix},
\label{eq14}
\end{eqnarray}
where $t^0$ is the generator of the U(1) color-singlet subgroup and
$t^1,t^2,...,t^8$ are the eight generators of the SU(3) color-octet
subgroup, with  
 $A_\qed^\mu(x)$=$ A^\mu_ 0
(x)t^0$ and 
$A_\qcd^\mu(x)$=$\sum_{a=1}^8 A^\mu_ a
(x)t^a$.  We use the
convention of summation over repeated indices, but the summation
symbol and indices are occasionally written out explicitly to avoid
ambiguities.  The current $j^\mu$ and gauge field $A^\mu$ also possess
the additional flavor label $f$ and the interaction label $\lambda$.
For brevity of notations, the indices $a$, $f$, and $\lambda$ in
various quantities are often implicitly understood except when they
are needed.  The coupling constants $g_f^a$ in (1+1)D are given
explicitly by
\begin{subequations}
\begin{eqnarray}
\label{qedcc} 
&&\hspace{-0.9cm}g_u^0\!=\!-Q_u\,g_{\2d}^{\qed}\!\!\!\!,\!
~~~g_d^0\!=\!-Q_d\,g_{\2d}^{\qed} {\rm~~~~for~QED},
\\ &&\hspace{-0.9cm}g_{\{u,d,s\}}^{\{1,..,8\}}=Q_{\{u,d,s\}}^{{}^{\qcd}}\,g_{\2d}^\qcd
   {\rm~~~~~~~~~~~~~~for~QCD},
\label{qcdcc}
\end{eqnarray}
\end{subequations}
where we have introduced the charge numbers
\begin{eqnarray}
&&Q_u^{\qed} =2/3, Q_d^{\qed}=-1/3, Q_s^{\qed}=-1/3, ~~~
Q_u^{\qcd} =Q_d^{\qcd}=Q_s^{\qcd} = 1,
\\
&&
Q_{\bar q} ^{\{\qcd,\qed\}} = -Q_q ^{\{\qcd,\qed\}}.
\nonumber
\end{eqnarray}
  The Lagrangian density for the system is \cite{Pes95}
\begin{subequations}
\begin{eqnarray}
{\cal L}&&=\bar \psi (i \sD)\psi - \frac{1}{4}F_{\mu \nu} F^{\mu \nu}-
m \bar \psi \psi, \\
i\sD &&=
 \gamma^\mu ( i\sd~ + g A_\mu),
\end{eqnarray}
\end{subequations}
where for the non-Abelian QCD dynamics ,
\begin{eqnarray}
F_{\mu \nu} &&= \partial_\mu A_\nu - \partial_\nu A_\mu -i g [A_\mu,
  A_\nu], ~~~~~~F_{\mu \nu}=F_{\mu \nu}^a t^a,
\end{eqnarray}
and the equation of motion for the gauge field $A_\mu$ is
\begin{eqnarray}
\label{Max4}
D_\mu F^{\mu \nu}&& = \partial_\mu F^{\mu \nu} -i g [A_\mu,
  F^{\mu\nu}] = g j^{\nu}, ~~~~~ j^\nu= j^{\nu \, a} t^a, ~~~~j^{\nu
  \,a} =2 ~{\rm tr}~ {\bar \psi}_f \gamma^\nu t^a \psi_f.
\end{eqnarray} 

We wish to search for bound states arising from the QED and QCD
interactions in (1+1)D by the method of bosonization
\cite{Col75,Col76,Cas74,Hal75,Wit84,Gep85,Abd01,Fri93,Gro96}, which
consists of introducing boson fields $\phi^a$ to describe an element
$u$ of the U(3) group and showing subsequently that these boson
fields lead to stable bosons with finite or zero masses.  As in any
method of bosonization, the method will succeed for systems that
contain stable and bound boson states with relatively weak residual
interactions between the bosons.  Thus, not all the degrees of freedom
available to the bosonization technique will lead to good boson states
with these desirable properties.

An element of the U(1) subgroup of the U(3) group can be represented
by the boson field $\phi^0$
\begin{eqnarray}
u = \exp\{ i 2 \sqrt{\pi} \phi^0 t^0 \}.
\end{eqnarray}  
Such a bosonization poses no problem as it is an Abelian subgroup.  It
will lead to a stable boson as in the Schwinger confinement mechanism .

On the other hand, the U(3) gauge interactions under consideration
contain the non-Abelian color SU(3) interactions.  Consequently the
bosonization of the color degrees of freedom should be carried out
according to the method of non-Abelian bosonization which preserves
the gauge group symmetry \cite{Wit84}.  While we use non-Abelian
bosonization for the U(3) gauge interactions, we shall follow
Coleman to treat the flavor degrees of freedom as independent degrees
of freedom \cite{Col75,Col76}.  This involves keeping the flavor
labels in the bosonization without using the flavor group symmetry.

In the non-Abelian bosonization, the current $j_\pm$ in the light-cone
coordinates, $x^{\pm}$=$(x^0 \pm x^3)/\sqrt{2}$, is bosonized as
\cite{Wit84}
\begin{subequations}
\label{jj}
\begin{eqnarray}
\label{jja}
j_+ & = & ~~(i/2\pi) u^{-1} (\partial_+ u),\\
\label{jjb}
j_- & = & -(i/2\pi) (\partial_- u) u^{-1}.
\end{eqnarray}
\end{subequations}

To carry out the bosonization of the color SU(3) subgroup, we need to
introduce boson fields $\phi^a$, with $a=1,..,8$, to describe an
element $u$ of SU(3) with the eight $t^a$ generators which provide
eight degrees of freedom as
\begin{eqnarray}
\label{uua}
u = \exp \left \{ i 2 \sqrt{\pi} \sum_{a=1}^8 \phi^a t^a \right \},
\label{uuu}
\end{eqnarray}
However, the QCD SU(3) gauge field and quark current dynamics in QCD
will be coupled in the color-octet space in the eight $t^a$
directions.  Such couplings in the non-Abelian degrees of freedom will
lead to color excitations, the majority of which will not lead to
stable collective excitations.  A general variation of the element
$\delta u /\delta x^\pm$ will lead to quantities that in general do
not commute with $u$ and $u^{-1}$, resulting in $j_\pm$ currents in
Eqs.\ ({\ref{jj}) that are complicated non-linear admixtures of the
  boson fields $\phi^a$.  It will be difficult to look for stable
  boson states with these currents.

We can guide ourselves to a situation that has a greater chance of
finding stable bosons by examining the bosonization problem from a
different viewpoint.  We can introduce a unit generator vector
$\tau^1$ randomly in the eight-dimensional SU(3) generator space,
\begin{eqnarray}\label{cosines}
 \tau^1=\sum_{i=1}^8 n_a t^a, ~~~\text{with}~~
n_a={\rm tr }\{\tau^1 t^a\}/2~~{\rm and~}
~\sqrt{n_1^1+n_2^2+...+n_8^2}=1.
\label{eq23b}
\end{eqnarray}
We can describe an SU(3) group element $u$ by an amplitude $\phi^1$
and the unit vector $\tau^1$.  To look for stable boson states, we
wish to restrict our considerations in the color-octet subspace with a
fixed orientation of $\tau^1$ but allowing the amplitude $\phi^1$
boson field to vary.  The boson field $\phi^1$ describes one degree of
freedom, and the direction cosines $\{n^a,a=1,..,8\}$ of the unit
vector $\tau^1$ describe the other seven degrees of freedom.  A
variation of the amplitude $\phi^1$ in $u$ while keeping the unit
vector orientation fixed will lead to a variation of $\delta u/\delta
x^\pm $ that will commute with $u$ and $u^{-1}$ in the bosonization
formula (\ref{jj}), as in the case with an Abelian group element.
Such a quasi-Abelian approximation will lead to simple currents and
stable QCD bosons with well defined masses, which will need to be
consistent with experimental QCD meson data.  On the other hand, a
variation of $\delta u/\delta x^\pm $ in any of the other seven
orientation angles of the unit vector $\tau^1$ will lead to $\delta
u/\delta x^\pm $ quantities along other $t^a$ directions with
$a$=$\{1,...,8\}$.  These variations of $\delta u/\delta x^\pm $ will
not in general commute with $u$ or $u^{-1}$.  They will lead to
$j_\pm$ currents that are complicated non-linear functions of the
eight degrees of freedom. We are therefore well advised to search for
stable bosons by varying only the amplitude of the $\phi^1$ field,
keeping the orientation of the unit vector fixed, and forgoing the
other seven orientation degrees of freedom.  For the U(3) group, there
is in addition the group element $u = \exp\{ i 2 \sqrt{\pi} \phi^0 t^0
\}$ from the QED U(1) subgroup.  Combining both U(1) and SU(3)
subgroups, we can represent an element $u$ of the U(3) group by
$\phi^0$ from QED and $\phi^1$ from QCD as \cite{Won10}
\begin{eqnarray}
\label{uua}
u = \exp\{ i 2 \sqrt{\pi} \phi^0 \tau^0 + i 2 \sqrt{\pi} \phi^1
\tau^1\},
\label{uuu}
\end{eqnarray}
where we have re-labeled $t^0$ as $\tau^0$ such that
\begin{eqnarray}
2{\rm tr}(\tau^\lambda\tau^{\lambda'}
)=\delta^{\lambda{\lambda'}}, ~~{\rm with} ~~ \lambda, \lambda' = 0,1.
\end{eqnarray}
The superscripts in the above equation is the interaction label with
$\lambda$=$\{ 0,1\}$ for QED and QCD, respectively.  When we write out
the flavor index explicitly, we have
\begin{eqnarray}
\label{uuaf}
u_f = \exp\{ i 2 \sqrt{\pi} \phi_f^0 \tau^0 + i 2 \sqrt{\pi} \phi_f^1
\tau^1\}.
\end{eqnarray}
From (\ref{jja}) and (\ref{jjb}), we obtain
\begin{subequations}
\begin{eqnarray}
j_{f\pm} &=& \mp \frac{1}{\sqrt{\pi}} \left [ (\partial _\pm\phi_f^0)
  \tau^0 + (\partial _\pm\phi_f^1)\tau^1\right ]~~~~~~~~~{\rm
  when~all} ~Q_f^\lambda=1, \\ &=& \mp \frac{1}{\sqrt{\pi}} \left [
  Q_f^0(\partial _\pm\phi_f^0) \tau^0 + Q_f^1 (\partial
  _\pm\phi_f^1)\tau^1\right ]~\text{when we include charge number}
~Q_f^\lambda.~~~~~~~~~~
\end{eqnarray}
\end{subequations}
The Maxwell equation in light-cone coordinates is
\begin{eqnarray}
\partial _\mu \partial ^\mu A^\pm - \partial _ \pm \partial _\mu A^\mu
&=& g j^ \pm .
\end{eqnarray}
We shall use the Lorenz gauge
\begin{eqnarray}
\partial _\mu A^\mu =0,
\end{eqnarray}
then the solution of the gauge field is
\begin{eqnarray}
A^\pm &=& \frac{g}{ 2\partial _+ \partial _-} j^ \pm.
\end{eqnarray}
Interaction energy ${ H}_{\rm int}$ is
\begin{eqnarray}
{ H}_{\rm int} &&= \frac{g}{2}\int dx^+dx^-~2\,{\rm tr}\, (j \cdot A
)=\frac{g}{2}\int dx^+dx^- ~2\,{\rm tr}\, ( j^+ A^- + j^- A^+)
\nonumber\\ &&=\frac{g}{2}\int dx^+dx^- ~2\,{\rm tr}\,\left ( j^+
\frac{g}{ 2\partial _+ \partial _-} j^- + j^- \frac{g}{ 2\partial _+
  \partial _-} j^+ \right ).
\end{eqnarray}
We integrate by parts, include the charge numbers and the interaction
dependency of the coupling constant, $g_{\rm 2D}^\lambda$=$g^\lambda$,
and we obtain the contribution to the Hamiltonian density from the
confining interaction between the constituents, $H_{\rm int}=\int dx^+
dx^- {\cal H}_{\rm int}(\phi_f^\lambda)$,
\begin{eqnarray}
{\cal H}_{\rm int}(\phi_f^\lambda) &=&\frac{1}{2} \biggl [
  \frac{(g_{\2d}^0)^2}{\pi} (\sum_f^{N_f} Q_{f}^0\phi_f ^0 )^2 +
  \frac{(g_{\2d}^1)^2}{\pi} (\sum_f^{N_f} Q_{f}^1\phi_f^1)^2 \biggr ],
\label{216}
\end{eqnarray}
which matches the results of \cite{Col76,Nag09}.  

For the mass bi-linear term, we follow Coleman \cite{Col76} and Witten
\cite{Wit84} and bosonized it as
\begin{eqnarray}
m_f :\bar \psi_f \psi_f :&& \to (-\frac{e^\gamma}{2\pi} ) \mu m_f
~2{\rm tr} \left ( \frac{u_f + u_f^{-1}}{2} \right ), \nonumber\\ &&=
(-\frac{e^\gamma}{2\pi} ) \mu m_f ~2{\rm tr} \cos \left (2 \sqrt{\pi}
\phi_f ^0\tau^0 + 2\sqrt{\pi} \phi_f^1\tau^1 \right ),
\end{eqnarray}
where $\gamma=0.5772$ is the Euler constant, and $\mu$ is a mass scale
that arises from the bosonization of the scalar density ${\bar \psi}
\psi $.  It is proportional to the quark condensate $\langle 0| {\bar
  \psi} \psi |0 \rangle $ and is interaction-dependent \cite{Col76}.
In the QCD case, we shall see later that through the
Gell-Mann-Oakes-Renner relation, $\mu$ is related to the quark
condensate $\langle 0 | \bar q q |0\rangle $ by Eq.\ (\ref{236}).

When we sum over flavors, we get the contribution to the Hamiltonian
density from quark masses,
\begin{eqnarray}
H_{\rm m}=\int dx^+ dx^- {\cal H}_{\rm  m}(\phi_f^\lambda),
\end{eqnarray}
where
\begin{subequations}
\begin{eqnarray}
{\cal H}_{\rm m} ( \phi_f^\lambda)\!&=&\!e^\gamma \mu \sum_f\! m_f \left [
  (\phi_f^0 )^2 + (\phi_f^1)^2 \! + \!...\right ]~~~\text{when $\mu$  is independent of interaction},~~~~~~~~~
\\
    &=&\!e^\gamma \sum_f\! m_f \left [ \mu^0 (\phi_f^0 )^2\! + \!\mu^1 (\phi_f^1)^2+ ...\right ]~\text{when \!$\mu$ depends on  interaction}.~~~~~~~~
\label{218} 
\end{eqnarray}
\end{subequations}
Finally, for the kinematic term, we bosonize it as \cite{Wit84,Gep85}}
\begin{eqnarray}
: \bar \psi i \sd \psi : ~~~~ \to~~~ \frac{1}{8\pi} \biggl  \{ 2~{\rm tr} \left [
    \partial_ \mu u )~(\partial ^\mu u ^{-1})\right ] \biggr  \} + n\Gamma,
\end{eqnarray}
where $n\Gamma$ is the Wess-Zumino term which vanishes for $u$ of
(\ref{uuu}) containing commuting elements $\tau^0$ and $\tau^1$.  We
get the kinematic contribution
 \begin{eqnarray}
&H_{_{\rm kin}}=\int dx^0 dx^1 {\cal H}_{_{\rm kin}} =\int dx^0 dx^1
   \sum_f \frac{1}{8 \pi} \biggl  \{~2\,{\rm tr} \left [ \partial_ \mu u_f
     )~(\partial ^\mu u_f ^{-1})\right ] \biggr  \}, 
\end{eqnarray}
where
\begin{eqnarray}
&{\cal
     H}_{_{\rm kin}}(\phi_f^\lambda)= \frac{1 }{2}\sum_f \left [
     \partial _\mu\phi_f ^0\partial ^\mu \phi_f ^0 + \partial
     _\mu\phi_f ^1\partial ^\mu \phi_f ^1\right ]= \frac{1 }{2}
   \sum_\lambda \sum_f \left [ (\Pi_f^\lambda)^2 + (\partial _x\phi_f
     ^\lambda)^2 \right ],
\label{220}
\end{eqnarray}
and $\Pi_f^\lambda$ is the momentum conjugate to $\phi_f^\lambda$. The
total Hamiltonian density in terms of $\phi_f^\lambda$ is
\begin{eqnarray}
{\cal H}(\phi_f^\lambda)= {\cal H}_{_{\rm kin}}(\phi_f^\lambda)+{\cal
  H}_{\rm int}(\phi_f^\lambda)+{\cal H}_{\rm m}(\phi_f^\lambda),
\label{221}
\end{eqnarray}
where $ {\cal H}_{_{\rm kin}}(\phi_f^\lambda)$, ${\cal H}_{\rm
  int}(\phi_f^\lambda)$, ${\cal H}_{\rm m}(\phi_f^\lambda)$ are given
by Eqs.\ (\ref{220}), (\ref{216}), and (\ref{218}) respectively.

\subsection{Orthogonal transformation to  $q\bar q$\, flavor eigenstates}

We consider $q\bar q$ systems with dynamical flavor symmetry that lead
to flavor eigenstates as a linear combination of states with different
flavor amplitudes.  Such eigenstates arise from additional
considerations of isospin invariance, SU(3) flavor symmetry, and
configuration mixing.  As a result of such considerations, the
physical $q\bar q$ composite eigenstates $i$ in the interaction type
$\lambda$, $\Phi_i^\lambda$, can be quite generally related to various
flavor components $\phi_f$ by a linear orthogonal transformation as
\begin{eqnarray}
\Phi_i^\lambda=\sum_ f D_{if}^\lambda \phi_f ^\lambda.
\label{eq40}
\end{eqnarray}
The orthogonal transformation matrix $D_{if}^\lambda$ obeys
$(D^\lambda)^{-1}=(D^\lambda)^\dagger$ with
$((D^\lambda)^\dagger)_{fi}$=$D_{if}^\lambda$.  The inverse
transformation is
\begin{eqnarray}
\phi_f^\lambda=\sum_ i D_{if}^\lambda \Phi_i^\lambda.
\end{eqnarray}
Upon substituting the above equation into (\ref{221}), we obtain the
total Hamiltonian density in terms of the physical flavor state
$\Phi_i^\lambda$ as
\begin{eqnarray}
{\cal H}(\Phi_f^\lambda)= [{\cal H}_{_{\rm kin}}(\Phi_i^\lambda)+{\cal
    H}_{\rm int}(\Phi_i^\lambda)+{\cal H}_{\rm m}(\Phi_i^\lambda)],
\end{eqnarray}
\begin{subequations}
\begin{eqnarray}~~
\hspace*{-2.5cm}\text{where}~~~~~~~~{\cal H}_{_{\rm kin}}
(\Phi_i^\lambda) &&= \frac{1 }{2} \sum_\lambda \sum_i \left [\partial
  _\mu\Phi_i ^\lambda\partial ^\mu \Phi_i ^\lambda \right ] = \frac{1
}{2} \sum_\lambda \sum_i \left [ (\Pi_i^\lambda)^2+ (\partial
  _x\Phi_i^\lambda)^2 \right ], \\ {\cal H}_{\rm int}(\Phi_i^\lambda)
&&=\frac{1}{2}\biggl [ \sum_\lambda \frac{(g_{\2d}^\lambda)^2}{\pi}
  (\sum_f Q_{f}^\lambda\sum_i D_{if}^\lambda \Phi_i ^\lambda )^2 \biggr ],
\label{225b}
\\ {\cal H}_{\rm m}(\Phi_i^\lambda)&&=e^\gamma \sum_f m_f \left [
  \sum_\lambda \mu^\lambda (\sum_i D_{if}^\lambda
  \Phi_i^\lambda )^2 \right ].
\label{225c}
\end{eqnarray}
\end{subequations}
We can get the boson mass $m_i^\lambda$ of the physical state
$\Phi_i^\lambda$ by expanding the potential energy term, ${\cal
  H}_{\rm int}(\Phi_i^\lambda)+{\cal H}_{\rm m}(\Phi_i^\lambda)$,
about the potential minimum located at $\Phi_i^\lambda=0$, up to the
second power in $(\Phi_i^\lambda)^2$, as
\begin{eqnarray}
{\cal H}(\Phi_i^\lambda) &&= \sum_\lambda \sum_i\left [ \frac{1 }{2}
  (\Pi_i ^\lambda)^2 +\frac{1 }{2} (\partial _x \Phi_i ^\lambda)^2
  +\frac{1}{2} (m_{i}^\lambda)^2 (\Phi_i ^\lambda)^2 \right ]+ ...,
\end{eqnarray}
\begin{eqnarray}
\hspace*{-3.0cm}\text{where}~~~~~~~~~~~~~~~~&&(m_{i}^\lambda)^2
=\biggl [ \frac{\partial^2}{\partial (\Phi_{i}^\lambda)^2} [{\cal
      H}_{\rm int} (\{\Phi_i^\lambda\})+ {\cal H}_{\rm
      m}(\{\Phi_i^\lambda\})]\biggr
]_{\Phi_0^\lambda,\Phi_1^\lambda=0} .
\label{eq45}
\end{eqnarray}
From Eqs.\ (\ref{225b}) and (\ref{225c}), we find the squared mass
$(m_i^\lambda)^2$ for the state $\Phi_i^\lambda$ of interaction $\lambda$
to be
\begin{eqnarray}
(m_{i}^\lambda)^2&&=\frac{(g_{\2d}^\lambda)^2}{\pi} \left [
    \sum_f^{N_f}  D_{if}^\lambda Q_{f}^\lambda\right ] ^2 + e^\gamma
  \sum_f^{N_f} m_f \mu^\lambda(D_{if}^\lambda)^2.
\label{eq43a}
\end{eqnarray}
This mass formula includes the mixing of the configurations, and is
applicable to the open string model in QCD and QED.  

The two terms on the right hand side of (\ref{eq43a}) receive
contributions from different physical sources.  The first term, the
``massless quark limit'' arises from the confining interaction between
the quark and the antiquark.  A comparison of this first term with the
mass formula for a single unit charge in Eq.\ (\ref{eq6}), $m^2 =
g_\2d^2 /\pi$, suggests that in the state $i$ with constituents $q$
and $\bar q$ interacting in the $\lambda$-type interaction with flavor
mixing, the $q\bar q$ composite particle behaves as if it consists
effectively of a quark constituent $q$ with an effective charge
$\tilde Q_q^\lambda(i)$ given by
\begin{eqnarray}
\tilde Q_q^\lambda(i) = 
    \sum_f^{N_f}  D_{if}^\lambda Q_{f}^\lambda ,
    \label{eq47a}
\end{eqnarray}
confined with an antiquark constituent $\bar q$ of the opposite
effective charge
\begin{eqnarray}
\tilde Q_{\bar q}^\lambda(i) = 
-\tilde Q_q^\lambda(i).
\end{eqnarray}
For brevity of notation of the quark effective charge, the label $(i)$
for the composite particle are often implicitly understood such that
$\tilde Q_{ q}^\lambda$$\equiv$$\tilde Q_{ q}^\lambda(i)$ and $\tilde
Q_{\bar q}^\lambda$$\equiv$$\tilde Q_{\bar q}^\lambda(i)$, if no
ambiguity arises.  The quark constituent in a $q\bar q$ meson
possesses both a effective color charge number $\tilde Q_q^1$ for the
QCD interaction and an effective electric charge $\tilde Q_q^0$ for
the QED interaction.  We shall discuss the dependence of the
quark-antiquark potential on the effective charges in subsection 4G.

The second term arises from quark masses and the quark condensate,
$\sum_f m_f \langle \bar \psi_f \psi_f\rangle $.  It can be called the
``quark mass term'', or the ``quark condensate term'' because it
depends on both.  If one labels the square root of the first term in
(\ref{eq43a}) as the confining interaction mass and the square root of
the second term as the condensate mass, then the hadron mass obeys a
Pythagoras theorem with the hadron mass as the hypotenuse and the
confining interaction mass and the quark condensate mass as two sides
of a right triangle.

Under the quasi-Abelian approximation of the non-Abelian QCD dynamics,
the Schwinger confinement mechanism leads naturally to the (1+1)D
open-string description of neutral confined bosons for both QCD and
QED, with masses that depend on the magnitudes of the gauge field
coupling constants $g_\2d^\lambda$ as given in Eq.\ (\ref{eq43a}).
For the QCD interaction, we mentioned earlier that such a
one-dimensional open-string solution of the lowest energy hadron
states was suggested early on by the dual-resonance model
\cite{Ven68}, the Nambu and Goto meson string model
\cite{Nam70,Nam74,Got71}, the 'tHooft's two-dimensional meson model
\cite{tHo74a,tHo74b}, the classical yo-yo string model \cite{Art74},
Polyakov's quantum bosonic string \cite{Pol81}, and the Lund model
\cite{And83}.  The open-string description of hadrons was supported
theoretically for QCD by lattice gauge calculations in (3+1)D in which
the structure of a flux tube shows up explicitly
\cite{Hua88,Bal05,Cos17,Car13,Bic18,Bic19}.  The open-string
description for $q\bar q$ QCD systems in high-energy hadron-hadron and
$e^+$-$e^-$ annihilation collisions provided the foundation for the
(1+1)D inside-outside cascade model for hadron production of Bjorken,
Casher, Kogut, and Susskind \cite{ Bjo73,Cas74}, the yo-yo string model
\cite{Art74}, the generalized Abelian Model \cite{Bel79}, the
projected Abelian model \cite{tHo80}, the Abelian dominance model
\cite{Sei07,Suz08}, and the Lund model in high-energy collisions
\cite{And83}.  The flux tube description of hadrons receives
experimental support from the limiting average transverse momentum and
the rapidity plateau of produced hadrons
\cite{Cas74,Bjo73,Won91,Gat92,Won09}, in high-energy $e^+$-$e^-$
annihilations \cite{Aih88,Hof88,Pet88,Abe99,Abr99} and $pp$ collisions
\cite{Yan08}.

While a confined open string in (1+1)D as the idealization of a stable
quark-antiquark system in (3+1)D is well known in QCD, not so
well-known is the analogous confined open string quark-antiquark
system in (1+1)D with a lower boson mass in QED, when we apply the
Schwinger confinement mechanism for massless fermions to light quarks
in QED in (1+1)D, as discussed in Section 3 \cite{Won10,Won11,Won14,Won20}.  We
have asked earlier in the introduction whether the confined
quark-antiquark one-dimensional open string in QED in (1+1)D could be
a reasonable idealization of a stable QED $q\bar q$ meson in (3+1)D,
just as the confined quark-antiquark Nambu-Goto open string in QCD in
(1+1)D can be the idealization of a stable QCD $(q\bar q)$ meson in
(3+1)D.

From the viewpoint of phenomenology, we note that no
fractionally-charged particles have ever been observed.  Thus, quarks
cannot be isolated.  The non-isolation of quarks is consistent with
the hypothesis that the open string $q\bar q$ boson solution in (1+1)D
is an idealization of a confining flux tube between the quark and the
antiquark in (3+1)D.  On such a hypothesis, the open string
description of a QED-confined $q\bar q$ boson is a reasonable concept
that can be the basis of a phenomenological description of QED mesons
whose validity needs to be constantly confronted with experiments.  We
can therefore study the question of quark confinement in QED from the
phenomenological point of view.  It is also necessary to examine
 the question of quark confinement in QED from the lattice
gauge calculations viewpoints in (3+1)D, to be taken up in Section 6.

\subsection{Relation between the coupling constants in (3+1)D and (1+1)D} 

In the phenomenological open string model for QCD and QED mesons, we
need an important relationship to ensure that the boson masses
calculated in the lower (1+1)D can properly represent the physical
boson masses in (3+1)D.  The one-dimensional $q\bar q$ open string in
(1+1)D can be considered as an idealization of a flux tube with a
transverse radius $R_T$ in the physical world of (3+1)D.  The flux
tube in (3+1)D has a structure with a transverse radius $R_T$, but the
coupling constant $g_\4d$ is dimensionless.  In contrast, the open
string in (1+1)D does not have a structure, but the coupling constant
$g_\2d$ has the dimension of a mass.  We proved previously that the
(1+1)D open string can be considered an idealization of a flux tube
with a transverse radius $R_T$ in the physical meson in (3+1)D, if the
coupling constants $g_\2d$ in (1+1)D and $g_\4d$ in (3+1)D are related
by \cite{Won09,Won10,Won20,Kos12,Won22}
\begin{eqnarray}
(g_{\2d})^2=\left ( \frac{1}{\pi
    R_T^2} \right ) (g_{\4d})^2
    =\frac{4\alpha_{\4d}}{R_T^2} ,
    \label{eq49}
\end{eqnarray}
whose qualitative consistency can be checked by dimensional analysis.
Thus, when the dynamics in the higher dimensional 3+1 space-time is
approximated as dynamics in the lower (1+1)D, information on the flux
tube structure is stored in the multiplicative conversion factor
$(1/\pi R_T^2)$ in the above equation that relates the physical
coupling constant square $(g_{4D})^2$ in (3+1)D to the new coupling
constant square $ (g_{2D})^2$ in (1+1)D, as we discussed earlier in
Section 3.  As a consequence, there is no loss of the relevant
physical information.  The boson mass $m$ determined in (1+1)D is the
physical mass in (3+1)D, when we relate the coupling constant $g_\2d$
in (1+1)D to the physical coupling constant $g_{\4d}$ in (3+1)D and
the flux tube radius $R_T$ by Eq.\ (\ref{eq49}).  Consequently, the
masses of the QED and QCD mesons in (3+1)D in the open-string
description are approximately
\begin{eqnarray}
m_\qcd^2=\frac{(g_\2d^\qcd)^2}{\pi}=\frac{4 \alpha_\qcd}{\pi R_T^2},~~~~~~~ m_\qed^2==\frac{(g_\2d^\qed)^2}{\pi}=\frac{4 \alpha_\qed}{\pi R_T^2}.
\label{eq19}
\end{eqnarray}
With $\alpha_{\4d}^{\qed}$\!=$\alpha_{{}_{\rm QED}}$=1/137,
$\alpha_{\4d}^{\qcd}$\!=$\alpha_s $ $\sim$ 0.68 from hadron
spectroscopy \cite{Won10,Won20} and $R_T$$\sim$0.4 fm from lattice QCD
calculations \cite{Cos17} and $\langle p_T^2 \rangle $ of produced
hadrons in high-energy $e^+e^-$ annihilations \cite{Pet88}, we
estimate the masses of the open string QCD and QCD mesons to be
\begin{eqnarray}
m^{\qcd}\sim 458{\rm ~ MeV}, ~~~ {\rm and}~~ m^{\qed}\sim 48 {\rm
  ~MeV}.
\label{eq23}
\end{eqnarray}
The above mass scales provide an encouraging guide for the present
task of a quantitative description of the QCD and QED mesons, using
QCD and QED gauge field theories in (1+ 1)D.  The QED mesons reside in
the region of many hundred of MeV whereas the QED mesons in the region
of many tens of MeV.

To get a better determination of the QCD and QED meson masses, it is
necessary to take into account the flavor mixtures $D_{ij}^\lambda $
and the quark color and electric charges $Q_{\{u,d,s\}}^{\lambda}$,
the quark masses $m_f$, and the chiral condensate, as discussed in
\cite{Won20}.

\subsection{ Open string model description of QCD mesons}

We consider $q\bar q$ systems with dynamical flavor symmetry with
physical flavor eigenstates $\Phi_i ^\lambda$ as a linear combination
of states with different amplitudes of $|q_f \bar q_f\rangle $ for
quark-antiquark pairs of different flavor $f$, $\Phi_i^\lambda=\sum_f D_{if}^\lambda |q_f \bar q_f \rangle = \sum_f D_{if}^\lambda \phi_f$.  Such eigenstates arise
from additional considerations of flavor SU(2) isospin invariance,
SU(3) flavor symmetry, and configuration mixing.  As a result of such
considerations in flavor symmetry and configuration mixing, the
physical eigenstates $\Phi_i^\lambda $ can be quite generally related
to various flavor components $\phi_f^\lambda $=$|q_f^\lambda \bar
q_f^\lambda \rangle $ by a linear orthogonal transformation as
$\Phi_i^\lambda=\sum_ f D_{if}^\lambda \phi_f ^\lambda$ of
Eq.\ (\ref{eq40}) leading to the meson mass formula in the open string
model in Eq.\ (\ref{eq43a}).

We study first the QCD case with the label $\lambda=1$, which is often
implicitly understood in this subsection.  Equation (\ref{eq23})
indicates that the mass scale $m^{\qcd}\!\!\!\!\sim$ 458 MeV $ \gg
m_u, m_d, m_s$.  It is necessary to include $u$, $d$, and $s$ quarks
with $N_f $=3 in the analysis of open strings QCD mesons.

It is generally known that QCD has an approximate SU(3)$_L\times$
SU(3)$_R$ chiral symmetry and also an approximate flavor
U(3)$\times$U(3) symmetry.  If the axial symmetry is realized as the
Goldstone mode as a result of the spontaneous chiral symmetry
breaking, then one would naively expect the singlet isoscalar $\eta'$
particle to have a mass comparable to the pion mass.  Experimentally,
there is the U$_A$(1) anomaly \cite{Kog74a,Kog75a,Kog75b,Wei79,Wit79}
in which the $\eta'$ mass of 957.8 MeV is so much higher than the
$\pi$ mass.  On the basis of the Schwinger confinement mechanism
\cite{Sch62,Sch63}, Kogut, Susskind, and Sinclair
\cite{Kog74a,Kog75a,Kog75b} suggested that such a U$_A$(1) anomaly
arises from the long-range confinement between the quark and the
antiquark, as the $\eta'$ acquires a large mass from the long-range
confining interaction between a quark and an antiquark.  The long
range gauge interaction affects not only $\eta'$ mass but also the
other pseudoscalar $\pi^0$, and $\eta$ masses, and there are
furthermore the effects of quark rest masses, the quark condensate, and the configuration
mixing between $\eta$ and $\eta'$.  When these effects are properly
taken into account, the pseudoscalar particles $\pi^0$, $\eta$, and
$\eta'$ can indeed be adequately described as open string QCD mesons.

We denote the flavor component states $\phi_i$, with $\phi_1=|u \bar u
\rangle$, $\phi_2=|d \bar d \rangle$, and $\phi_3=|s \bar s \rangle$,
and assume the standard quark model description of physical states
$\Phi_j$ with $|\pi^0\rangle$=$\Phi_1$, $|\eta\rangle$=$\Phi_2$, and
$|\eta'\rangle$=$\Phi_3$ in terms of flavor octet and flavor singlet
states.  The physical states of $|\pi^0\rangle$, $|\eta\rangle$, and
$|\eta'\rangle$ can be represented in terms of the flavor states
$\phi_1$, $\phi_2$ and $\phi_3$ by
\begin{subequations}
\begin{eqnarray}
|\pi^0\rangle&=&\Phi_1=\frac{\phi_1-\phi_2}{\sqrt{2}},
\\ |\eta~\rangle&=&\Phi_2= |\eta_8\rangle \cos \theta_P- |
\eta_1\rangle \sin \theta _P,
\label{229b}
\\ |\eta'\rangle&=&\Phi_3= |\eta_8 \rangle \sin \theta_P+|
\eta_1\rangle \cos \theta _P,
\label{229c}
\end{eqnarray}
where the mixing of the $|\eta\rangle$ and $|\eta'\rangle$ is
represented by a mixing angle $\theta_P$, and the flavor-octet state
$|\eta_8\rangle$ and the flavor-singlet state $|\eta_1\rangle$ are
\begin{eqnarray}
&&|\eta_8\rangle
=\frac{\phi_1+\phi_2-2\phi_3}{\sqrt{6}}, \\ & &\,|\eta_1\rangle
=\frac{\sqrt{2}(\phi_1+\phi_2+\phi_3) }{\sqrt{6}}.
\end{eqnarray}
\end{subequations}
The physical states $\Phi_i$=$\sum_f D_{if}\phi_f$ and the flavor
component states $\phi_f$, are then related by
\begin{eqnarray}
\begin{pmatrix}
 \Phi_1\\ \Phi_2\\ \Phi_3
\end{pmatrix}
\!\!=\!\!\begin{pmatrix} \frac{1}{\sqrt{2} } & - \frac{1}{\sqrt{2} } &
0 \\ \frac{1} {\sqrt{6}} \{ \cos \theta_P \!-\!\sqrt{2} ~ \sin
\theta_P\} & \frac{1} {\sqrt{6}} \{ \cos \theta_P \!-\!\sqrt{2} ~ \sin
\theta_P\} & \frac{1} {\sqrt{6}}\{-2\cos \theta_P \!-\! \sqrt{2}\sin
\theta_P\} \\ \frac{1}{\sqrt{6}}\{\sin \theta_P \!+\! \sqrt{2}\cos
\theta_P\} & \frac{1}{\sqrt{6}}\{\sin \theta_P \!+\! \sqrt{2}\cos
\theta_P\} & \frac{1}{\sqrt{6}}\{-2\sin \theta_P\! +\! \sqrt{2}\cos
\theta_P\} \\
\end{pmatrix}
\!\!
\begin{pmatrix}
 \phi_1\\ \phi_2\\ \phi_3
\end{pmatrix}\!\!,
\nonumber
\end{eqnarray}
with the inverse relation $\phi_f= \sum_{i=1}^3 D_{if}\Phi_i$,
\begin{eqnarray}
\begin{pmatrix}
 \phi_1\\ \phi_2\\ \phi_3
\end{pmatrix}=
\begin{pmatrix}
 \frac{1}{\sqrt{2} } & ~~~ \frac{1} {\sqrt{6}} \{ \cos \theta_P
 -\sqrt{2} ~ \sin \theta_P\}~~~ & \frac{1} {\sqrt{6}}\{ \sin \theta _P
 + \sqrt{2} ~\cos \theta _P \} \\ - \frac{1}{\sqrt{2} } & \frac{1}
 {\sqrt{6}}\{ \cos \theta_P - \sqrt{2} ~\sin \theta_P\} & \frac{1}
 {\sqrt{6}}\{ \sin \theta _P + \sqrt{2} ~\cos \theta _P \} \\ 0
 &\frac{1} {\sqrt{6}}\{-2\cos \theta_P \!-\! \sqrt{2}\sin \theta_P\} &
 \frac{1}{\sqrt{6}}\{-2\sin \theta_P\! +\! \sqrt{2}\cos \theta_P\} \\
\end{pmatrix}
\begin{pmatrix}
 \Phi_1\\ \Phi_2\\ \Phi_3
\end{pmatrix}.~~~~~~
\end{eqnarray}

The mass formula (\ref{eq43a}) gives
\vspace*{-0.2cm}
\begin{eqnarray}
(m_i^\qcd)^2=\frac{(g_\2d^\qcd)^2}{\pi} (\sum_{f=1}^{N_f}  D_{if}^\qcd Q_f^\qcd)^2 +
\sum_{f=1}^{N_f} m_f (D_{if}^\qcd)^2 e^\gamma \mu ^{\qcd} ,~~~~~~ \frac{(g_\2d^\qcd)^2}{\pi}=\frac{4\alpha_s}{ \pi R_T^2} ,
\label{231}
\end{eqnarray}
yielding an effective color charge $\tilde Q_q^\qcd$=$\sum_{f=1}^3
D_{if}^\qcd Q_f^\qcd$.

With color charge $Q_{\{u,s,d\}}^{\qcd}$=1,  the pion state
$|\pi^0\rangle$=$\Phi_1$ has an effective charge $\tilde
Q_q^\qcd(\pi^0)$ given by  $\sum_{f=1}^3 D_{1f}^\qcd Q_f^\qcd$=$1/\sqrt{2}-1/\sqrt{2}$=0. 
Consequently 
the first term in the mass formula (\ref{231}), the ``massless
quark limit'' term, 
 gives a zero effective
color charge, $\tilde Q_q^\qcd(\pi^0)$=0,  and a value of zero for the 
pion mass.   The only contribution to the pion mass  comes from the second
``quark condensate'' term in (\ref{231}) where 
$(D_{if}^\qcd)^2$=1/2. 
 The mass formula
(\ref{231}) for the pion then gives
\begin{eqnarray}
m_\pi^2= (m_u + m_d)\frac{ e^\gamma \mu^\qcd}{2},
\end{eqnarray}
which is in the same form as the Gell-Mann-Oakes-Renner relation
\cite{Gel68},
\begin{eqnarray}
m_\pi^2= (m_u + m_d) \frac{ |\langle 0|\bar q q |0 \rangle|}{F_\pi^2},
\label{233}
\end{eqnarray}
where $|\langle0| \bar q q |0\rangle|$ is the light $u$ and $d$
quark-antiquark condensate and $F_\pi$ is the pion decay constant.  Consequently, we can infer that the unknown mass scale
$\mu^\qcd$ in the bosonization formula for QCD has indeed the physical
meaning of the quark condensate.  We can therefore identify $\mu^\qcd$
in the bosonization mass formula (\ref{231}) for QCD as
\begin{eqnarray}
\frac{e^\gamma \mu^\qcd}{2} = \frac{ |\langle 0|\bar q q|0
  \rangle|}{F_\pi^2}.
\label{236}
\end{eqnarray}
By such an identification and calibrating the pion mass to be the
experimental mass $m_\pi$, the mass formula (\ref{231}) for the
pseudoscalar QCD mesons can be re-written as \cite{Won20}
\begin{eqnarray}
(m_i^\qcd)^2=
\frac{(g_\2d^\qcd)^2}{\pi} 
(\sum_{f=1}^{N_f} D_{if}^\qcd) ^2 +
m_\pi^2 \sum_{f=1}^{N_f} \frac{m_f}{m_{ud}} (D_{if}^\qcd)^2,
~~~~~~ \frac{(g_\2d^\qcd)^2}{\pi}=\frac{4\alpha_s}{ \pi R_T^2} ,
\label{qcd}
\end{eqnarray}
where $m_{ud}=(m_u+m_d)/2$.  

We are ready to test whether the QCD mesons $\pi^0$, $\eta$ and
$\eta'$ can be appropriately described as open string states in the
1+1 dimensional bosonization model.  For these QCD neutral mesons,
there is a wealth of information on the matrix $D_{if}^\qcd$ that
describes the composition of the physical states in terms of the
flavor components, as represented by the mixing angle $\theta_P$
between the flavor octet and flavor singlet components of the SU(3)
wave functions in $\eta$ and $\eta'$ in (\ref{229b}) and
(\ref{229c}). The ratio of the strange quark mass to the light $u$ and
$d$ quark masses that is needed in the above mass formula is also
known.  From the tabulation in PDG \cite{PDG19}, we find
$\theta_P=-24.5^o$ and ${m_s}/m_{ud}$= 27.3$_{-1.3}^{+0.7}$.  The only
free parameters left in the mass formula (\ref{qcd}) are the strong
interaction coupling constant $\alpha_s$ and the flux tube radius
$R_T$.

For the value of $\alpha_s$, previous works on the non-perturbative
potential models use a value of $\alpha_s$ of the order of $0.4-0.6$
in hadron spectroscopy studies \cite{Bar92,Won00,Won01,Bal08,Deu16}.
However, these potential models contain a linear confining
interaction, in addition to the one-gluon exchange interaction
involving $\alpha_s$.  In contrast, the present simplified 1+1
dimensional treatment uses only a single attractive gauge interaction,
involving $\alpha_s$ and playing dual roles.  We should be prepared to
allow for a larger value of the strong coupling constant $\alpha_s$ in
our case.  We find that $\alpha_s$=0.68 gives a reasonable description
of the masses of the mesons considered, and we can take the difference
between this $\alpha_s$ value and the $\alpha_s$ value of $0.6$ used
for the lowest meson masses in earlier hadron spectroscopy studies
\cite{Bar92,Won00,Won01,Bal08,Deu16} as a measure of the degree of
uncertainties in $\alpha_s$, resulting in $\alpha_s$=0.68 $\pm$ 0.08.

\begin{table}[H]
\centering
\caption { Comparison of experimental and theoretical masses of
  neutral, $I_3$=0, and $S$=0 QCD and QED mesons obtained with the
  semi-empirical mass formula (\ref{qcd}) for QCD mesons and
  (\ref{qed}) for QED mesons, with $\alpha_s$=0.68$\pm$0.08,
  $R_T$=0.40$\pm$0.04 fm, and $\alpha_{{}_{\rm QED}}$=1/137.  }
\vspace*{0.2cm}
\begin{tabular}{|c|c|c|c|c|c|c|}
\cline{3-6} \multicolumn{2}{c|}{} & &Experimental& Semi-empirical
&Meson mass\\ \multicolumn{2}{c|}{}& $J^\pi I$ & mass & mass
& in massless \\ \multicolumn{2}{c|}{}& & & formula & quark limit
\\ \multicolumn{2}{c|}{}& & (MeV) & (MeV) & (MeV) \\ 
\hline
QCD&\,\,$\pi^0$ & 0$^- $1 &\!\!134.9768$\pm$0.0005\!\!& 134.9$^\ddagger$ & 0 \\ 
\!\!meson\!\!&\!$\eta$ &0$^-$0 &\!\!547.862$\pm$0.017\!\!&498.4$\pm$39.8~~ & 329.7$\pm$57.5~ \\ 
& $\eta'$  &$0^-0$ & 957.78$\pm$0.06& 948.2$\pm$99.6~ &723.4$\pm$126.3 \\ \hline
QED& isoscalar  &0$^-0$& & 17.9$\pm$1.5 & 11.2$\pm$1.3 \\ 
\!\!meson\!\!& isovector  &$0^-1$ & & 36.4$\pm$3.8&33.6$\pm$3.8 \\ 
\hline
                & X17& (1$^+$) &\!\!16.70$\pm$0.35$\pm$0.5$^\dagger$~~\!\!& & \\ 
Possible& X17 & (0$^-$)  &\!\!16.84$\pm$0.16$\pm$0.20$^\#$\!\!& & \\ 
QED& X17 & (1$^-$)  &\!\!16.86$\pm$0.17$\pm$0.20$^\square$\!\!& & \\ 
meson& E38 &                &  37.38$\pm$0.71$^\oplus$& & \\ 
\!candidates\!& E38 &                &  40.89$\pm$0.91$^\ominus$& & \\ 
                               & E38 &                &  39.71$\pm$0.71$^\otimes$& & \\ \hline
\end{tabular}
\vspace*{0.1cm}

\hspace*{-5.19cm}$^\ddagger$ Calibration mass~~~~~~~~~~~~~~~~~~~~~~~\\
\hspace*{-0.10cm}$^\dagger$\,A. Krasznahorkay $et~al.$, Phys.Rev.Lett.116,042501(2016), $^8$Be$^*$ decay\\
\hspace*{-1.98cm}$^\#$A. Krasznahorkay $et~al.$, arxiv:1910.10459, $^4$He$^*$ decay~~~~\\
\hspace*{-1.98cm}$^\square$A. Krasznahorkay $et~al.$, arxiv:2209.10795, $^{12}$C$^*$ decay~~~~\\
\hspace*{-0.70cm}$^\oplus$\,K. Abraamyan $et~al.$, EPJ Web Conf       204,08004(2019),$d$Cu$\to$$\gamma \gamma X$~\\
\hspace*{-0.70cm}$^\ominus$\,K. Abraamyan $et~al.$, EPJ Web Conf       204,08004(2019),$p$Cu$\to$$\gamma \gamma X$~\\
\hspace*{-0.70cm}$^\otimes$\,K. Abraamyan $et~al.$, EPJ Web Conf       204,08004(2019),~$d$C$\to$$\gamma \gamma X$~\\
\label{tb1}
\end{table}
For the value of $R_T$, lattice gauge calculations with dynamical
fermions give a flux tube root-mean-square-radius $R_T$=0.411 fm for a
quark-antiquark separation of 0.95 fm \cite{Cos17}.  The experimental
value of $ \langle p_T^2\rangle $ of produced hadrons ranges from 0.2
to 0.3 GeV$^2$ for $e^+$-$e^-$ annihilations at $\sqrt{s}$ from 29 GeV
to 90 GeV \cite{Pet88}, corresponding to a flux tube radius
$R_T$=$\hbar/\sqrt{\langle p_T^2\rangle }$ of 0.36 to 0.44 fm.  It is
reasonable to consider flux tube radius parameter to be $R_T=0.4 \pm
0.04$ fm.  This set of parameters of $\alpha_s$=0.68$\pm$0.08 and
$R_T$=0.40$\pm$0.04 fm give an adequate description of the $\pi^0$,
$\eta$ and $\eta'$ masses as shown in Table I.

From our comparison of the experimental and theoretical masses in
Table \ref{tb1}, we find that by using the method of bosonization and
including the confining interaction and the quark condensate, the mass
formula (\ref{qcd}) in the 1+1 dimensional open string model can
indeed describe the masses of $\pi^0$, $\eta$, and $\eta'$,
approximately within the limits of the uncertainties of the
theory. The formulation can be used to extrapolate to the unknown
region of open string $q\bar q$\, QED mesons.

In order to infer the importance of the second quark condensate term
relative to the massless quark limit arising from the confining
interaction in (\ref{qcd}), we tabulate in Table I the results of the
hadron mass values obtained in the massless quark limit.  We observe
that for the pion mass, the massless quark limit is zero, and the pion
mass arises only from the second quark condensate term.  The
importance of the quark condensate diminishes as the hadron masses
increases to $\eta$ and $\eta'$.

\subsection{Open string model description of QED mesons}

Having confirmed the approximate validity of the (1+1)D open string
description of neutral QCD mesons which exist in (3+1)D, we proceed to
extrapolate to the unknown region of the open string $q\bar q$ QED
mesons in (3+1)D.  In such a discussion involving the QED interaction
in this subsection, the interaction label superscript is implicitly
$\lambda$=0 for QED unless noted otherwise.  The mass scale in
(\ref{eq23}) gives $m^{\qed}\!\!  \sim 48 {\rm ~MeV}\gg m_u, m_d$, but
$m^{\qed}$ is less than $m_s$.  In the treatment of QED mesons, it is
only necessary to include $u$ and $d$ quarks and antiquarks, with $N_f
=2$.  It is also necessary to include the fractional quark electric
charges $Q_f^\qed$.
 
We denote the flavor $q\bar q$ component states $\phi_1=|u \bar u
\rangle$, $\phi_2=|d \bar d \rangle$, and construct the physical
isoscalar $I$=0 $|\Phi_1^{\qed}\rangle$ and the isovector
$(I=1,I_3=0)$ $|\Phi_2^{\qed}\rangle$ states as
\begin{eqnarray}
|\text{(isoscalar)}I=0,I_3=0\rangle&=\Phi_1=({\phi_1+\phi_2})/{\sqrt{2}},
\nonumber\\ |\text{(isovector)}I=1,I_3=0\rangle&=\Phi_2=({\phi_1-\phi_2})/{\sqrt{2}}.
\end{eqnarray}
They are related by $\Phi_i^\qed $=$\sum_f D_{if}^\qed \phi_f$ and
$\phi_i$=$\sum_f D_{if}^\qed \Phi_i$,
\begin{eqnarray}
\begin{pmatrix}
 \Phi_1\\ \Phi_2
\end{pmatrix}
\!\!=\!\!\begin{pmatrix} \frac{1}{\sqrt{2} } & +\frac{1}{\sqrt{2} }
\\ \frac{1}{\sqrt{2} } & - \frac{1}{\sqrt{2} }
\end{pmatrix}
\!\!
\begin{pmatrix}
 \phi_1\\ \phi_2
\end{pmatrix}
, ~~~~~
\begin{pmatrix}
 \phi_1\\ \phi_2
\end{pmatrix}
\!\!=\!\!\begin{pmatrix} \frac{1}{\sqrt{2} } & + \frac{1}{\sqrt{2} }
\\ \frac{1}{\sqrt{2} } & - \frac{1}{\sqrt{2} }
\end{pmatrix}
\!\!
\begin{pmatrix}
 \Phi_1\\ \Phi_2
\end{pmatrix}.
\end{eqnarray}
The mass formula (\ref{qcd}) for the mass of the physical QED meson
$\Phi_i$ becomes
\begin{eqnarray}
(m_I^\qed)^2&=\frac{(g^\qed_\2d)^2}{\pi}\left [ \frac{Q_u^\qed+(-1)^IQ_d^\qed}{\sqrt{2}} \right ]^2
+ m_{ud} e^\gamma \mu^\qed,  ~~~ {\rm with}~~  \frac{(g^\qed_\2d)^2}{\pi}=\frac{4\alpha_{{}_{\rm QED}}}{ \pi R_T^2} ,
\label{240}
\end{eqnarray}
where the mass scale $\mu^\qed$ for QED mesons is not known.  From the
results for QCD mesons in (\ref{231})-(\ref{233}), we expect an
analogous relationship relating the mass scale $\mu^\qed$ and the
quark condensate for QED mesons,
 \begin{eqnarray}
e^\gamma \mu^\qed \propto |\langle 0|\bar q q|0 \rangle|_{_{\qed}},
\end{eqnarray}
where $|\langle 0|\bar q q|0 \rangle|_{_{\qed}}$ is the quark
condensate in the presence of QED gauge interactions between the quark
and the antiquark.

The chiral condensate depends on the interaction type $\lambda$,
specifically, on the coupling constant.  We note that the chiral
current anomaly in the chiral current depends in (3+1)D on the
coupling constant as $e^2=(g_\4d^\qed)^2 $ as given in Eq. (19.108) of
\cite{Pes95}
\begin{eqnarray}
\partial _\mu j^{\mu 53 }=-\frac{e^2}{32\pi} \epsilon^{\alpha \beta \gamma\delta } F_{\alpha \beta} F_{\gamma \delta},
\end{eqnarray}
which shows that the degree of non-conservation of the chiral current
is proportional to $e^2$.  It is therefore reasonable to infer that
the chiral condensate term in Eqs. (58) and (61) depends on the coupling constant as
$(g_\4d^\lambda)^2$ or $\alpha_\lambda$.  Hence, we have a general
mass formula for the mass $m_I^\lambda$ of a neutral QCD and QED meson $I$
in the $\lambda$-type interaction given by
\cite{Won20}
\begin{eqnarray}
(m_{ I}^\lambda)^2=
\frac{(g_\2d^\lambda)^2 }{\pi}
\left [\sum_{f=1}^{N_f} 
D^\lambda _{If} Q^\lambda_f  \right]^2 
\!\!\!+ m_\pi^2\frac{\alpha_\lambda}{\alpha_\qcdd} 
 \frac{\sum_f^{N_f}  (D^\lambda_ {If})^2 m_f  }{m_{ud}},  ~~~
\frac{(g_\2d^\lambda)^2 }{\pi}= 
\frac{4\alpha_\lambda}{ \pi R_T^2},
 ~ ~~ \lambda=\begin{cases} 0 & \text{QED} \cr 1  & \text{QCD} \cr \end{cases}.
\label{qed}
\end{eqnarray}
Here, the first term is the massless quark limit arising from the
confining interaction between the quark and the antiquark, with
$Q_u$=2/3, $Q_d$=$-$1/3, and $\alpha_{{}_{\rm QED}}=$1/137.  The
second term arises from the quark masses and the quark condensate in
the presence of the QED interaction.

In applying the above mass formula for QED mesons, we extrapolate from
the QCD sector to the QED sector by using those $R_T$ and $\alpha_s$
parameters that describe well the $\pi^0$, $\eta$ and $\eta'$ QCD
mesons.  We do not know the flux tube radius $R_T$ for the QED
interaction.  The meager knowledge from the possible QED meson
spectrum and the anomalous soft photons suggests that the assumption
of an intrinsic quark flux tube radius $R_T$ appears to be a
reasonable concept as it gives a good description of the X17,
E38, and anomalous soft photon masses as described in \cite{Won20},
subject to future amendments as more data become available.  So, for
the QED interaction with $\alpha_\qed$=1/137 and we use the same $R_T$
as used in the QCD mesons.

We list the theoretical masses of the neutral, $I_3$=0, QED mesons
obtained by Eq.\ (\ref{qed}) in Table I.  We find an $I$=0 isoscalar
QED meson at $m_{\rm isoscalar}^{\qed}$=17.9$\pm$1.5 MeV and an
$(I$=1,$I_3$=0) isovector QED meson at $m_{\rm
  isovector}^{\qed}$=36.4$\pm$3.8 MeV.  As the $I^G(J^{PC})$ quantum
numbers of the QCD mesons are known, we can infer the quantum numbers
of the corresponding QED mesons with the same $I$ and $S$ by
analogy. Such an inference by analogy provides a useful tool to
determine the quantum numbers and some electromagnetic decay
properties of QED mesons. Using such a tool, we find that the
isoscalar QED meson has quantum numbers $I^G(J^{PC})$=$1^-(0^{-+})$
and the isovector $I_3$=0 QED meson has quantum numbers
$I^G(J^{PC})$=$0^+(0^{-+})$.  Within the theoretical and experimental
uncertainties, the matching of the $I(J^\pi)$ quantum numbers and the
mass may make the isoscalar QED meson a good candidate for the X17
particle and the isovector QED meson for the E38 particle.
 It will
be of great interest to confirm or refute the existence these
hypothetical particles by independent experimental investigations, as
a test of the QED meson concepts.

In order to show the effects of the second quark condensate term
relative to the massless quark limit in (\ref{qed}), we tabulate in
Table I the results of the QED meson masses values obtained in the
massless quark limit.  One observes that the mass of the isoscalar QED
meson with the quark condensate is 17.9$\pm$1.5 MeV but it is reduced
to 11.2$\pm$1.3 MeV in the massless quark limit without the quark
condensate. 

\subsection{Effective charge numbers and quark space-time classical trajectories}

With the solutions of the the phenomenological QCD and QED open string
models well at hand, it is illuminating to examine in some detail the
effective charge numbers for the QCD and QED mesons.  The effective
charges are useful concepts as they present a simple and physical way
to study the dynamics and the interaction of the constituents in these
mesons in the presence of flavor symmetry and flavor mixing.  The
dynamics occur in such a way that even though quarks of different
charges in different flavors are involved, they can be equivalently
described as having a quark constituent interacting with its antiquark
constituent with an opposite effective charge.  With the introduction
of effective charge, the flavor dynamics is subsumed and greatly
simplified.  We list the effective color and electric charges of
$\pi^0$, $\eta$ and $\eta'$ in Table \ref{tb2}.  For other hadrons
when there is no flavor mixing, there is no need to introduce
effective charges, and their the color and electric quark charges are
those given by the standard quark model.

\begin{table}[h]
\centering
\caption { The effective color charge 
$\tilde Q_{q}^\qcd(i)$
and the effective electric charge
$\tilde Q_{q}^\qed(i)$
of the quark constituent in the  QCD or QED meson  $i$, in the presence of flavor mixing.
The mixing angle $\theta_p=-24.5^o$ is  from \cite{PDG19}.}
\vspace*{0.2cm}
\begin{tabular}{|c|c|c|c|c|c|c| c|}
\cline{2-7} \multicolumn{1}{c|}{}& & &  & & & \\ 
\multicolumn{1}{c|}{}&  particle $i$ & $J^\pi I$ & $\tilde Q_{q}^\qcd(i)$ &  $\tilde Q_q^\qcd(i)$
& $\tilde Q_{q}^\qed(i)$ & $\tilde Q_{q}^\qed(i)$ \\ 
\multicolumn{1}{c|}{}& & & & $\theta_p=-24.5^o$&  & $\theta_p=-24.5^o$\\ 
\hline
QCD&\,\,$\pi^0$ & 0$^- $1 & 0 &  0 &$1/\sqrt{2}$  & 0.7071 \\ 
\!\!meson\!\!&\!$\eta$ &0$^-$0 &~~$ -\sqrt{3}\sin \theta_p $~~ &0.7182 & 
~~$3\cos \theta_p/\sqrt{6}$~~ & 1.1144 \\ 
& $\eta'$  &$0^-0$ & $ ~~\sqrt{3}\cos \theta_p  ~~$& 1.576 &$~~ \sin \theta_p/\sqrt{6}$~~ &  -0.1693 \\ \hline
QED& isoscalar (X17)&0$^-0$& &  & $1/(3\sqrt{2})$  & \\ 
 \!\!meson\!\!& isovector(E38)  &$0^-1$ & & &   $1/\sqrt{2}$ & \\ 
\hline
\end{tabular}
\label{tb2}
\end{table}

It is interesting to note that for the isovector $\pi^0$ with isospin
$I=1, I_3=0$, the effective quark color charge $\tilde Q_{q}^\qcd$ is
zero and thus, $\pi^0$ is a Goldstone boson whose mass would be zero
if the mass comes only from the confining interaction.  That is, 
the contribution to the pion mass arising  from
the confining interaction is zero because the effective charge that
will respond to the confining force is zero.   The mass for $\pi^0$
arises only from the quark condensate, the second
term of Eq.\ (\ref{qed}), corresponding to the variational response of
the quark condensate due to the variation of the boson field as
indicated by Eq.\ (\ref{eq45}).  As noted by Georgi \cite{Geo22}, the
case of quarks with two flavors with a zero effective color charge is
a remarkable occurrence, because it leads to a confined boson with a
zero mass from the confining potential and is an example of an
``unparticle'' system without a mass scale or length scale, a case with
conformal symmetry.  There appears to be an automatic fine-tuning in
the two-flavor Schwinger model \cite{Geo20a}.  The $\pi^0$ in QCD is
however not massless and does not become an unparticle on account of
the quark condensate in the second term of Eq.\ (\ref{qed}), which
depends on the quark rest masses and the strength of the coupling
constant.  It is the forces associated with the presence of the
condensate and its variation with respect to the boson field that
gives the second term and the mass of the pion (Eq.\ (\ref{eq45})).
\begin{figure}[H]
\centering
\includegraphics[scale=0.30]{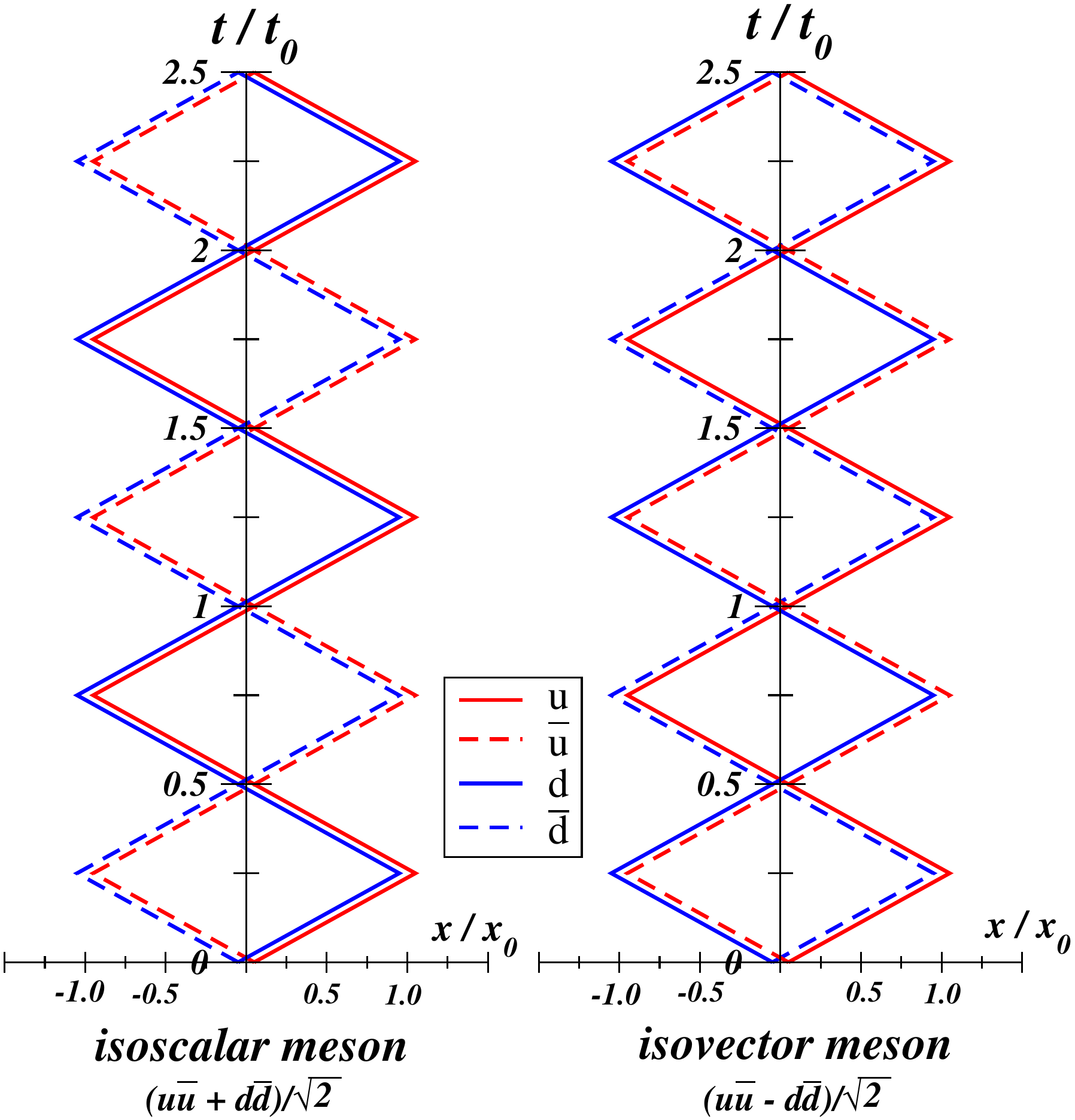}

\vspace*{0.2cm}\hspace*{-0.2cm}($a$)\hspace*{2.2cm}($b$)

\vspace*{-0.2cm} 
\caption{($a$) Classical space-time trajectories of massless quarks in
  the yo-yo motion of a meson with two flavors in the isospin singlet
  state with $I=0, I_3=0$.  The motion of the $u$ and $d$ quarks are
  in phase, and the quark charges effectively add together.  ($b$) Classical
  pace-time trajectories of massless quarks in yo-yo motion of a meson
  with two flavors in the isospin triplet state with $I=1, I_3=0$.
  The motion of the $u$ and $d$ quarks are out of phase, and the
  quark charges effectively subtract from each other.  }
\label{fig4a}
\end{figure}

 It would be intuitively illuminating to see how the effective charges
 in Table \ref{tb2} are related to the dynamics of quark constituents
 for the case with two flavors.  The effective charge $\tilde
 Q_i^\lambda$ for the quark constituent $q$ interacting with its
 antiquark constituent $\bar q$ in particle $i$ with isospin $I$
 interacting in the interaction of type $\lambda$ is
\begin{eqnarray}
\tilde Q_q^\lambda (i) =
    \sum_f^{N_f}  D_{if}^\lambda Q_{f}^\lambda=\frac{Q_u^\lambda+(-1)^IQ_d^\lambda}{\sqrt{2}}, 
\end{eqnarray}
for which we have $Q_u^\qcd=Q_d^\qcd=1$, and $Q_u^\qed=2/3$, and
$Q_d^\qed=-1/3$.  As indicated in Table \ref{tb2}, the effective color
charge $\tilde Q_{q}^\qcd(\pi^0)$=0, the effective electric charges
$\tilde Q_q^\qed$(isoscalar)=$1/(3\sqrt{2})$ and $\tilde
Q_q^\qed$(isovector)=$1/\sqrt{2}$.  We can examine and study the
space-time dynamics of the up and down quarks.  Treating the quarks as
massless and studying the classical trajectories of the constituents
in the yo-yo model in (1+1)D \cite{Art74}, we describe the quark
trajectories as described in Chapter 7 of \cite{Won94}.  We show 
 in
Fig.\ref{fig4a}($a$)
the yo-yo trajectories of the up and
down quarks in the flavor singlet $I=0$ state, and in Fig.\ \ref{fig4a}($b$)
the flavor triplet $I=1,I_3=0$ state.  We observe that in the
isoscalar $I$=0 case, the classical trajectories of the quark motion
is in phase, and the charges of the two quarks add together and in the
isovector $I$=1 case, the classical trajectories are out of phase and
the two charges subtract from each other.  On the other hand, the color
charges are of the same sign in QCD, but they are of different
magnitudes and signs in QED.  Therefore, there is a complete
cancellation of the color-charges for the case of QCD isovector case,
but an additive reinforcement in the QCD isoscalar case.  Thus, the
quark effective color charge for $\pi^0$ is zero, $\tilde
Q_q^\qcd(\pi^0)$=0.  On the other hand, for the case of electric
charges in QED, because the up quark and the down quark have different
magnitudes of electric charges with  opposite signs, the effective electric charge of
QED meson is less than the effective electric charge of the isovector QED
meson.  The factor of $1/\sqrt{2}$ come from the wave functions,
resulting in the effective electric charges $\tilde
Q_q^\qed$(isoscalar)=$1/(3\sqrt{2})$ for the $I$=0  QED meson and $\tilde
Q_q^\qed$(isovector)=$1/\sqrt{2}$ for the $(I,I_3)$=$(1,0)$ QED meson.

\section{ Experimental evidence for the  Possible Existence of the QED mesons  }

\subsection{Different modes of QED meson decays}

We have discussed in Section 2 the mechanism of the production 
of $q \bar q$ pairs in
many low- and high-energy $e^+$-$e^-$, hadron-hadron, and
nucleus-nucleus collisions.  A $q\bar q$ pair will be produced and
materialize as a QED or QCD meson final state, when the center-of-mass
energy $\sqrt{s}(q\bar q)$ of the pair coincides with the eigenenergy
of a confined QED or QCD meson.  In the energy range $(m_q + m_{\bar
  q}) <\sqrt{s}(q\bar q) < m_\pi$, a $q\bar q$ will be produced and
materialize as a QED meson when $\sqrt{s}(q\bar q)$ coincides with the
eigenenergy of a confined QED meson.  At energies different from the
eigenenergies of QCD and QED mesons, no $q\bar q$ pair will be
produced, because quarks cannot be isolated.

\begin{figure}[h]
\centering
\includegraphics[scale=0.90]{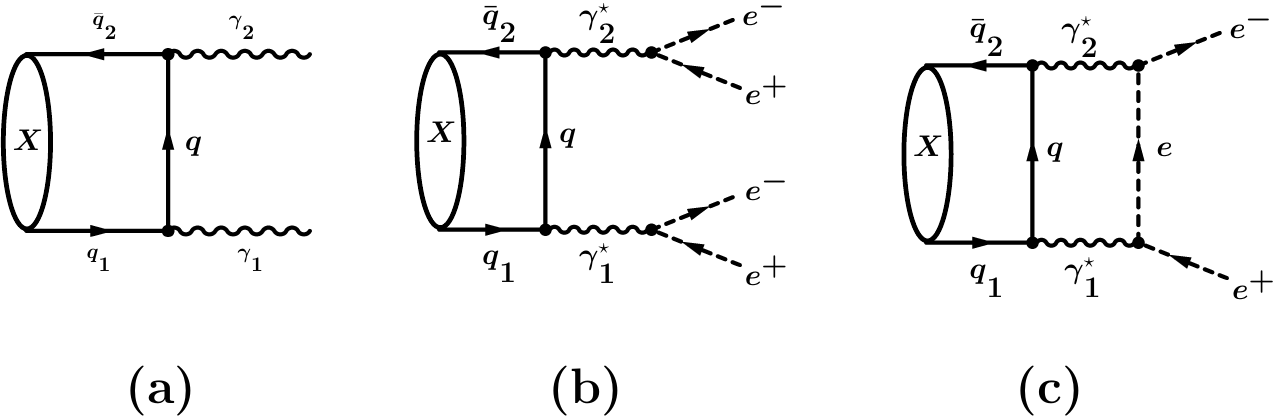}
\caption{($a$) A QED meson $X$  can decay into two real
  photons $X \to \gamma_1 + \gamma_2$, ($b$) It can decay into two
  virtual photons each of which subsequently decays into a $(e^+ e^-)$ pair,
  $X \to \gamma_1^* + \gamma_2^* \to (e^+ e^-) + ( e^+ e^-) $, and
  ($c$) it can decay into a single $(e^+ e^-)$ pair, $X \to \gamma_1^*
  + \gamma_2^* \to e^+ e^-$.  }
\label{fig4}
\end{figure}

Methods for the detection of various QCD mesons are well known and we
do not need to belabor them again.  On the other hand, a QED meson can
be detected by its decay products from which its invariant mass can
be measured.  It is necessary to know its decay modes.

In (1+1)D, a QED meson with massless quarks cannot decay as the quark
and the antiquark execute yo-yo motion along the string.  The exchange
photon in (1+1)D between the quark and the antiquark constituent leads
to longitudinal confinement, and there are no free transverse photons.
The open string in one dimension is however only an idealization of a
flux tube.  In the physical (3+1)D, the structure of the flux tube and
the transverse photons must be taken into account.  In (3+1)D
space-time, the quark and the antiquark at different transverse
coordinates in the flux tube traveling from opposing longitudinal
directions in a QED meson can make a turn to the transverse direction
by which the quark and the antiquark can meet and annihilate, leading
to the emission of two real transverse photons 
$\gamma_1\gamma_2$
as depicted in  the
Feynman diagram Fig.\ \ref{fig4}(a).  A QED meson can decay into
two virtual photons 
$\gamma_1^*\gamma_2^*$ 
each of which subsequently decays into an $e^+e^-$
pair as
$(e^+e^-)$$(e^+e^-)$  shown in Fig.\ \ref{fig4}($b$).  The coupling of the
transverse photons to an electron pair leads further to the decay of
the QED meson into an electron-positron pair 
$e^+e^-$
as shown in
Fig.\ \ref{fig4}(c).  The mass of a QED meson can be obtained by
measuring the invariant mass of its decay products.

\subsection{The observation of the anomalous soft photons}

In exclusive high-energy $K^+ p$ \cite{Chl84,Bot91}, $\pi^+ p$
\cite{Bot91}, $\pi^- p$ \cite{Ban93,Bel97,Bel02pi}, $pp$ collisions
\cite{Bel02}, and $e^+$$e^-$ annihilations
\cite{DEL06,DEL08,Per09,DEL10} at the $Z^0$ resonance at
$\sqrt{s}$=91.1876 GeV \cite{Per09,DEL06,DEL08,DEL10}, a large number
of hadrons are produced.  Photons are also produced and are detected 
as converted $e^+ e^-$ pairs  in these
reactions.  In these exclusive measurements in which the secondary
photons from hadron decays can be tracked and separated out from the 
direct photons, 
attention can be focused on the produced direct photons and their correlation
with the hadron products.

 Upon choosing the longitudinal axis as the incident hadron axis in
 hadron-hadron collisions, or the jet axis of one of the jets in
 $e^+$-$e^-$ annihilations, the associated direct photons can be
 further classified as direct soft  photons and direct hard photons.  The (direct) soft photons are
 those with a transverse momentum $p_T$ less than about 60 MeV/c, while
 the (direct) hard photons are those with $p_T$ greater than about 60 MeV/c.  The
 bremsstrahlung process with the production of   both soft and hard
 photons, but no hadrons, has been  studied in $e^+$-$e^-$ elastic
 scattering and in $e^+ + e^- \to \mu^+ + \mu^-$ reactions as a purely
 electroweak processes, and such bremsstrahlung process has been found to
 agree well with QED considerations \cite{DEL08}.

Experimental measurements of the (direct) soft photons associated with
hadron production is a interesting problem because it involves the
soft photon production in QED, the hadron production in QCD, and the
interplay between QCD and QED particle production processes.
Exclusive measurements of the soft photon production over a period of
many decades since 1984 consistently exhibit the anomalous soft photon
anomaly as we shall describe in detail below.

\subsection{The Low Theorem}

Because the Low Theorem \cite{Low58} plays an important role in
quantifying the anomalous soft photon anomaly, it is worthwhile to
review its theoretical foundation and its contents.

\begin{figure} [H]
\centering
\includegraphics[scale=0.65]{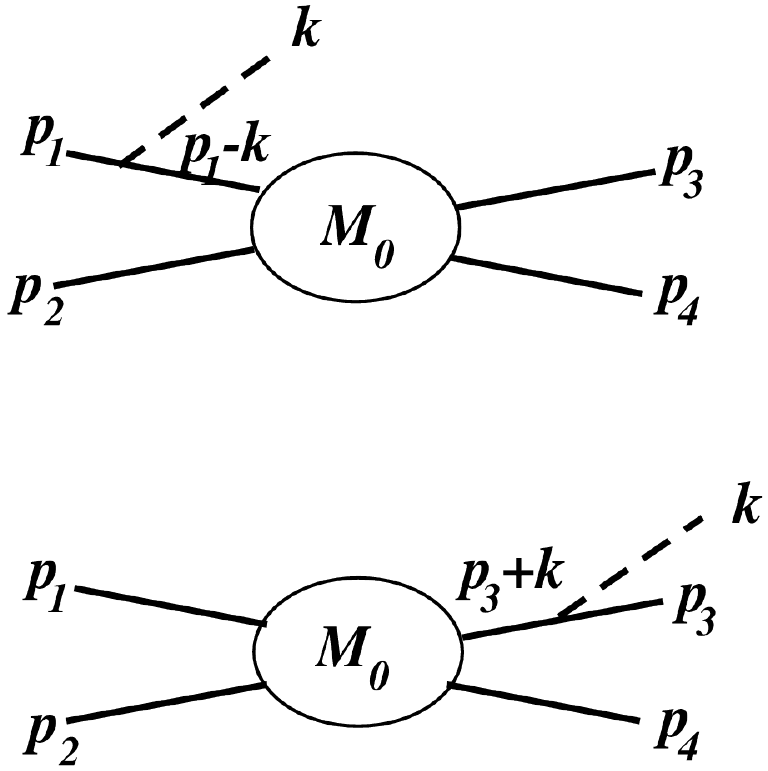}
\hspace{4.0cm}
\caption{ Feynman diagrams 
for~$p_1$+$p_2$$\to$$ p_3$+$ p_4$+$ k$.\hspace*{5.0cm}
  }
\label{fig5}
\end{figure}

\vspace*{0.20cm} We consider the process $p_1+p_2 \to p_3 + p_4 + k$
where $p_i$ represents a hadron and its momentum, and $k$ represents a
photon and its momentum.  For the simplest case with neutral $p_2$ and
$p_4$, and charged $p_1$ and $p_3$ hadrons, the Feynman diagrams are
shown in Fig.\ 6.  The amplitude for the production of a photon with a
polarization $\epsilon$ is

\begin{eqnarray}
M(p_1p_2; p_3p_4 k)& =& M_0(p_1p_2; p_3p_4 )\left (\frac{e_1 p_1\cdot \epsilon}{(p_1-k)^2}
+\frac{e_3 p_3\cdot \epsilon}{(p_3+k)^2}\right ) ~~~~~~~\nonumber \\
 &  &\hspace*{-1.1cm}=M_0(p_1p_2; p_3p_4 )\left (
\sum_i^{\rm all~charged~particles}\frac{\eta_i e_i p_i\cdot \epsilon}{2p_i \cdot k}
\right )\!,
\label{2.1}
\end{eqnarray}
where $e_i$ is the charge of $p_i$, and $\eta_i$ is +1 for an outgoing
hadron and -1 for an incoming hadron respectively.  In obtaining the
above equation, we have assumed soft photons with small $k$ values 
so that the production amplitude satisfies
\begin{eqnarray}
M_0(p_1-k~~ p_2;~ p_3 ~p_4)\sim  M_0(p_1 ~p_2; ~p_3+k  ~~p_4)\sim M_0(p_1 ~p_2; ~p_3 ~p_4).
\end{eqnarray}
This is a reasonable assumption in high energy processes in which
$|{\bf p_1}|$ and $|{\bf p_3}|$ in the C.M.  frame are much greater
than the transverse momentum of the soft photon, $k_T$.  The amplitude
$M_0$ with the production of a soft photon is then approximately
independent of $k$ and can be adequately represented by $M_0(p_1p_2;
p_3p_4 )$, the Feynman amplitude for the production of only hadrons.

We can generalize the above Eq.\ (\ref{2.1}) to the process $p_1+p_2
\to p_3 + p_4 +...+ p_N+ k$ where $p_i$ is a hadron and $k$ is a soft
photon.  The Feynman amplitude is
\begin{eqnarray}
M(p_1p_2; p_3p_4... p_N k)& =& M_0(p_1p_2; p_3p_4...p_N )\left (
\sum_i^{\rm all~charged~particles}\frac{\eta_i e_i p_i\cdot \epsilon}{2p_i \cdot k}
\right ),
\end{eqnarray}

\noindent 
From the relation between Feynman amplitudes and cross sections, the
above equation gives \cite{Low58}
\begin{eqnarray}
\frac{  dN_\gamma}{ d^3 k}= \frac{\alpha}{2\pi k_0 } 
\int d^3p_1  d^3p_2d^3p_3 ... d^3p_N
\sum_{i,j=1}^{N}\eta_i \eta_j e_i e_j \frac{- (p_i \cdot p_j)} 
{(p_i \cdot k )(p_j \cdot k)}
\frac{dN_{\rm hadrons}}{d^3p_1  d^3p_2d^3p_3 ... d^3p_N},
\label{eq43}
\end{eqnarray}
where the sum over $i$ and $j$ include all incoming and outgoing
primary charged particles.  In exclusive measurements, the
distribution of all participating incoming and outgoing charged
particles are track and detected, the quantity ${dN_{\rm
    hadrons}}/{d^3p_1 d^3p_2d^3p_3 ... d^3p_N}$ is measured, and the
sum in Eq.\ (\ref{eq43}) can be carried out.  Thus, the spectrum of
soft photons arising from QED bremsstrahlung can be calculated from
exclusive measurements on the spectrum of the produced hadrons.

\subsection{
Experimental Measurements of Anomalous Soft Photon in Hadron production}

In which part of the photon spectrum are soft photons expected to be
important?  Upon choosing the beam direction as the longitudinal axis,
Eq.\ (\ref{eq43}) indicates that the contributions are greatest when
the transverse momentum of the photon, $k_T$ $\propto$ $p_i \cdot k$,
is small, as pointed out by Gribov \cite{Gri67}.  Hence, it is of
great interest to measure the yield of soft photons with small
transverse momenta.
 
Many high-energy experiments were carried out to obtain the spectra
of soft photons with transverse momenta of the order of many tens of MeV/c,
with the soft photons detected as converted $e^+ e^-$ pairs. 
The soft photon spectra  were then compared with what would be  expected from
electromagnetic bremsstrahlung of the hadrons as given by
Eq.\ (\ref{eq43}). 

Anomalous soft photon production in excess of QED bremsstrahlung
predictions by a factor of 4.0$\pm$0.8 in association with hadron
production was first observed in 1984 by the WA27 Collaboration at
CERN using the BEBC bubble chamber in $K^+ +p$ collisions at $p_{\rm
  lab}(K^+)$=70 GeV \cite{Chl84}.  For over several decades since that
time, in many exclusive measurements in high-energy $K^+ p$
\cite{Bot91}, $\pi^+ p$ \cite{Bot91}, $\pi^- p$ \cite{Ban93,Bel97,Bel02pi},
$pp$ collisions \cite{Bel02}, and $e^+$$e^-$ annihilations
\cite{DEL06,DEL08,Per09,DEL10}, it has been consistently and repeatedly observed that
whenever hadrons are produced, anomalous soft photons in the form of
excess $e^+e^-$ pairs, about 4 to 8 times of the bremsstrahlung
expectations, are proportionally produced, and when hadrons are not
produced, these anomalous soft photons are also not produced as shown
in Fig.\ \ref{fig6a} \cite{DEL08}.   
The transverse momenta of these 
excess $e^+$$e^-$ pairs lie in the range of a few MeV/c to many tens
of MeV/c, corresponding to a mass scale of the $e^+e^-$ pair from a
few MeV to many tens of MeV (Fig.\ \ref{fig7}).  The anomalous soft photon measurements
are summarized well by Perepelitsa in \cite{Per09} in Table \ref{tb3}.

\begin{table}[h]
\caption { The ratio of the soft photon yield associated with hadron
  production to the bremsstrahlung yield in high-energy hadron-hadron
  collisions and $e^+$-$e^-$ annihilations, compiled by
  V. Perepelitsa \cite{Per09}.  }
\vspace*{0.3cm}\hspace*{0.1cm}
\begin{tabular}{|l|l|c|c|}
\cline{1-4}
~~~~~~~~~~Experiment    &   Collision   &  Photon $k_T$    &     Photon/Brem
  \\
                     	&	Energy               &                           &    Ratio
 \\ \hline
     $K^+ p$, CERN,WA27,  BEBC (1984)          &~~70 GeV/c    &  $k_T <$ 60 MeV/c & 4.0 $\pm$0.8
 \\ \hline
     $K^+ p$, CERN,NA22,  EHS (1993)          & 250 GeV/c    &  $k_T <$ 40 MeV/c & 6.4 $\pm$1.6
 \\ \hline
     $\pi^+ p$, CERN,NA22,  EHS (1997)          & 250 GeV/c    &  $k_T <$ 40 MeV/c & 6.9 $\pm$1.3
 \\ \hline
     $\pi^- p$, CERN,WA83,OMEGA (1997)          & 280 GeV/c    &  $k_T <$ 10 MeV/c & 7.9 $\pm$1.4
 \\ \hline
      $\pi^+ p$, CERN,WA91,OMEGA (2002)          & 280 GeV/c    &  $k_T <$20 MeV/c & 5.3 $\pm$0.9
 \\ \hline
      $p p$, CERN,WA102,OMEGA (2002)          & 450 GeV/c    &  $k_T <$20 MeV/c & 4.1 $\pm$0.8 
 \\ \hline
      $e^+$$e^-$$ \to$hadrons, CERN,DELPHI   & $\sim$91 GeV(CM)    &  $k_T <$60 MeV/c &~ 4.0   
 \\ 
      with hadron production (2010)       &  &   & 
 \\ \hline
      $e^+$$e^-$$ \to$$\mu^+$$\mu^-$, CERN,DELPHI   & $\sim$91 GeV(CM)    &  $k_T <$60 MeV/c &~ 1.0   
 \\ 
      with no hadron production (2008)       &  &   & 
 \\ \hline
\end{tabular}
\label{tb3}
\end{table}
These experimental measurements indicate that low-$k_T$ soft photons
are produced in excess of what is expected from the QED
electromagnetic bremsstrahlung process.  In particular, in DELPHI
measurements in high-energy $e^+$-$e^-$ annihilations in $Z^0$
hadronic decay, the ratio of the soft photon yield to the
bremsstrahlung yield associated with hadron production is about 4
\cite{DEL10}, whereas the ratio of soft photon yield to the
bremsstrahlung yield in the corresponding $e^+ +e^- \to \mu^+ +\mu^-$
reaction is about 1 \cite{DEL06}.  This indicates clearly that
anomalous soft photons are present only when hadrons are produced.
The QCD hadron production is accompanied by an additional 
soft photon source appearing as excess $e^+e^-$ pairs.  The anomalous soft photons provide an
interesting window to examine non-perturbative aspects of QCD and QED
particle production processes.

\subsection{DELPHI measurements of anomalous soft photon in high-energy $e+$-$e^-$ annihilations}

\begin{figure} [H]
\vspace*{-0.3cm}\hspace*{2.5cm}
\includegraphics[scale=0.70]{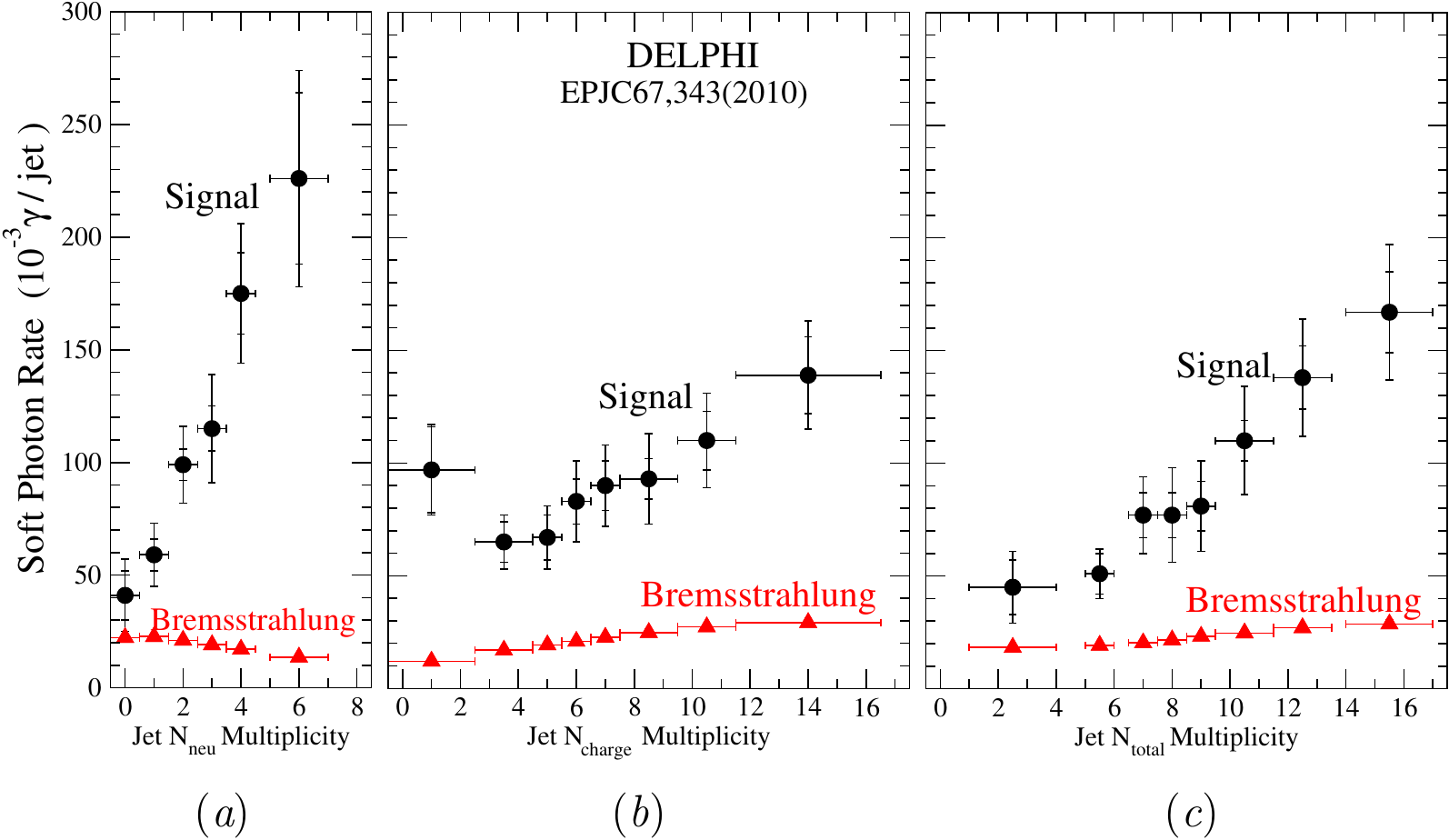}
\vspace*{-0.3cm}
\caption{ Number of soft photons $N_\gamma$ in units of $(10^{-3}
  \gamma$/ jet),
   shown as solid circular points, 
  as a function of ($a$) neutral hadron multiplicity 
  $N_{\rm neu}$, ($b$) charged hadron multiplicities, $N_{\rm
    charge}$, and ($c$) the total hadron multiplicity, $N_{\rm
    total}$, from the DELPHI Collaboration \cite{DEL10}.  The number of soft photons from the bremsstrahlung process is shown as triangular data points.   }
\label{fig6a}
\end{figure}

DELPHI carried out a series of quantitative measurements to study the
anomalous soft photons in the hadronic decay of the Z0 resonance at
$\sqrt{s}$=91.19 GeV at CERN.  In addition to measuring the overall
ratio of soft photon production, the DELPHI Collaboration measured the
soft photon yields in different regions of the phase space in coincidence with
various hadron production variables.  They provide a wealth of
information on the characteristics of the anomalous soft photons in
association with hadron production \cite{DEL06}-\cite{DEL10}.  The
main features of the observations from the DELPHI Collaboration can be
summarized as follows:
\begin{enumerate}
\item
Anomalous soft photons in the form of excess $e^+e^-$ pairs 
are produced in association with hadron
production at high energies.  They are absent when there is no hadron
production as demonstrated in Ref. \cite{DEL08} and in Fig. \ref{fig6a}(c).  The
anomalous soft photon yield is proportional to the hadron yield as
shown in Fig.\ \ref{fig6a}(c).

\item
The anomalous soft photon yield increases approximately linearly with
the number of produced neutral or charged particles, but, the
anomalous soft photon yields increases much faster with increasing
neutral particle multiplicity than with charged particle multiplicity,
as shown in Fig.\ \ref{fig6a}(a) and \ref{fig6a}(b).

\item
The transverse momenta of the anomalous soft photons are in the region
of many tens of MeV/c, as shown in Fig.\ \ref{fig7}($b$).
\end{enumerate}

\subsection{Models of Anomalous Soft Photons}

Many different model have been proposed to explain the anomalous soft
photon production in association with hadron production.  As the
transverse momenta of the anomalous photons are of order of many tens
of MeV, Van Hove and Lichard suggested that there is a source of these
low-energy photons in the form of a glob of cold quark-gluon system of
low temperature with $T \sim$10 to 30 MeV at the end of parton
virtuality evolution in hadron production \cite{Van89,Lic94}.  In such a
cold quark-gluon plasma, soft photons may be produced by $q+\bar q \to
\gamma +g$ or $g+q \to \gamma + q$, and will acquire the
characteristic temperature of the cold quark-gluon plasma.

Kokoulina $et~al.$ followed similar idea of a cold quark-gluon plasma
as the source of soft photon production \cite{Kok07}.  Collaborative
evidence of a cold quark-gluon plasma with $T\sim$ 10 to 30 MeV from
other sources however remain lacking.

Barshay proposed that pions propagate in pion condensate and they emit
soft photons during the propagation.  Rate of soft photon emission
depends on the square of pion multiplicity \cite{Bar89}.  The concept
of a pion condensation in high-energy $e^+$-$e^-$ annihilations in
$Z^0$ hadronic decay has however not been well established.

Shuryak suggested that soft photons are produced by pions reflecting
from a boundary under random collisions. Hard reflections lead to no
effect, but soft pion collisions on wall leads to large enhancement in
soft photon yield \cite{Shu89}.

Balek, Pisutova, and Pisut presented a review of the data analysis,
the corrections to the Low Theorem, and the proposed models up to 1989
\cite{Bal90}.

Czyz and Florkowski proposed that soft photons are produced by
classical bremsstrahlung, with parton trajectories following string
breaking in a string fragmentation.  They suggested that photon
emissions along the flux tube agree with the Low limit whereas photon
emissions perpendicular to the flux tube are enhanced over the Low
limit \cite{Czy94}.

Nachtmann $et~ al.$ suggested that soft photons produced by
synchrotron radiation from quarks in the stochastic QCD vacuum
\cite{Nac94,Leb22}.  Hatta and Ueda suggested that soft photons are
produced in ADS/CFT supersymmetric Yang-Mills theory \cite{Hat10}.
Darinian $et~al.$ suggested that they arose from the Unruh radiation
of quarks \cite{Dar91}.  Simonov suggested that soft photons arose
from closed quark-antiquark loop \cite{Sim08}.  Kharzeev and Loshaj
suggested that the soft photon arise from the continuous spectrum from
the induced currents in the Dirac sea \cite{Kha14}.

\subsection{QED-confined  $q \bar q$ meson description of the anomalous soft photons}

We would like to focus our attention on 
 an 
open-string description of QED-confined $q \bar q$ meson as the source
of the anomalous soft photons because it  has the prospect of 
 linking  many
different anomalies together in a consistent framework.   As the anomalous soft photons arise
from excess electron-positron pairs, the parent particles of the
anomalous soft photons must be neutral objects.  The results from
DELPHI and Fig.\ 7 indicate that they are proportionally produced only
as hadrons are produced, and their rate of production in association
with neutral hadrons is much greater than the  production rate in
association with a charged hadron.  The simultaneous and correlated
production alongside with hadrons suggest that the neutral parent
particle of the anomalous soft photons is likely to contain some
elements of the hadron sector, such as a light quark and a light
antiquark of the same flavor.  The elements of the hadron sector
comprise of $u$, $d$, $c$, $s$, $b$, $t$ quarks, antiquarks, and
gluons.  The mass scale of the anomalous soft photons excludes all but
the $u$ and $d$ quarks and antiquarks as possible constituents. 

 The
quark and antiquark constituents carry color and electric charges and
they interact mutually with the QCD and QED interactions.  A parent
particle of the anomalous soft photons cannot arise from the
quark-antiquark pair interacting with the QCD interaction
non-perturbatively, because such an interaction will endow the pair
with a mass much greater than the mass scale of the anomalous soft
photons.  We are left with the possibility of the quark and the
antiquark interacting with the QED interaction.  We note that 
light $u$ and $d$ 
quarks  have small masses, they
reside predominantly in (1+1)D, and they interact in the QED interaction.
They fit the requirements  for which the Schwinger confinement mechanism \cite{Sch62,Sch63} is applicable.
We can therefore apply the Schwinger confinement mechanism to quarks to 
conclude that
 a light quark $q$ and its antiquark $\bar q$ will be confined as
a $q\bar q$ boson in the Abelian U(1) QED gauge interaction in (1+1)D,
as in an open string, with a mass 
proportional to the coupling constant of the interaction. 
  From the work of Coleman, Jackiw, and Susskind \cite{Col75,Col76},
we can infer further that the Schwinger confinement mechanism persists
 even for massive quarks in (1+1)D. 
Such a possibility of confined $q\bar q$  is
further reinforced by the special nature of a confining gauge
interaction, for which the greater the coupling constant of the
attractive confining interaction, the greater will be the mass of the
composite particle it generates (see Eq.\ (\ref{eq6})), in contrast to
a non-confining interaction in which the effect is just the
opposite\footnote{ For the confining QED interaction in (1+1)D in the
Schwinger mechanism, $m=g_\2d/ \sqrt{\pi}$, the mass $m$ increases
with an increase in the coupling constant $g_\2d$, say, from the QED to
the QCD interaction. In contrast, in a non-confining QED interaction
between an electron and a positron, the mass of a positronium is
$m_{\rm positronium}=2m_e - m_e \alpha_c^2/4 n^2$, for which the
mass of the positronium  decreases as the coupling constant $\alpha_c$
increases.}.  Relative to the QCD interaction, the QED interaction
will bring the quantized mass of a $q\bar q$ pair to the lower mass
range of the anomalous soft photons, as Eq.\ (6) indicates.
It was therefore proposed in \cite{Won10,Won11,Won14} that a quark and
an antiquark in a $q\bar q$ system interacting with the QED
interaction may lead to new open string bound states (QED-meson
states) with a mass of many tens of MeV. These QED mesons may be
produced simultaneously with the QCD mesons in the string
fragmentation process in high-energy collisions
\cite{Chl84,Bot91,Ban93,Bel97,Bel02pi,Bel02,DEL06,DEL08,Per09,DEL10}, and the
excess $e^+e^-$ pairs may arise from the decays of these QED mesons.

The phenomenological open string description of the the QCD and QED
mesons in the last Section indicates that $\pi^0, \eta$, and $\eta'$
particles can be adequately described as open string $q\bar q$ QCD
mesons.  By extrapolating into the $q\bar q$ QED sector in which a
quark and an antiquark interact with the QED interaction, we find an
open string isoscalar $I(J^\pi)$=$0(0^-)$ QED meson state at
17.9$\pm$1.5 MeV and an isovector $(I(J^\pi)$=$1(0^-), I_3$=0) QED
meson state at 36.4$\pm$3.8 MeV as listed in Table I.  We can make an
approximate consistency check on these extrapolated masses using the
transverse momentum distributions of the anomalous soft photons.

\subsection{Transverse momentum distributions of anomalous soft photons }

\begin{figure} [H]
\centering
\includegraphics[angle=0,scale=0.70]{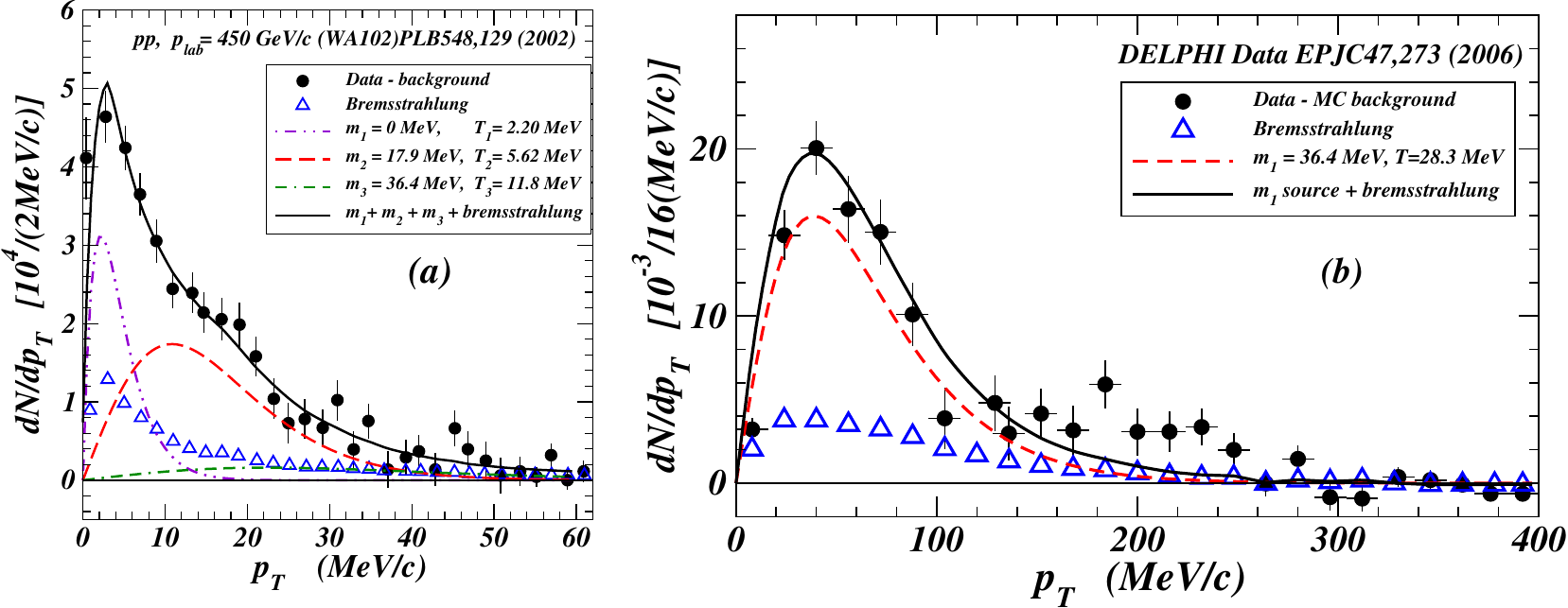}
\caption{ (a) Anomalous soft photon $dN/dp_T$ data from $pp$
  collisions at $p_{\rm lab}$=450 GeV/c obtained by Belogianni
  $et~al.$ \cite{Bel02}.  (b) Anomalous soft photon $dN/dp_T$ data
  from the DELPHI Collaboration for $e^+e^-$ annihilation at the $Z_0$
  mass of 91.18 GeV \cite{DEL06}.  The solid circular points represent
  the experimental data after subtracting the experimental background,
  and triangle points represent the deduced bremsstrahlung
  contributions.   The component yields from different
  masses of the thermal model are shown as separate curves.  The total theoretical yields   from the different masses and the QED bremsstrahlung contribution are shown as the
  solid curves. }
\label {fig7}
\end{figure}

The WA102 Collaboration \cite{Bel02} and the DELPHI collaboration
\cite{DEL10} measured the transverse momentum distributions of the
anomalous soft photons which provide valuable information on the
masses of the parent bosons of the anomalous soft photons.  They are
exclusive measurements in which the momenta of all participating
charged particles are measured.  The knowledge of the momenta of all
initial and final charged particles allows an accurate determination
of the QED bremsstrahlung $dN/dp_T$ distributions shown as triangular
points in Fig.\ \ref {fig7}.  As shown in Fig.\ \ref {fig7}, the
observed yields of soft photons exceed the bremsstrahlung
contributions substantially in both $pp$ collisions in \ref {fig7}(a)
and $e^+e^-$ annihilations in \ref {fig7}(b).  The solid circular
points in Fig.\ \ref{fig7}  show the experimental $dN/dp_T$ data of
soft photons after subtracting the experimental background.
Fig.\ \ref {fig7}(a) gives the WA102 data for $pp$ collisions at
$p_{\rm lab}$=450 GeV/c from Belogianni $et~al.$ \cite{Bel02}, and
Fig.\ \ref {fig7}(b) is the DELPHI results for $e^+e^-$ annihilation
at the $Z_0$ mass of 91.18 GeV \cite{DEL06}.

We would like to inquire how the production of the QED mesons may be
consistent with the transverse momentum distribution of the anomalous
soft photons.  For such an investigation, we rely on the thermal model
which describes well the transverse momentum distributions in the
production of hadrons of different masses in high-energy $pp$
collisions \cite{Hag65,Abe07,Abe09,Ada11}.  We assume that the thermal
model can be extended from the production of QCD mesons to the
production of QED mesons whose decay products are assumed to appear as
anomalous soft photons.  In such a thermal model, the transverse
momentum distribution of the
produced QED mesons is related to the produced
QED meson mass $m$ by \cite{Hag65}
\begin{eqnarray}
\frac{dN}{ p_T dp_T}=A e^{-\sqrt{m^2+p_T^2}/T}.
\end{eqnarray}
The contribution to the total $dN/dp_T$ from each boson of mass $m$ is
proportional to $p_{{}_T}e^{-\sqrt{m^2+p_T^2}/T}$ which is zero at
$p_T$=0 and has a peak at the location $p_T$ given by
\begin{eqnarray}
p_T^2 = \frac{1}{2} [ T^2 + \sqrt{T^4 +4T^2m^2}].
\end{eqnarray}
 If $m=0$, then $dN/dp_T$ peaks at $p_T=T$.  If $m$ is much greater
 than $T$, then $dN/dp_T$ peaks at $p_T \sim$ $\sqrt{mT}$.  Hence, for
 each contributing boson mass component, the thermal model gives a
 distribution that starts at zero at $p_T$=0 and reaches a peak of
 $dN/dp_T$ and decreases from the peak.  The total $dN/dp_T$ is a sum
 of contributions from different QED mesons and boson components,
\begin{eqnarray}
\frac{dN}{p_T dp_T}\!=\!  \sum_{i}A_i e^{-{\sqrt{m_i^2+p_T^2}/T_i}}.
\end{eqnarray}
There can be as many contributing bosons as the number of underlying
peaks in the $dN/dp_T$ spectrum.  While different decompositions of
the spectrum into different masses (and peaks) are possible, the
structure of the $dN/dp_T$ data appears to require many components in
Fig.\ \ref{fig7}(a) and only a single component in Fig. \ref {fig7}(b).  In the
thermal model analysis of the $pp$ data in Fig. \ref {fig7}(a), we
note that there appears to be a boson component of real photons with
$m_1$=0.  Because of the $m_i T_i$ ambiguity\footnote{In fitting the
thermal model, the $m_iT_i$ ambiguity gives different values of $m_i$
for different values of $T_i$ without changing significantly the
overall quality of the fitting.} associated with the product of the
meson mass $m_i$ and the temperature $T_i$, we are content with only a
consistency analysis.  We assume that QED isoscalar and isovector
mesons with masses as given by Table \ref{tb1} are produced in the collision,
and their subsequent decay into $e^+ e^-$ pairs give rise to the
excess $e^+ e^-$ pairs observed as anomalous soft photons.  Allowing
other parameters to vary, the thermal model fit in Fig.\ \ref
{fig7}(a) is obtained with parameters $A_1$=$3.85\times10^4$$/(2{\rm
  MeV}/c$), $T_1$=2.20 MeV, $A_2$=6.65$\times10^4$$/(2({\rm MeV}/c$),
$T_2$=5.62 MeV, $A_3$=0.266$\times10^4$$/(2({\rm MeV}/c$), and
$T_3$=11.8 MeV, where the different components also shown as separate
curves.  Adding the contributions from the three components onto the
bremsstrahlung contributions yields the total $dN/dp_T$ shown as the
solid curve.  The comparison in Fig.\ \ref {fig7}(a) indicates that
the $pp$ data are consistent with a photon component and a boson
component with a mass around 17 MeV.  The magnitude of the $m_3$=36.4
MeV component is of the same order as the bremsstrahlung contribution
or the noise level, and is rather uncertain  in Fig.\ \ref{fig7}($a$).  In Fig.\ \ref {fig7}($b$),
the addition of the single component with
$A_1$=2.7$\times$10$^{-3}$$/(16{\rm MeV}$/c), $m_1$=36.4 MeV, and
$T_1$=28.3 MeV onto the bremsstrahlung contributions gives a
consistent description of the soft photon data in $e^+e^-$
annihilations shown as the solid curve.

The component with $m_1$=0 in Fig.\ \ref {fig7}(a) may be associated
with the decay of the QED mesons into two photons.  If so, it will be
of interest to measure the $\gamma \gamma$ invariant mass to look for
diphoton resonances, as carried out in
\cite{Abr12,Abr19,Ber11,Ber14,Sch11,Sch12}.  The $m_2$=17.9 MeV
components in Fig.\ \ref {fig7}(a) and the $m_3$=36.4 MeV component in
Fig.\ \ref {fig7}(b) may be associated with the predicted isoscalar
and isovector QED mesons of Table \ref{tb1}.  If so, a measurement on
the invariant masses of the $m_2$ and $m_3$ components will be of
great interest to confirm the existence of these QED mesons.  The
recent reports of the observation of a hypothetical E38 boson at 38
MeV and the structures in the $\gamma \gamma$ invariant masses at
10-15 MeV and 38 MeV \cite{Abr12,Abr19,Ber11,Ber14,Sch11,Sch12}
provide encouraging impetus for further studies.

In a recent study, D'yachencho
and collaborators  re-analyzed
the transverse momentum distributions of anomalous soft photons of the
WA102 Collaboration \cite{Bel02} and the momentum distribution of
the reaction of  $p+C
\to 2\gamma + X$  \cite{Abr12} 
and reached similar  conclusions on possible
production of neutral X17 and E38 bosons in these reactions
\cite{Dya21,Dya21a}.

\subsection{Observation of the anomalous X17 particle in 
  $^{3}$H($p$,$e^+e^-$)$^{4}$He$_{\rm g.s.}$
  and $^{7}$Li($p$,$e^+e^-$)$^{8}$Be$_{\rm g.s.}$ 
  }

The anomalous soft photons discussed in the last subsection are
produced in highly-relativistic reactions in which the produced
particles travel with velocities close to the speed of light with a
transverse momentum $p_T$ of the order of many tens of MeV/c.  Hence
the parent particles of the anomalous soft photon that decay into
electron-positron pairs would have a rest mass of order $m\sim p_T/
$(speed of light), or many tens of MeV/c$^2$.  They indicate possible
existence of neutral particles in the mass regions of many tens of
MeV.  The extraction the anomalous particle masses from the $p_T$
spectrum in the thermal model analysis has some degrees of
uncertainty.  The particle masses can be better ascertained by direct
measurements.  Independent of the anomalous soft photons, two other
methods have been employed in the search of light neutral bosons.  One
uses the decay of the neutral boson into an $e^+e^-$ pair
\cite{deB96,deB97,deB01,Vit08,deB11,Kra16,Kra19,Kra21,Kra21a,Sas22,Kra22}, and the other
uses the decay into two real photons \cite{Abr09,Abr12,Abr16,Abr19}.
 
In search of possible candidate particles for the axion \cite{Don78}, de Boer and
collaborators initiated a program to study the $e^+ e^-$ spectrum in
low-energy proton fusion of light nuclei
\cite{deB96,deB97,deB01,Vit08}.  The E1 $e^+ e^-$ decay of the
17.2 MeV state in $^{12}$C, and the M1 $e^+ e^-$ decay of the 17.6 MeV
state in $^8$Be to their respective ground states were examined to
look for short-lived neutral bosons with masses between 5 and 15
MeV/c$^2$. Whereas for the E1 decay at large correlation angles
exhibits no deviation from internal pair conversion, the M1 angular
correlation in the decay of the excited 1$^+$ state of $^8$Be at 18.15
MeV surprisingly deviates from internal pair conversion at the
4.5$\sigma$ level \cite{deB96, deB97,deB01}.  In collaboration with
ATOMKI, the group reported a neutral particle at 12 $\pm$ 2.5 MeV in
the decay of the excited 1$^+$ isovector 17.64 state of $^8$Be
\cite{Vit08}.
 
Krasznarhokay and collaborators at ATOMKI has been continuing the
search for a neutral boson with an improved $e^+e^-$ spectrometer.  In
a subsequent experiment in the reaction of $p$+$^7$Li$\to e^+$+$e^-$+$^8$Be$_{\rm
  ground\, state}$ at proton energies about 1 MeV, the 18.15
MeV $J^\pi I$=$1^+ 0 $ excited $^8$Be$^*$ state was observed to decay
to the $^8$Be ground state by the emission of a hypothetical neutral
``X17'' boson with a mass of 16.70$\pm$0.35(stat)$\pm$0.5(syst) MeV
\cite{Kra16}.  The ATOMKI observation of such an X17 boson has
generated a great deal of interest.  Even though a neutral isoscalar
QED-confined $q\bar q$ state was predicted earlier to have a mass of
12.8 MeV in Table I of \cite{Won10}\footnote{ The prediction of the
mass of 12.8 MeV for the isoscalar QED meson in Table I of
\cite{Won10} was calculated in the massless quark limit.  When the
correction of the non-zero quark mass has been taken into account, the
predicted mass of the isoscalar QED meson increases to 17.9 MeV
\cite{Won20}}, the observed neutral boson with a mass of 17 MeV led to
many speculations inside and  outside the known 
families  of particles of 
the Standard Model
\cite{Won10,Won20,Zha17,Fen16,Fro17,Bat15,Ros17,Bor19,Bor19a,Cha22, Ell16,Alv18,Mun18,Ban18,Pad18,Kub22,Viv22,Viv22a,Bar22}.
 
Supporting experimental evidence for the X17 particle in the decay of
the excited 18.05 MeV $J^\pi I$=$0^- 0$ state of $^4$He
\cite{Kra19,Kra21} and the decay of the excited 17.23 MeV $J^\pi
I$=$1^- 1$ state of $^{12}$C \cite{Kra22} have been reported by the
ATOMKI Collaboration.  Earlier observations of similar $e^+$$e^-$
pairs with invariant masses between 3 to 20 MeV in the collision of
nuclei with emulsion detectors have been reported
\cite{El88,El96,Jai07,deB11}.  There are furthermore
possible $\gamma\gamma$ invariant mass structures at an energy around 10 to 15 MeV and 38
MeV in $pp$, and $\pi^- p$ reactions in COMPASS experiments
\cite{Ber11,Ber14,Bev12,Sch11,Sch12,Ber12,Bev20}.  Different
theoretical interpretations, astrophysical implications, and
experimental searches have been presented, including the fifth force
of Nature, the extension of the Standard Model, the QCD axion, dark
matter, the QED mesons, 12-quark-state and many others
\cite{Zha17,Fen16,Fro17,Bat15,Ros17,Ell16,Alv18,Mun18,Ban18,Pad18,Won10,Won20,Kub22,Viv22,Viv22a,Bar22}.
As reviewed  in the Proceedings to the X17 Workshop \cite{X1722},
the confirmation of the 
X17 particle is actively pursued by many laboratories, 
including  ATOMKI \cite{x17Kra}, Dubna \cite{x17Abr},
STAR \cite{x17STAR}, MEGII \cite{x17MEG}, TU Prague \cite{x17Prague},
NTOF \cite{x17NTOF}, NA64 \cite{x17NA64}, INFN-Rome
\cite{x17INFNRome}, NA48 \cite{x17NA48}, Mu3e \cite{x17Mu3e},
MAGIX/DarkMESA \cite{x17MAGIX}, JLAB PAC50 \cite{x17JLAB,x17JLAB1},
PADME \cite{x17PADME},  DarkLight \cite{ARIEL,Tre22},  LUXE \cite{Hua22}, and Montreal Tandem \cite{Mon22}.

\begin{figure} [H]
\hspace*{4.4cm}
\includegraphics[scale=0.60]{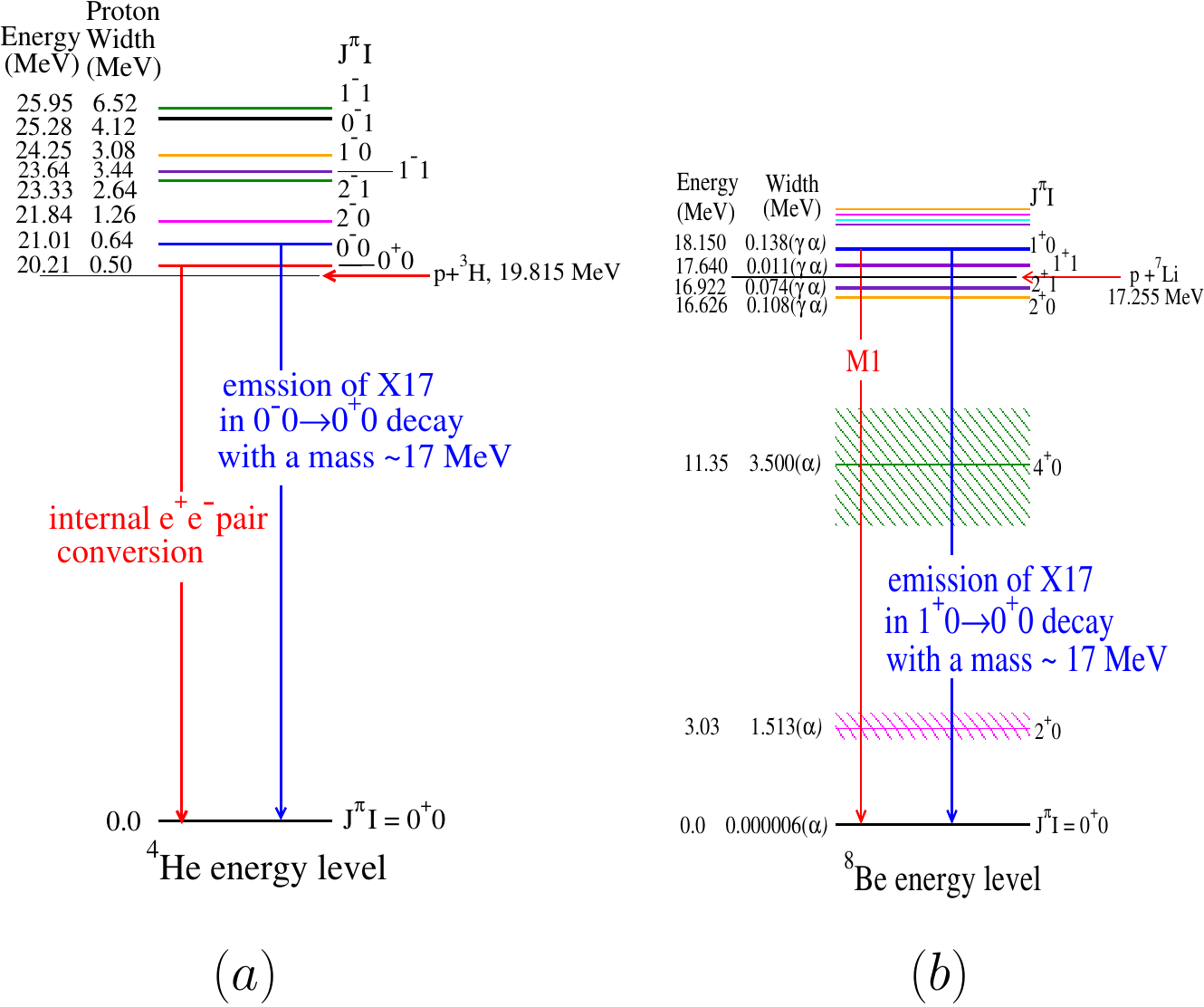}
\caption{ (a) The energy levels and widths of $^4$He states showing
  the transition in the decay of the 21.01 MeV $J^\pi I$=$0^-0$ $^4$He
  excited state to the $^4$He ground state with the emission of the
  X17 particle at $\sim$17 MeV, as interpreted in \cite{Kra19}.   (b) The energy levels and widths of
  $^8$Be states showing the transition in the decay of the excited
  18.15 MeV $J^\pi I$=$1^+0$ state to the $^8$Be ground state with the
  emission of the X17 particle, as interpreted in \cite{Kra16}.
  }
\label{fig7a}
\end{figure}

In the ATOMKI experiments, proton beams at a laboratory
  energy of 0.5  to 2.5 MeV were used to fuse
with $^3$H, $^7$Li, and $^{11}$B target nuclei to form
excited states of $^4$He, $^8$Be, and $^{12}$C, respectively
\cite{Kra16,Kra19,Kra21,Kra21a,Kra22,Sas22}.  The target nuclei have been so
chosen that the ground states of the product nuclei by proton fusion
are closed-shell nuclei.  Specifically, the $^4$He ground state in the
 reaction of 
$^3$H$(p,e^+e^-)^4$He$_{\rm g.s.}$ is a spherical
closed-shell nucleus.  The $^8$Be ground state in the reaction of 
$^7$Li$(p,e^+e^-)^8$Be$_{\rm g.s.}$  is an prolate close-shell
nucleus with a longitudinal to transverse radius ratio of about 2:1.   The
$^{12}$C ground state in the reaction of $^{11}$B$(p,e^+e^-)^{12}$C$_{\rm g.s.}$
 is an oblate closed-shell nucleus with a longitudinal to
transverse radius ratio of  about 1:2 \cite{Won70}.  There is consequently a
large single-particle energy gap between 
particle states above the
closed shell and the hole states at the top of the fermi surface below
the closed shell.  The captured proton in the product nucleus in the
proton fusion reaction  populate  proton single-particle states above the
large closed-shell energy gap, with the proton hole single-particle
states of the triton core $^3$H much below at the top of the Fermi
surface.  The transition of the proton from the proton 
particle states above the closed-shell energy gap to the proton 
hole state at
the top of the fermi surface to reach the closed-shell $^4$He ground
state in the reaction will release the large closed-shell-gap energy
that is of order about 17-20 MeV, as shown in Fig.\ \ref{fig7a}($a$)
and \ref{fig7a}($b$) .  Such a large closed-shell-energy gap may be
sufficient to produce a neutral boson, if there would exist such a
stable neutral boson particle with the proper energy,  quantum numbers, 
and other conditions appropriate for its production.
 
The picture is simplest for the $p$+$^3$H$\to ^4$He$^*$$\to$
$(e^+$+$e^-)$+$^4$He$_{\rm g.s.}$ reaction for which the ground state
of the product $^4$He nucleus is spherical.  The captured proton in
the lowest few excited $^4$He$^*$ state must occupy a $p$-shell
proton single-particle state that is significantly stretched outside the
triton $^3$H core with a proton and two neutrons  in the lower occupied 
$s$-states.  In the
collision of the incident proton on the target $^3$H nucleus, there is
the fusion barrier which comprises of the Coulomb and centrifugal
barrier for a single-particle with an angular momentum $l$,
\begin{eqnarray}
E_l= \frac{Z_p Z_T \,e^2 }{ R}+ \frac{l(l+1) \hbar^2} {2   \mu   R^2},
\end{eqnarray}
where $R=r_0 (A_p^{1/3}+ A_T^{1/3})$, $\mu = [(A_p A_T)/(A_p+A_T)]
m_{\rm nucleon}$, $(Z_p, A_p)$, and $(Z_T, A_T)$ are the projectile
and target charge and atomic numbers, respectively.  For $r_0$=1.3 fm,
the $p$+$^3$H$\to ^4$He$^*$ fusion barrier for $l=1$ is about 6 MeV
which places the low energy beam of around 1   MeV to lie below the
fusion barrier for the $l=1$ partial waves.

\begin{figure} [H]
\hspace*{1.4cm}
\includegraphics[scale=0.85]{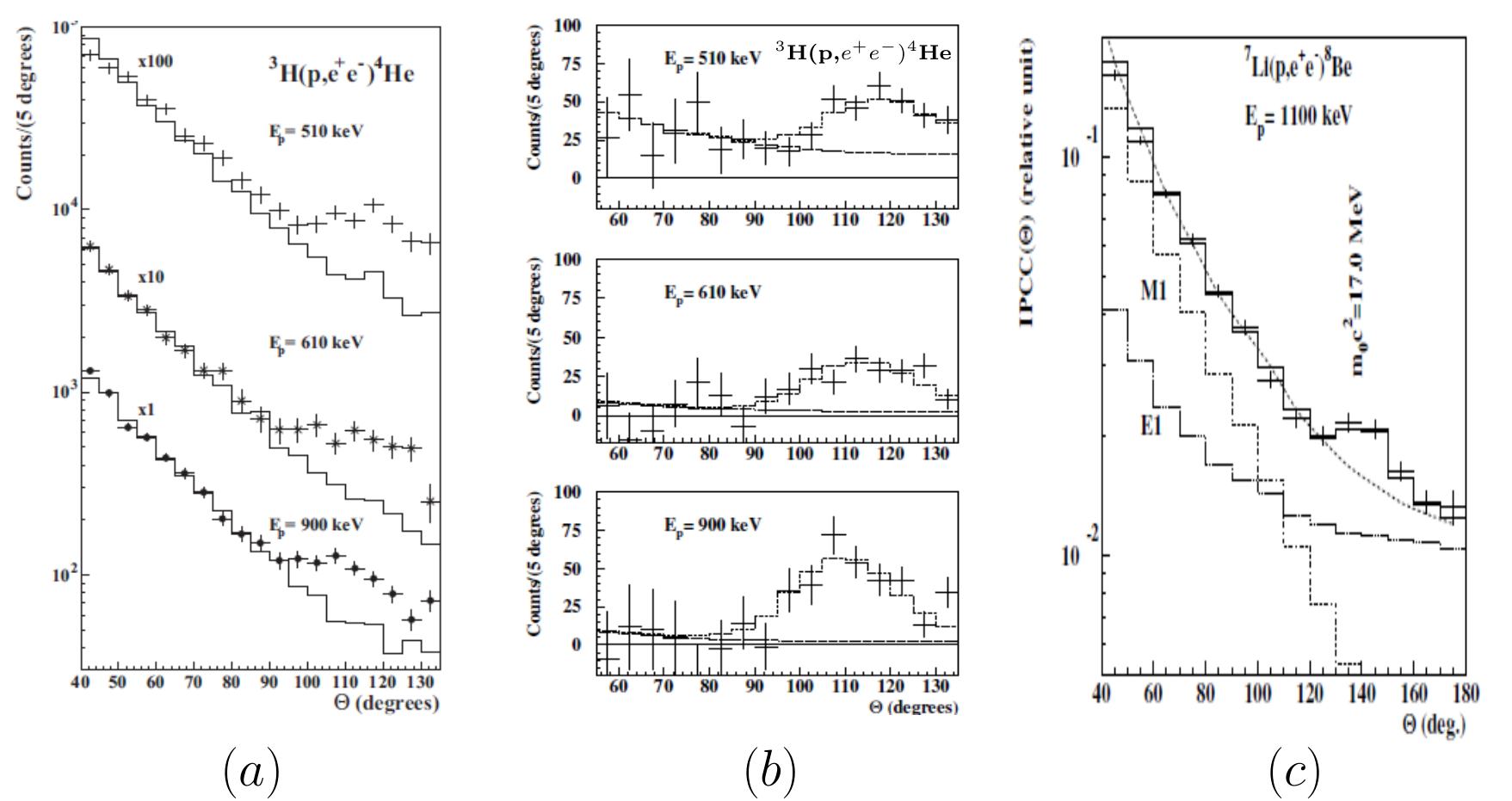}
\caption{ 
 (a) ATOMKI experimental $e^+$ and $e^-$ angular correlations at
  different proton energies in the reaction of $^4$He$(p,e^+e^-)$$^4$He$_{\rm g.s.}$
  \cite{Kra19,Kra21}.  The solid curves are contributions
  from known sources of the $e^+ e^-$ cross sections from the decay of
  the 20.21 MeV $0^+ 0$ $^4$He$^*$ excited state to $0^+0$(ground
  state) and the E1 decay from non-resonant capture into an excited
  $^4$He continuum state, displaying an excess of $e^+e^-$ production
  at large correlation angles \cite{Kra19,Kra21}.  ($b$) The data of an
  excess $e^+e^-$ production at large correlation angles can be
  described by the de-excitation of the 21.02 MeV $0^+ 0$ $^4$He$^*$
  state to the ground state with the emission of a hypothetical X17
  particle with a mass of 16.94 MeV, as described by the solid
  histograms \cite{Kra19,Kra21}.  (c) Similar $e^+e^-$ excess occurs in the reaction of 
  $^7$Li$(p,e^+e^-)$$^8$Be$_{\rm g.s.}$ \cite{Kra21a}.  The data of excess
  $e^+e^-$ production at large correlation angles can be described as
  arising from the decay of the excited 18.15 MeV $J^\pi I$=$1^+0$
  state of $^8$Be to the ground state with the emission of the X17
  particle at $\sim$17 MeV (solid histogram) \cite{Kra21a}.
}
\label{fig9}
\end{figure}

The capture of a proton of $E_{\rm lab}$ will lead to an excited
$^4$He$^*$ nucleus at an excitation of $E_x$=$ (A_T/(A_T+1))E_{\rm
  lab} + Q$ where the $Q$ value for the p+$^3$H$\to ^4$He$^*$ reaction
is 19.815 MeV.  The ATOMKI experimental proton energies of $E_{\rm
  lab}$=$\{$510, 610, and 900$\}$ keV in \cite{Kra21} in Fig.\ \ref{fig9}(a),
correspond to the $^4$He$^*$ excitation energies of $E_x$=$\{20.20,
20.27, 20.49\}$ MeV.  We can compare these excitation energies $E_x$
with the energy levels of $^4$He$^*$ states in Fig.\ \ref{fig7a}($a$)  \cite{Til92}, 
where the lowest
states are likely pocket resonances in the potential pocket \cite{Lee20}, as
evidenced by their narrow proton  widths:  The $J^\pi I=0^+ 0 $ state at 20.21
MeV has a proton width of 0.64 MeV while the $J^\pi I=0^- 0 $ state at
21.02 MeV has a proton width of 0.50 MeV.

Depending on the collision energy, there will be probabilities for
capturing into the pocket resonances inside the potential pocket with
various proton capture widths.  These resonances will subsequently
decay to the ground state by internal conversion with the emission
of an $e^+ e^-$ pair.  For example, capturing to the $0^+0$
resonance state at 20.21 MeV will decay to the ground state by
internal conversion with the emission of an $e^+ e^-$ pair, as
indicated in Fig.\ 9($a$). There is also a finite probability for the
direct non-resonant $l$=1 capture into a continuum state which tunnels
through the Coulomb and centrifugal barrier to the interior region and
subsequently decays into the ground state,
 with the emission an $e^+ e^-$ pair by
the E1 electromagnetic radiation.  The range of the collision energies
in the ATOMKI experiment lead to the population of both the $J^\pi
I=0^+ 0 $ state at 20.21 MeV and the $J^\pi I=0^- 0 $ state at 21.02
MeV.  The de-excitation of the $J^\pi I=0^+ 0 $ state at 20.21 MeV can
be accounted for by the internal pair conversion from $0^+ 0 \to 0^+ 0
$ final state.  The solid curves in Fig.\ \ref{fig9}($a$) represent the sum of
(i) the $0^+ 0 \to 0^+ 0 $ internal conversion contribution and (ii) the E1
non-resonant capture contribution to $e^+ e^-$ \cite{Kra19}.  As one
observes in Fig.\ \ref{fig9}($a$), there appears an excess of the $e^+e^-$
counts at large correlation angles.  The excess $e^+e^-$ yield can be
described as arising from the decay of the $J^\pi I$=$0^- 0$ excited
state of $^4$He$^*$ at 21.02 MeV to the $^4$He ground state with the
emission of a neutral ``X17'' boson with a mass of 16.94$\pm$0.12$\pm$
MeV as indicated in Fig.\ \ref{fig9}($b$) \cite{Kra21}.  In such an
interpretation, the reaction for the production of the X17 state is
\begin{eqnarray}
^4{\rm He}^*{\rm (21.02~MeV, }J^\pi  I= 0^- 0)  \to  ~ {\rm X17}~ + ~  ^4{\rm He}( 
{\rm ground ~state},J^\pi I=0^+0),
\end{eqnarray}
which implies that the quantum numbers of the emitted X17 particle is
$J^\pi I =0^- 0$.

In another experiment in the reaction of $^7$Li+$p \to ^8$Be$^*$$\to$
$(e^+$+$e^-)$+ $^8$Be$_{\rm g.s.}$  at ATOMKI
\cite{Kra16,Kra19} at a proton beam energy of a few MeV, the 18.15 MeV
$J^\pi I$=$1^+ 0$ state of $^8$Be$^*$ was populated together with
its neighboring states.  From
the nuclear reaction viewpoints, the captured proton populates
single-particle states in the deformed Nilsson single-particle
$p$-shell states.  The fusion barrier for $l=1$ partial waves lies at
about 4.5 MeV if the target nucleus were spherical.  Even though the
fusion barrier will be modified because of the target deformation
\cite{Won73}, the fusion barrier for $l=1$ would likely place the low
energy beam of around 1 to 2 MeV to be below the fusion barrier for
$l=1$ partial waves, as evidenced by their narrow resonance widths.  The capture
of a proton of $E_{\rm lab}$ will lead to an excited $^8$Be$^*$
nucleus at an excitation energy of $E_x= (A_T/(A_T+1))E_{\rm lab} + Q$
where the $Q$ value for s the p+$^7$Li$\to ^8$Be$^*$ reaction is
17.255 MeV.  The proton fusion experiments at ATOMKI with energies
$E_{\rm lab}$=$\{$0.800,1.04,1.10,1.20$\}$ MeV in
$p$+$^7$Li$\to$$^8$Be$^*$ in \cite{Kra16} correspond to the production
of an excited $^8$Be$^*$ nucleus at $E_x=$$ \{$17.955,18.165,18.2175,18.305$\}$ MeV, respectively.

We can compare these excitation energies $E_x$ with the energy levels
of $^8$Be$^*$ states in Fig.\ \ref{fig7a}($b$) \cite{Til04}.  Of particular interest are the 17.640 MeV
$J^\pi I=$$1^+1$ isovector state with a total width of 0.011 MeV
decaying by $\gamma$ and $\alpha$ emissions, and the 18.150 MeV $J^\pi
I=$$1^+0$ isoscalar state with a total width of 0.138 MeV decaying by
$\gamma$ and $\alpha$ emissions.  Depending on the collision energy,
there will be probabilities for capturing into the pocket resonance
states inside the potential pocket with various proton capture widths.
These resonances will subsequently decay into the ground state by
internal conversion with the emission of an $e^+ e^-$ pair. There will
also be a finite probability for the non-resonant $l$=1 capture into a
continuum state which tunnels through the Coulomb and centrifugal
barrier and decays to the ground state with an $e^+ e^-$ pair by an E1
electromagnetic radiation. 
The $J^\pi I=1^+ 0 $
state at 18.15 MeV can make a transition to the $0^+ 0 $ final ground state
by the M1 $e^+e^-$ internal
pair conversion.  The dotted curve
in Fig.\ 10(c) represents the sum of (i) the M1 $e^+e^-$ internal
conversion contribution and  (ii) the E1 non-resonant capture contribution
to $e^+ e^-$ \cite{Kra21a}.  As one observes in Fig.\ \ref{fig9}($c$), there appears an
excess of the $e^+e^-$ counts at large correlation angles around
$\theta\sim 140^o$.  The excess $e^+e^-$ yield can be described by the
solid curve as arising from the decay of $J^\pi I$=$1^+0 $ excited
state of $^8$Be$^*$ at 18.15 MeV to the $^8$Be ground state with the
emission of a neutral ``X17'' boson with a mass of 17.11$\pm$0.12 MeV
\cite{Kra21a}.

The approximate equality of the masses of the hypothetical neutral
boson in $^4$He$^*$ and $^8$Be$^*$ decays suggests that they are
likely to be the same particle, the X17 particle with $J^\pi I =0^- 0
$, emitted in the decay of the excited $0^-0$ state at 21.02 MeV of
$^4$He$^*$,  in the $l=1$ partial wave from the excited $1^+0$ state at
18.05 MeV of $^8$Be$^*$.  From Wheeler's molecular viewpoints of the
nuclei structure of light nuclei \cite{Whe37}, the ground state of
$^8$Be is likely to be in the form of two $\alpha$ clusters because of
the strong binding of the alpha particle.  So, in the excited 18.05
MeV $J^\pi I=$1$^+$0 excited $^8$Be$^*$ system, the emission would
likely come from one of the two $\alpha$ clusters of $^8$Be$^*$ and
the emitting cluster has a non-zero angular momentum relative to the
$^8$Be$^*$ nucleus center of mass.  From this viewpoint, it is
reasonable to assign the quantum numbers of the X17 to be $J^\pi I
=0^- 0$ from the $^8$Be$^*$ decay, the same as those from the
$^4$He$^*$ decay.

Recently, a question was raised concerning the 
background subtraction of the $e^+e^-$ signals
from the M1 and E1 contributions
in the analysis of the anomalous X17 particle in the
$^{7}$Li+$p$$\to$Be$^*$ decay \cite{Kra16}.  Hayes $et~al.$ \cite{Hay22}
re-examined the angular correlations in the $e^+e^-$ decay of those
excited states in $^8$Be  in the ATOMKI
experiment.  In the range of ATOMKI proton energies, the ratio of the
E1 and M1 contributions to the $e^+e^-$ production was found to be  a sensitive
function of energy.  
They questioned
the earlier assumption in \cite{Kra16} of an energy-independent
admixture of the E1 and M1 contributions for the $e^+e^-$ production.  
They found that the existence of a `bump'
in the measured angular distribution depended  strongly  on the
assumed M1/E1 ratio and 
the measured large-angle contributions to the $e^+ e^-$ angular
distribution to be lower than expectation.  
They cast doubts on the evidence for the existence of the anomaly in
the present analysis.  

However, there was another recent  measurement of the angular correlation of $e^+ e^-$ pairs produced in the
$^7$Li($p,\gamma$)$^8$Be$_{g.s}$ reaction at the
sharp $E_p$= 441 keV resonance and  650 keV, 800 keV and 1100
keV proton beam energies \cite{Sas22}. The spectra measured at the resonance 
could be
understood through the M1 internal pair creation process, but in the
case of the off-resonance regions (direct proton capture) significant
anomalies were observed in the $e^+ e^-$ angular correlations
supporting the X17 hypothetical particle creation and decay. Clearly,
many more measurements in search of the X17 particles and
many theoretical analyses need to be carried out to clarify the situation.

\subsection{Observation of the anomalous X17 particle in 
$^{11}$B(p,$e^+e^-$)$^{12}$C$_{\rm g.s.}$}

Subsequent to the observation of the X17 particle, it was suggested
that the electromagnetic E1 decay of the excited $J^\pi I $=$1^- 1$
$^{12}$C$^*$ state at 17.23 MeV to the ground state should be studied
in order to determine whether the X17 particle has a vector or
axial-vector characteristics \cite{Fen20}.  ATOMKI measurements with
proton energies between 1.5 to 2.5 MeV showed that the X17 particle
was generated predominantly by the E1 radiation.  It was concluded in
\cite{Kra22} that the association of the observation of the X17
particle with the decay of the 17.23 MeV $J^\pi I$=$1^-1$ $^{12}$C$^*$
state supports the vector character of the X17 particle, as suggested
by Feng and collaborators \cite{Kra22}.  In such an interpretation,
the reaction for the production of the X17 state is
\begin{eqnarray}
^{12}{\rm C}^* ~{\rm (17.23~MeV, }J^\pi I= 1^- 1) \to ~ {\rm X17}~ + ~
  ^{12}{\rm C}({\rm ground ~state},J^\pi I=0^+0),
\end{eqnarray}
which implies that the quantum numbers of the emitted X17 particle is
$J^\pi I =1^- 1$.

In the fusion of $p$ with $^{11}$B target nucleus, the captured proton
populates a Nilsson oblate orbital originating from the spherical
$p$-shell states.  The Coulomb and centrifugal barrier for a spherical
$l=1$ partial waves would lie at about 4.3 MeV.  Even though the
fusion barrier will be modified because of the target deformation, the
collision energy of $E_{\rm lab}$ from 1.5 to 2.5 MeV are much below
the fusion barrier.   So most of the the lowest $^{12}$C$^*$ excited
states populated by the proton fusion reaction are likely pocket
resonances in the potential pocket with small proton decay widths.
The capture of a proton of $E_{\rm lab}$ will lead to an excited
$^{12}$C$^*$ nucleus at an excitation of $E_x= (A_T/(A_T+1))E_{\rm
  lab} + Q$ where the $Q$ value for the $p$+$^{11}$B$\to ^{12}$C$^*$
reaction is 15.957 MeV \cite{Kel17}.  The proton-$^{11}$B fusion
experiments at ATOMKI with energies of $E_{\rm lab}$=$\{$1.5,1.70,1.88, 2.10, 2.5$\}$ MeV in \cite{Kra22} correspond to a the
production of an excited $^{12}$C at energies of  $E_x$=$\{17.33, 17.52, 17.68,
17.88, 18.25 \}$ MeV, respectively.  They can be compared with the
energy levels and their widths of $^{12}$C$^*$ states listed in Table
IV  \cite{Kel17}.
\begin{table}[h]
\caption { Listed are the $^{12}$C energy levels accessible to
  low-energy proton fusion with $^{11}$B and their widths, from
  \cite{Kel17}.  The $Q$ value for the fusion of
  $p$+$^{11}$B$\to$$^{12}$C$^*$ is 15.975 MeV.}
\vspace*{0.3cm}\hspace*{0.1cm}
\begin{tabular}{|c|c|c|c|}
\cline{1-4}
      &     &  &     \\
 $E_x$ in $^{12}$C     &  ~~~ $J^\pi I$~~~   &  $\Gamma$    &     Decay
  \\
        (MeV)              	&	                       &              (MeV)              &    
 \\ \hline
    16.1060$\pm$0.0008 & $2^+$ 1  & 0.0053  &  $\gamma, p ,\alpha$ 
 \\ \hline
    16.62$\pm$0.050& $2^-$ 1  & 0.28  &  $\gamma, p ,\alpha$ 
 \\ \hline 
   17.23 & $1^-$ 1  & 1.150  &  $\gamma, p ,\alpha$ 
 \\ \hline
    17.76$\pm$0.020 & $0^+$ 1  & 0.096  &  $ p, \alpha$ 
 \\ \hline
   18.16$\pm$0.070 & $(1^+$ 0)  & 0.24  &  $ \gamma, p$ 
  \\ \hline
   18.35$\pm$0.050 & $3^-$ 1  & 0.22  &  $ \gamma, p, \alpha$ 
   \\ \hline
  18.35$\pm$0.050 & $2^-$ 0+1  & 0.35  &  $n, p, \alpha$ 
   \\ \hline
   18.60$\pm$0.100 & $(3^-)$   & 0.30  &  
  \\ \hline
   18.71     & $(I=1)$   & 0.10  &  
 \\ \hline
  18.80$\pm$0.040 & $2^+$   & 0.10  &  $\gamma, n, p$ 
   \\ \hline
   19.20$\pm$0.600 & $(2^-,1)$   & 0.49  &  $\gamma,  p,\alpha$
   \\ \hline
   19.40 $\pm$0.025& $(1^-$ 1)  & 0.10  &  $  \gamma, p, \alpha$ 
   \\ \hline
   19.555$\pm$0.025 & $(1^-$ 1)  & 0.10  &  $  \gamma, p, \alpha$ 
   \\ \hline
  19.69           & $1^+$  & 0.23  &  $  n, p$ 
   \\ \hline
  20.00$\pm$0.100 & $2^+$ & 0.375  &  $\gamma,  n, p$ 
   \\ \hline
\end{tabular}
\label{tb4}
\end{table}
One observes that there are five $^{12}$C$^*$ excited states which
will be populated within the experimental range of proton energies of
the ATOMKI experiments.  Three of these states, the 17.23 MeV $1^-1$
state, the 18.15 MeV $1^+0$ state, and the 18.35 MeV $3^-1$ state
emits gamma radiation which will contribute to the $e^+e^-$ internal
conversion.

The interpretation of the X17 as possessing the $J^\pi I = 1^-1$
isovector characteristics as presented in \cite{Kra22} is subject to
serious questions.  First of all, such an interpretation of the X17 as
an isovector $J^\pi I=1^-1$ particle differs from the the $J^\pi I=
0^-0$ isoscalar interpretation of the X17 particle as suggested
earlier in $^4$He$^*$ and $^8$Be$^*$ decays \cite{Kra16,Kra19}.  The
observed boson or bosons in the two measurements have about the same
mass, and are likely the same particle. If there is only a single
neutral boson with a mass of about 17 MeV, then one the two
assignments of the $J^\pi I$ quantum numbers of the neutral X17 particle may be incorrect.
Secondly, the large number of states in the neighborhood of the 17.23
MeV $1^+$1 $^{12}$C state (Table \ref{tb4}) that are populated within the
experimental range of proton energies of the ATOMKI experiments make
it clear that it may be necessary to take into account the possibility
that the X17 may arise from other states.  Of particular interest is
the 18.16 MeV $1^+0$ state with a width of 0.24 MeV, which decays by
$\gamma$ and $p$ emissions.  This 18.16 MeV $1^+0$ $^{12}$C state may
be the analogue of the 18.15 MeV $1^+0$ $^{8}$Be state because they
have the same quantum numbers, excitation energies, decay widths,
decay channels, and they may represent a proton populating a $p$
orbital away from a triton core in one of the $\alpha$ particles in
the picture of Wheeler's alpha-particle model.  The X17 particle
observed in the $^{11}$Li$(p,e^+e^-)^{12}$C$_{g.s.}$ reaction in
\cite{Kra22} may arise also from the decay of the 18.15 MeV 1$^+ 0$
state in $^{11}$Li$(p,e^+e^-)^{12}$C$_{g.s.}$ collisions.  If the
transition from the 18.15 MeV $J^\pi I=1^+0$ state to the ground state
contributes dominantly to the X17 production instead of the transition
from the 17.23 MeV $J^\pi I=1^- 1$ state as suggested in \cite{Kra22},
the X17 data of $^4$He$^*$ decay, the $^8$Be$^*$ decay and the
$^{12}$C$^*$ decay will be consistent with each other, whereas the
$J^\pi I=1^-1$ interpretation of $^{12}$C contradicts the the $J^\pi
I=0^-0$ interpretation of $^4$He and $^8$Be.

In order to discriminate the two possible assignments of $J^\pi I=1^-
1$ or $0^-0$, we can suggest the search for two-photon decay of the
X17 particle.  Such a decay is possible if the X17 is a $J^\pi I=0^-
0$ particle.  However, because of the Landau-Yang theorem
\cite{Lan48,Yan50}, the decay of a $1^-$ state into two photons is
impossible.  If X17 is found to decay into two photons, then the X17
can only be a $0^-0$ particle and not a $1^-1$ particle.

Another way to study the $^{12}$C$^*$ system and the X17 particle is
to excite a $^{12}$C target ground state by direct photon excitation
to various states, such as the $1^-1$ at 17.23 MeV by E1 excitation
and the $1^+0$ at 18.15 MeV by M1 excitation.  After the nucleus
excited to these specific states, their decay by the emission of an
X17 particle or a internal conversion photon in the form of $e^+ e^-$
can be examined.

There is one more way to study the X17 particle and its quantum number
assignments, if the X17 indeed has the $J^\pi I= 1^-1$ property as
proposed by \cite{Fen20,Kra22}.  Because the strength of the giant
dipole state of $^{12}$C resides at 25 MeV and not at 17.23 MeV and
the 17.23 represents only a very small fraction of the giant dipole
strength \cite{Ber75,He16}, the X17 particles would be copious
produced at $E_x$=25 MeV, and and such a copious X17 production would
allow the study of the X17 properties, if the X17 indeed has the
$J^\pi I= 1^-1$ property.

\subsection{The observation of the E38 Particle}

Abraamyan and collaborators at Dubna have been using the two-photon
decay of a neutral boson to study the resonance structure, the
dynamics, and the interaction of the lightest hadrons near their
thresholds.  Their experiments have been carried out by using the
proton, deuteron, and other light-ion beams of the Nucleotron at JINR,
Dubna on fixed internal fixed targets of C, Cu, and other nuclei, with
the production of hadrons and photons.  Their PHONTON2 photon detector
consists of two arms placed at 26$^o$ and 28$^o$ degrees from the beam
direction, with each arm equipped with 32 lead-glass photon detectors
as shown in Fig.\ \ref{fig10}($a$).  The photon detectors measure the
energies and the emission angles of the photons.  From the energies
and the opening angle $\theta_{12}$ of two photons $\gamma_1$ and
$\gamma_2$ in coincidence, the invariant mass of the pair,
$M_{\gamma_1 \gamma_2}=E_{\gamma 1} E_{\gamma 2}(1-\cos \theta_{12})$,
can be evaluated and the invariant mass distribution determined.
Neutral bosons with large invariant masses can be searched by
selecting photon pairs with large opening angles from different arms.
On the other hand, by selecting photon pairs from the same arm with
small opening angles, it is possible to study neutral bosons with
small invariant masses such as those below the pion mass gap $m_\pi$, if they ever
could be stable.  Thus, the PHOTON2 detector can be used to probe the
possible existence of neutral bosons over a large dynamical range.
Previously, along with the detection of the $\pi^0$ and $\eta$ mesons,
the Dubna Collaboration uncovered a new resonance structure at
$M_{\gamma \gamma}$=360 MeV (the R360 resonance) \cite{Abr09}.  The
resonance was subsequently confirmed by repeated measurements
{\cite{Abr16} with higher statistics.  We shall explore a possible model 
for R360 in Section 7F
in terms of a molecular state of two pions and two E38 QED mesons.

\begin{figure} [H]
\centering
\includegraphics[scale=0.80]{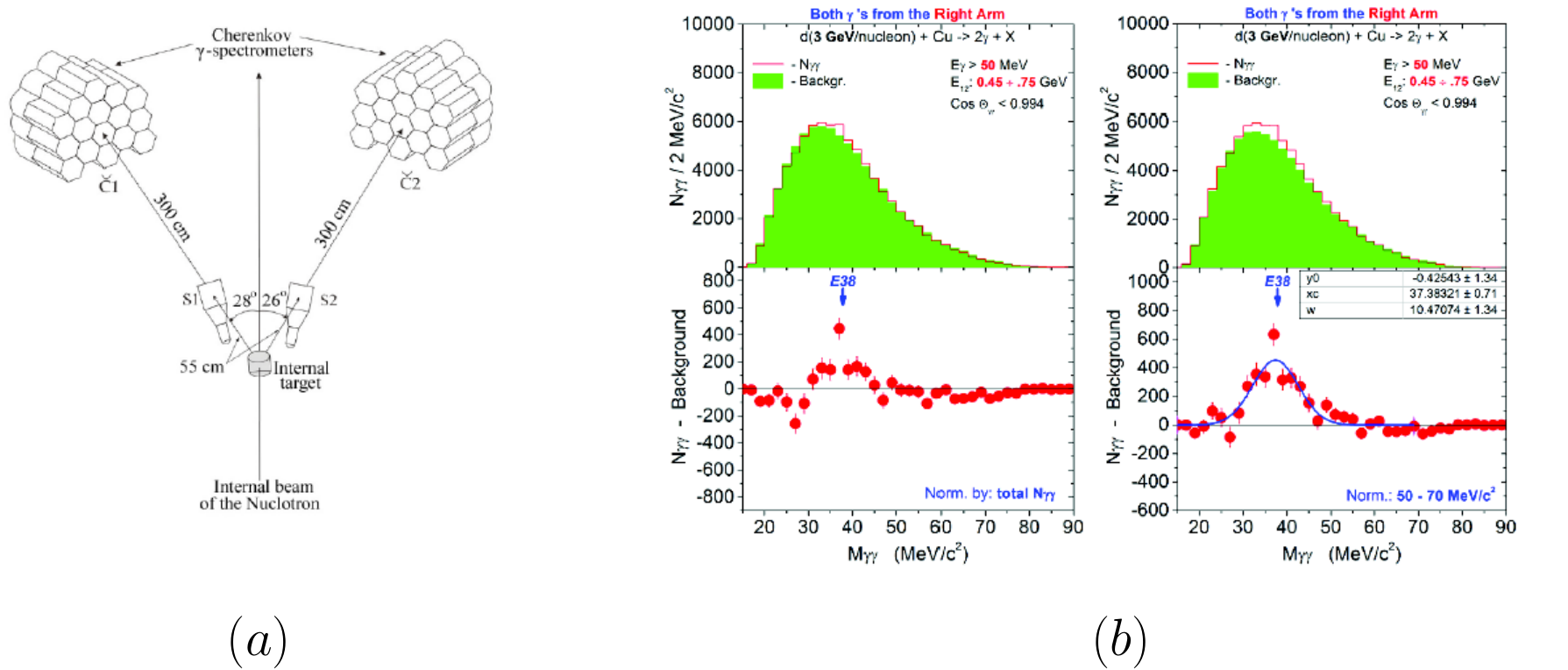}
\caption{ ($a$) Dubna PHOTON2 detector arrangement to study the
  two-photon decay of reaction products in the collision of proton and
  light ions on fixed internal targets.  ($b$) The invariant mass
  distribution of photon pairs from the same event are shown as the
  solid curves with the invariant mass distribution of photon pairs
  from the mixed events shown as the shaded regions.  The signal
  of the invariant mass of the correlated pair after subtracting the
  mixed-event gubernatorial background shows a resonance structure at
  38 MeV.
  }
  \label{fig10}
\end{figure}

In 2011, Van Beveren and Rupp studied $e^+e^-$$\to \pi^+\pi^-$
\cite{Aul05}, $p\bar p$$\to J/\psi \pi^+\pi^-$ \cite{Aal09}, and
$e^+e^-$$\to \Upsilon \pi^+\pi^-$ cross sections and observed that the
deviation of the cross section from a global fit oscillates with a
period of 73 to 79 MeV \cite{Bev11}.  An analysis of the invariant
mass distribution of $M_{\Upsilon(1S)}$ as determined from $e^+e^- \to
\Upsilon (2 ^3S_1) \to \pi^+\pi^- \to \pi^+\pi^- \mu^+\mu^-$ and the
invariant mass distribution of $M_{\Upsilon(1S)}$ as determined from
$\Upsilon (2 ^3S_1) \to \mu^+\mu^-$ \cite{Gui09} suggested a shift of the excitation function by 38 MeV.
  Van Beveren and Rupp therefore speculated on the
possible existence of a light boson with a mass of 38 MeV
\cite{Bev11}.  Their examination of the two photon spectrum of the
COMPASS Collaboration data \cite{Sch11} also suggested a possible
resonance at 38 MeV.  There ensued a debate on the background
subtraction of the COMPASS data and the reliability of the
identification of the E38 resonance using the raw COMPASS data
\cite{Ber11,Ber14,Bev12,Sch11,Sch12,Ber12,Bev20}.

Stimulated by the possibility of a light neutral boson with a mass of
about 38 MeV, the Dubna Collaboration undertook a search for the E38
particle using the two-photon decay.  The search was carried out in
the $d$(2.0 GeV/n)+C, $d$(3.0 GeV/n)+Cu and $p$(4.6 GeV)+C reactions
with internal targets at the JINR Nuclotron.  It was reported in
Version 1 of Ref.\ \cite{Abr12} that the invariant masses of the
two-photon distribution in these reactions exhibit a resonance
structure at around 38 MeV.  Unaware of the earlier theoretical
prediction of an isoscalar $J^\pi I=0^- 0$ QED-confined $q\bar q$
state at 38.4 MeV in Table I of \cite{Won10} that could be a good
candidate for such a particle, Version 2 of Ref.\ \cite{Abr12} stated,
however, ``due to non-ordinariness of the obtained results (standing
out of The Standard Model) and at the request of co-authors, the first
version of the article is withdrawn for further verification and more
detailed description of the experiment and data analysis.''  Subsequent
repeated checking of the results of their analysis over a period of
many years confirmed their findings, and the updated results were then
reported in \cite{Abr19}.

In the latest Dubna experiment, the  two-photon
invariant masses spectra were studied in nuclear reactions of  $d$C, $d$Cu, and $p$C  at
2.75, 3.83, and 5.5 GeV/c per nucleon, respectively.    In different momentum  windows,
photon pairs  in the same event from  the
PHOTON2 detector are selected and their invariant masses calculated to
construct the invariant mass distribution shown as the solid curves in
Fig.\ 11($b$).  The corresponding combinatorial backgrounds
are obtained in the event-mixing method by calculating the invariant
masses of two photons from two different events, shown as the shade
regions in Fig.\ 11($b$).  The difference between the
correlated invariant mass distribution and the mixed-event invariant
mass distribution gives the signal invariant mass distribution, as
shown in the lower panels of Fig.\ 11($b$) for different phase
space windows.  The resonance structure in Fig.\ 11(b) indicates
a neutral boson, the E38 particle, at about 38 MeV \cite{Abr19}.
Signals from other phase space windows show a similar resonance
structure.

The observation of E38 at Dubna completes an important piece of the
anomalous particle puzzle as the isoscalar X17 and the isovector E38
come in a pair, and  they are orthogonal linear combinations of the
$|u\bar u\rangle $ and $|d \bar d\rangle$ components.  The agreement
of their masses with those predicted by the phenomenological
open-string model of QED-confined $q\bar q$ model of Section V lends
support to the description that a quark and an antiquark
can be confined and bound as stable QED-confined 
mesons interacting in the Abelian U(1) QED interaction.  This is a rather unusual
and unfamiliar concept.  The confirmation of these anomalous particle
observations will be  therefore of great interest by detecting the X17, the
E38, and $\pi^0$ in the same experimental set up.

\subsection{Anomalous Particle Production mechanisms}  

It is necessary to understand the mechanisms how the anomalous
particles may be be produced in order to fully support the
interpretation on their production as QED mesons.  For the low-energy
production of the X17 particle in the decays of the excited states of
$^4$He$^*$ and $^8$Be$^*$ at ATOMKI \cite{Kra16,Kra21}, we envisage
the scenario as described in Section 2 that the excited states of
$^8$Be(1$^+ \,$18.15 MeV) and $^4{\rm He}(0^-\,$20.02 MeV) are formed
by pulling a proton out of one of the alpha-particles of the
$(\alpha)^n$-nucleus core and by placing the proton on an orbital that
is considerably outside the corresponding tritium core as shown in
Fig.\ \ref{fig3}($a$).  The stretched string-like interaction between
the proton and the tritium core polarizes the vacuum so much that the
proton may emit a virtual gluon which fuses with the virtual gluon
from the $^3$H core to lead to the production of a color-singlet  $q\bar q$ pair by
the reaction $g + g \to q+ \bar q$ as shown in Fig.\ \ref{fig3}($a$).
At the appropriate $\sqrt{s}(q\bar q)$ eigenenergy, the QED
interaction between the $q$ and the $\bar q$ may result in the
formation of the QED-confined $q$$\bar q$ bound state X17
\cite{Won10,Won20}, which subsequently decays into $e^+$-$e^-$.

For the production of E38 in high-energy nucleus-nucleus collisions at
Dubna Collaboration \cite{Abr12,Abr19}, and the anomalous soft photons
in the high-energy hadron-hadron collisions at CERN
\cite{Chl84,Bot91,Ban93,Bel97,Bel02pi,Bel02,Per09}, we envisage the
production of many $q\bar q$ pairs as described in Fig. 3($b$) of
Section I2.  In such high-energy processes, many $q\bar q$ pairs are
produced.  Even though most of the produced $q\bar q$ will materialize
as QCD mesons, some of the $q\bar q$ pairs may materialize as QED
mesons, as the E38 particle that decays into diphotons at Dubna
\cite{Abr19} and as the X17 particle and the E38 particle that decay
into the anomalous soft photons in the WA102 experiment \cite{Bel02}.

In the other process for the production of the anomalous soft photons 
 in high-energy $e^+e^-$ annihilation
at $Z^0$ mass at DELPHI, we envisage the production $q\bar q$ pairs as
described in Fig.\ \ref{fig1}($c$) of Section 2 in which the $e^+e^-$
annihilation process leads to the production of a large number of
$q\bar q$ pairs.  While most produced $q$$\bar q$ pairs will lead to
hadron production, there may however be $q$$\bar q$ pairs with
$(m_q+m_{\bar q})$$<$$\sqrt{s}$$(q\bar q)$$<$$ m_\pi$ for which the
QED interaction between the quark and the antiquark may lead to the
production of the X17 and E38 particles at the appropriate energies
and the anomalous particles subsequently decay into $e^+e^-$ pairs.

\section{Questions on quark confinement in compact QED in (3+1)D from lattice gauge
calculations}

\subsection{Lattice gauge calculations prediction of deconfined static quark and antiquark  in compact QED in (3+1)D}

We mentioned in Section 4 that in lattice gauge theory a static
fermion and a static antifermion in (3+1)D in compact QED interaction
has a strong-coupling confined phase and a weak-coupling deconfined
phase, as shown by Wilson, Kogut, Susskind, Mandelstam, Polyakov,
Banks, Jaffe, Drell, Peskin, Guth, Kondo and many others
\cite{Wil74,Kog75,Man75,Pol77,Pol87,Ban77,Gli77,Pes78,Dre79,Gut80,Kon98,Mag20}.
The transition from the confined phase to the deconfined phases occurs
at the coupling constant $\alpha_{\rm crit}=g_{\rm
  crit}^2/4\pi$=0.988989481 \cite{Arn03,Lov21}.  The magnitude of the
QED coupling constant as given by the fine-structure constant
$\alpha_c$=1/137 is less than $\alpha_{\rm crit}$.  That is, the QED
interaction between a quark and an antiquark belongs to the
weak-coupling deconfined regime.  Therefore, a static quark and a
static antiquark are deconfined in lattice gauge calculations in
compact QED in (3+1)D.

The deconfined static quark and static antiquark in the lattice gauge
results in (3+1)D poses a serious question.  There are experimental
circumstances in which a quark and an antiquark can be produced and
they can interact in QED alone, without the QCD interaction, as we
discussed in Section 2.  For example, we can study the reactions $e^+
$+$ e^- $$\to$$ \gamma^*$$ \to$$ q $+$\bar q$ and $e^+ $+$ e^- $$\to$$
\gamma^* \gamma^* $$\to$$ q $+$\bar q$ with a center-of mass energy
range $(m_q+ m_{\bar q}) < \sqrt{s}(q\bar q) < m_\pi$, where the sum
of the rest masses of the quark and the antiquark is of order a few
MeV and $m_\pi\sim 135$ MeV \cite{PDG19}.  The incident $e^+$+ $e^-$
pair is in a colorless color-singlet state, and thus the produced $q$
and $\bar q$ pair and the quanta mediating their interactions must
also combine together into a color-singlet final state.  In this
energy range below the collective QCD excitation mass gap of $m_\pi$,
there is insufficient energy to excite a  collective QCD excitation to produce a QCD meson.
A $q$ and $\bar q$ can be produced and interact in the QED
interaction alone.  
In the color-singlet $(q\bar
q)^1$ configuration, the produced $q$ and $\bar q$ 
can 
interact with the colorless Abelian U(1) QED
interaction to form a color-singlet $[(q \bar q)^1 \gamma^1]^1$ final
state, if there is a QED  meson
eigenstate at this eigenenergy.  At energies other than the QED meson
eigenenergies in this energy range below $m_\pi$, the $e^+$ + $e^-$
collision will probe the dynamics of a quark and antiquark interacting
in QED alone, without the QCD interaction.  The absence of fractional
charges in collisions in this energy range in $e^+$ + $e^-$ collisions
 indicates the absence of the continuum states of an  isolated quark and
an antiquark in the interaction of a quark and an antiquark in the
QED interaction.

On the other hand, the solution of deconfined static quark and static
antiquark in the lattice gauge calculations in QED in (3+1)D predicts
that the $q$ and $\bar q$ produced in $e^+$ + $e^-$ collisions in the
range of energy below the QCD collective excitation mass gap $m_\pi$
will not be confined and would appear as fraction charges, when the
quark interact with the antiquark in QED alone in (3+1)D.  However, no
such fractional charges have ever been observed.  Furthermore, the
phenomenological open-string QCD and QED meson model with the
hypothesis of a confined $q\bar q$ pair in QED in (3+1)D leads to
QED meson and QCD meson spectra in agreement with experimental data,
as discussed in Section 5.  There is a confined regime in (1+1)D QED
for dynamical massless quarks in the Schwinger confinement
mechanism\cite{Sch62,Sch63}, and there is also a confined regime for
quarks in compact QED in (2+1)D  \cite{Pol77,Pol87} in Polyakov's
transverse confinement, for all gauge coupling interaction strengths.
They indicate that the present-day lattice gauge calculations for
compact QED in (3+1)D may not be complete and definitive, because the
important Schwinger dynamical quark effects associated with light
quarks has not been included.  Future lattice gauge calculations with
dynamical quarks in compact QED interactions in (3+1)D will be of
great interest in clarifying the question of quark confinement in QED.

\subsection{Compact and non-compact U(1) QED gauge interactions in lattice gaige calcualtions }

Whatever the theoretical predictions on the confinement of quarks in
QED in (3+1)D may be, in the final analysis, the question whether a
$q\bar q$ pair is confined in QED in (3+1)D can only be settled by
experiment.  In the meantime, in the presence of the two opposing
theoretical conclusions on quark confinement in QED in (3+1)D and the
agreement of the experimental spectrum of the anomalous particles
with the QED meson predictions, it is possible that the two
conclusions can still be consistent with each other, if the
confinement conclusions arise from the inclusion of the Schwinger
confinement mechanism \cite{Sch62,Sch63} in the works of
\cite{Won10,Won11,Won14,Won20,Won22,Won22a,Won22b,Won22c}, while and
the deconfinement conclusion arises from the absence of the Schwinger
confinement mechanism in lattice calculations \cite{Arn03,Lov21}.
Therefore, it is worth constructing a plausible ``stretch (2+1)D''
flux tube model to demonstrate the importance of the Schwinger
confinement mechanism in compact QED in (3+1)D.

In such a demonstration, we note first of all that the Schwinger
confinement mechanism occurs in (1+1)D space-time, whereas the
physical world is in (3+1)D.  For the Schwinger confinement mechanism
to be operative, the quark-QED system must possess transverse
confinement as a flux tube in the (3+1)D space-time before the flux
tube can be idealized as a one-dimensional string in the (1+1)D
space-time.  In this regard, we note from Polyakov's previous results
that electric charges of opposite signs interacting in compact QED in (2+1)D are
confined and that the confinement persists for all non-vanishing
coupling constants, no matter how weak \cite{Pol77,Pol87}.  We can
combine Schwinger's longitudinal confinement in (1+1)D QED with
Polyakov's transverse confinement in (2+1)D compact QED to study QED
confinement in a flux tube environment in (3+1)D.
 
Before we come to the ``stretch (2+1)D'' flux tube model of $q\bar q$
production in compact QED in (3+1)D, it is necessary to clarify the
concept of compactness in the QED interaction.

There are two different types of QED U(1) gauge interactions
possessing different confinement properties \cite{Pol77,Pol87,Dre79}.
There is the compact QED U(1) gauge theory in which the gauge fields
$A^\mu$ are angular variables with a periodic gauge field action which
allows transverse photons to self-interact among themselves.  The
gauge field action in the compact QED U(1) gauge theory, in the
lattice gauge units and notations of
Ref.\ \cite{Pol77,Pol87,Dre79}, is
\begin{eqnarray}
S=\frac{1}{2g^2}\sum_{x,\alpha \beta} (1-\cos F_{x,\alpha \beta}),
\label{eq76}
\end{eqnarray}
where $g$ is the coupling constant and the gauge fields $F_{x, \alpha \beta}$ are 
\begin{eqnarray}
F_{x, \alpha \beta}\!=\!A_{x,\alpha}\!\!+\!\! A_{x+\alpha,\beta}\!\! - \!\! A_{x+\beta,\alpha} \!\! - \!\! A_{x,\beta},{\rm with} -\!\!  \pi\!\!  \le A_{x,\alpha} \!\! \le\!\!  \pi.~
\label{eq77}
\end{eqnarray}
There is also the non-compact  QED U(1) gauge theory with the gauge
field action \cite{Pol77,Pol87,Dre79}
\vspace*{-0.1cm}
\begin{eqnarray}
S=\frac{1}{4g^2}\sum_{x,\alpha \beta} F_{x,\alpha \beta}^2,  ~~~{\rm with}~~~~- \infty  \le A_{x,\alpha} \le  +\infty  .
\label{3}
\end{eqnarray}
In non-compact QED gauge theories, the photons do not
interact with other  photons and the fermions in   non-compact QED gauge theories are always de-confined.
  In compact QED in (2+1)D, the photons interact among themselves because the gauge fields are angular variables.  
Static 
charges of opposite signs are confined for all strength of the coupling constant
in compact QED in (2+1)D.  In
compact QED in (3+1)D, however, static opposite charges are confined only for
strong coupling but de-confined for weak coupling
\cite{Pol77,Pol87,Dre79}.  
Even though the compact and
the non-compact QED gauge theories in Eqs.\ (\ref{eq76}) and
(78) have the same continuum limit, they have different
confinement properties.
On the other hand, in the Schwinger confinement mechanism in QED in (1+1)D, 
massless fermions are confined for all coupling strengths 
in the continuum limit,   
which does not distinguish between compact and  non-compact QED \cite{Sch62,Sch63}.

We need to ascertain the type of the QED U(1) gauge interaction
between a quark and an antiquark in a QED meson.  As pointed out by
Yang \cite{Yan70}, the quantization and the commensurate properties of
the electric charges of the interacting particles imply the compact
property of the underlying QED gauge theory.  Because (i) quark and
antiquark electric charges are quantized and commensurate, (ii) quarks
and antiquarks are confined, and (iii) there are pieces of
experimental evidence for possible occurrence of confined $q\bar q$
QED meson states as we mentioned in the Introduction, it is therefore
reasonable to propose that quarks and antiquarks interact in the
compact QED U(1) interaction.

In compact QED, Polyakov \cite{Pol77,Pol87} showed previously that a
pair of opposite electric charges and their gauge fields in
(2+1)D$_{\{x^1,x^2,x^0\}}$ are confined, and that the confinement
persists for all non-vanishing coupling constants, no matter how weak.
As explained by Drell and collaborators \cite{Dre79}, such a
confinement in (2+1)D${}_{\{x^1,x^2,x^0\}}$ arises from the
angular-variable property of $A_\phi$ and the periodicity of the gauge
field action as indicated in Eq.\ (\ref{eq76}).  The gauge action
periodicity in the neighborhood of the produced opposite electric
charges leads to self-interacting transverse gauge photons.  These
transverse gauge photons interact among themselves, they do not
radiate away, and they join the two opposite electric charges and
their associated gauge fields by a confining interaction.

\subsection{The stretch (2+1)D model}

To construct a ``stretch (2+1)D'' flux tube model \cite{Won22b,Won22c}
for the production of a quark and an antiquark in a QED meson in
(3+1)D, we take the Polyakov's transverse confinement configuration in
compact QED in (2+1)D as input.  We envisage the production of the
nascent $q$$\bar q$ pair at the origin $(x^1,x^2,x^3)$=0 in the
center-of-mass system, at the eigenenergy $\sqrt{s}(q\bar q)$ of a
QED meson\footnote{
As discussed  in the Introduction, states of a quark and an antiquark do not exist
except in a confined $q\bar q$ eigenstate, so we can study the confinement dynamics of a quark and an antiquark only when the quark and the antiquark under consideration here are constituents of a confined system such as a QED meson.  Such a logical circularity arises because of the peculiar property of quark confinement.
}.  We take for simplicity the quark charge to be
positive, which can be easily generalized to other cases of
negatively-charged quark and flavor mixing.  
At birth and born to be in a QED meson state, the quark and
the antiquark must possess an equal and opposite momenta in the center-of-mass
system$^8$ as shown in Fig. 12(a).  The quark momentum defines the longitudinal direction from
which we can define the transverse coordinates $(x^1,x^2)$ and the
transverse planes at $x^3$=(constant).  The quark and the antiquark
produced at birth are located in the vicinity of the origin, and they
are separated by an infinitesimal longitudinal separation $\Delta x^3$
and an infinitesimal transverse separation $\Delta {\bb r}_\perp$ as in Fig.\ \ref{fig11}($a$).
Polyakov's transverse confinement of the quark and the antiquark in
compact QED in (2+1)D is realized on the transverse $(x^1,x^2)$-plane
at $x^3\sim 0$, at the birth of the $q$-$\bar q$ pair.  The creation
of the charge $q$-$\bar q$ pair will be accompanied by the associated
creation of their gauge fields $\bb A$, $\bb E$, and $\bb
B=\nabla\times A$, which by causality can only be in the neighborhood
of the created charges initially with the created $\bb E$ and $\bb B$
fields lying along the longitudinal $x^3$ direction.

\begin{figure} [H]
\centering
\vspace*{-0.cm}\hspace*{-0.3cm}
\resizebox{0.40\textwidth}{!}{
\includegraphics{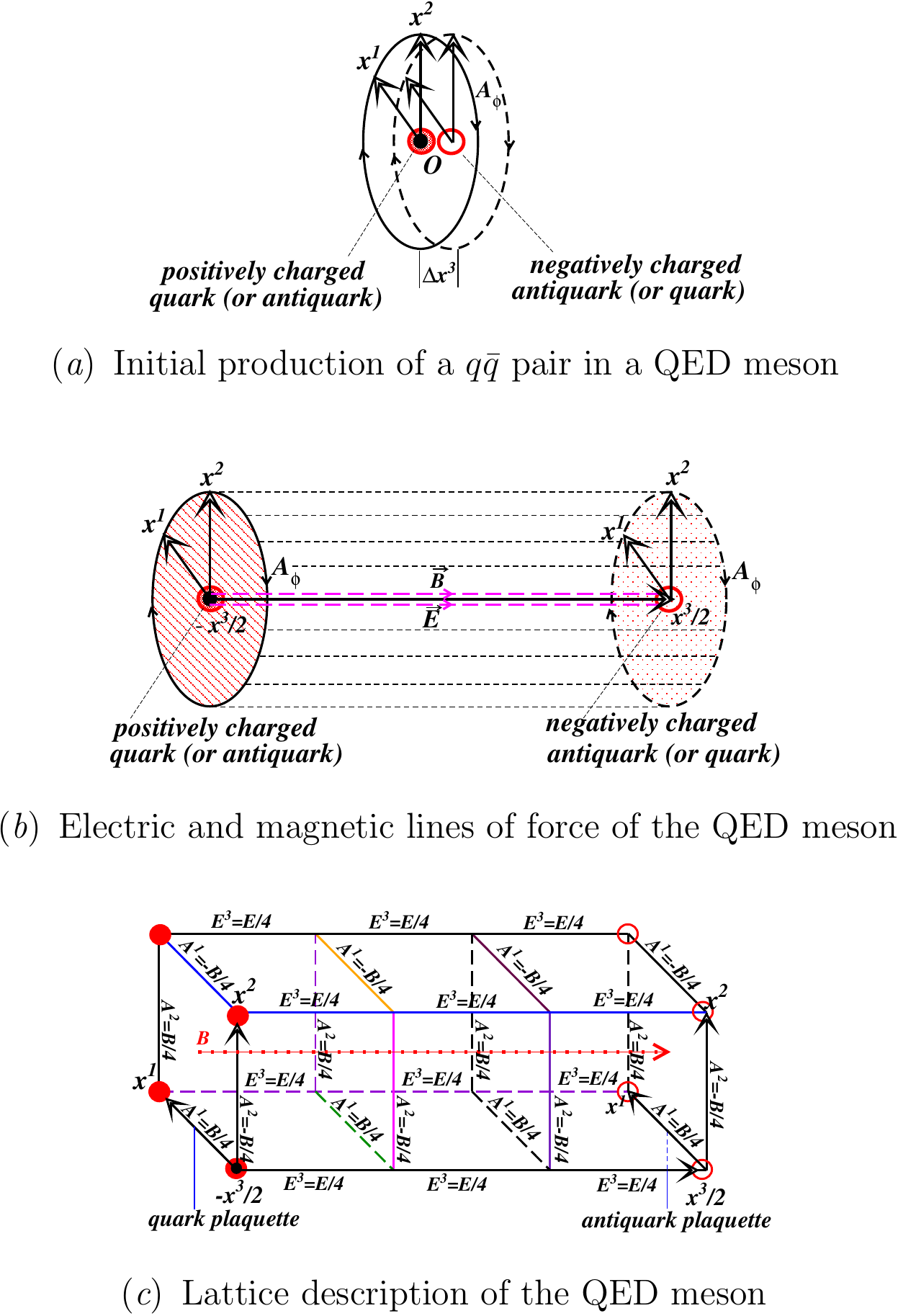}}
\caption{ Fig.\ \ref{fig11}(a) depicts the initial production of the
  $q\bar q$ pair of a QED meson, with an infinitesimal longitudinal
  separation $\Delta x^3$.  Fig.\ \ref{fig11}($b$) gives a snapshot of
  the configuration of the quark charge, the antiquark charge, gauge
  fields $\bb A, \bb E$, and $\bb B$=$\nabla $$\times$$ \bb A$ during
  the dynamical longitudinal yo-yo motion of the $q\bar q$ pair
  interacting in compact QED in a QED meson, starting from a (2+1)D
  transversely confined $q\bar q$ system.  Fig.\ \ref{fig11}($b$) is
  the ``stretch (2+1)D'' configuration.  Fig.\ \ref{fig11}($c$) is the
  corresponding lattice transcription, following the Hamiltonian
  formulation of Drell $et~al.$ \cite{Dre79}. }
\vspace*{-0.3cm}
\label{fig11}
\end{figure}
\vspace*{-0.1cm}

Subsequent to their births, the quark and the antiquark will execute
stretching and contracting ``yo-yo'' motion along the longitudinal
$x^3$ direction, appropriate for the QED meson bound state in
question.  As the quark and antiquark stretch outward in the
longitudinal $x^3$ directions, we can construct a longitudinal tube
structure of gauge fields in the stretch (2+1)D configuration by
duplicating longitudinally the transversely-confined gauge fields that
exist on the transverse $(x^1,x^2)$-plane at $x^3 \sim 0$ initially at
their birth, for the longitudinal region between the stretching quark
and antiquark.  A snap shot of the stretch (2+1)D flux tube
configurations at an early moment in the longitudinal stretching
motion is shown in Fig. \ref{fig11}($b$).  The transcription of
Fig.\ \ref{fig11}($b$) in terms of the lattice link and plaquette
variables is shown in Fig.\ \ref{fig11}($c$), by following the
Hamiltonian formulation and the notations of Drell $et~al.$
\cite{Dre79}.  Specifically, in the $A^0$=0 gauge we specify the
canonical conjugate gauge fields $\bb A$ and $\bb E$ at the links in
Fig.\ \ref{fig11}($c$), where we display only the $A^1,A^2$ and $E^3$
values of the conjugate gauge fields.  The $\bb B$ and $\bb E$ fields
are aligned along the longitudinal $x^3$ axis as shown in
Fig.\ \ref{fig11}($b$).  These gauge field configurations are the ones that
lead to Polyakov's transverse confinement of the quarks.  They
maintain the cylindrical symmetry of the stretch (2+1)D flux tube.
They will lead to the desirable full confinements of the quark and the
antiquark, as the magnetic field $\bb B$ sends the quark and antiquark
charges into the appropriate Landau orbitals to execute confined
transverse zero-mode harmonic oscillator zero-point motions on their
respective $\{x^1,x^2\}$ planes.  At the QED meson eigenenergy, the
electric field $\bb E$ along the
longitudinal $x^3$ direction sends the quark and the antiquark in
longitudinal 3D longitudinal stretching and contracting ``yo-yo''
motion.  The electric charge densities obey the Gauss law associated
with the divergence of the electric field $\bb E$.  The
positive electric quark charge fractions (solid circles in
Fig.\ \ref{fig11}($c$)) reside at the -$x^3$/2 plaquette vertices and
the negative electric antiquark charge fractions (open circles in
Fig.\ 1($c$)) at the antiquark plaquette vertices at their $x^3$/2
planes.  In the stretch (2+1)D configuration, the transverse gauge
fields $\bb A$ on the transverse links are copies of those on the
quark and the antiquark plaquettes at $x^3(q)$=$-x^3/2$ and at
$x^3(\bar q)=x^3/2$ respectively, and they are unchanged in the
stretching motion, while the longitudinal links are all $E^3$=$|\bb
E|/4$.

  By the duplication of the initial transversely-confined gauge fields
  along the longitudinal direction, we obtain a longitudinal tube with
  a cylindrical symmetry in (3+1)D.  The property of the transverse
  confinement at one longitudinal coordinate at birth is thereby
  extended to the whole tube, and the longitudinal tube will likely be
  a transversely-confining flux tube.

We show in the stretch (2+1)D model \cite{Won22c} that the
longitudinal $\bb B$ field that is present initially to confine quarks
and antiquarks on the transverse plane at their birth continues to
confine quarks and antiquarks transversely, because of the Landau
level dynamics.  The cloud of transverse gauge fields continue to
interact with each other to maintain the transverse confinement on
their transverse planes.  As a consequence, quarks, antiquarks, and
gauge fields will be transversely confined in the stretch (2+1)D flux
tube.

With the attainment of transverse confinement and $\bb E$ and $\bb B$
aligned along the longitudinal direction in the flux tube, it remains
necessary to examine the question of longitudinal confinement.
Therefore, we idealize the flux tube in the stretch (2+1)D
configuration as a longitudinal string in (1+1)D$_{\{x^3,x^0\}}$ and
approximate the quarks to be massless, with the information on the
transverse degrees of freedom stored as input parametric quantities in
the idealized (1+1)$_{\{x^3,x^0\}}$ space-time.  In the longitudinal
dynamics with massless quarks in QED in (1+1)D$_{\{x^3,x^0\}}$, there
is a gauge-invariant relation between the quark current $j^\mu$ and
the gauge field $A^\mu$ as given by \cite{Sch62,Sch63,Won94}
\vspace*{-0.2cm}
\begin{eqnarray}
j^\mu = \frac{g_\2d }{{\pi}} ( A^\mu 
- \partial^\mu \frac{1}{\partial_\lambda \partial^\lambda} \partial_\nu A^\nu),
\label{eq79}
\end{eqnarray}
where $g_\2d$ is the coupling constant in (1+1)D space-time which is
related to the coupling $g_\4d$ in (3+1)D bt Eq.\ (\ref{eq11}).  On
the other hand, the gauge field $A^\mu$ depends on the quark current
$j^\nu$ through the Maxwell equation,
\begin{eqnarray}
\partial_\nu(\partial^\nu A^\mu- \partial^\mu A^\nu)= -g_\2d j^\mu.
\label{eq80}
\end{eqnarray}
For the longitudinal motion, Eqs.
(\ref{eq79}) and (\ref{eq80}) lead to the Klein-Gordon equations
in $j^\mu$ and $A^\mu$
 for a boson with a mass
$m=g_\2d /\sqrt{\pi}$ ,
\begin{eqnarray}
- \partial_\nu \partial ^\nu  j^\mu=\frac{g_\2d^2}{\pi}j^\mu,~~~~\text{and} ~~~~
- \partial_\nu \partial ^\nu  A^\mu=\frac{g_\2d^2}{\pi} A^\mu.
\end{eqnarray}
Thus, the quark current $j^\mu$ and the longitudinal gauge fields
$A^\mu$ self-interact among themselves and build a longitudinal
confining interaction between the quark and the antiquark in (1+1)D.
As a consequence, in accordance with the Schwinger confinement mechanism for
massless fermions in QED in (1+1)D \cite{Sch62,Sch63}, the light quark
and the light antiquark interacting in QED will be longitudinally
confined just as well.  Possessing both transverse and longitudinal
confinements as in an open-string, the quark and the antiquark will be
confined and bound in a QED meson in (3+1)D.

\section{Implications of quark confinement in the QED interaction}

As the theoretical expositions and the accompanying experimental
evidence presented in the preceding sections strongly suggest possible
quark confinement in the QED interaction, it is worth examining the
implications of such a new concept in order to explore further new physics
on the frontier.  Such an exploration will necessarily be speculative
in nature.  However, some educated expectations may stimulate deeper
explorations into the unknown frontier, and some other expectations have
measurable experimentally consequences to merits further
considerations.

\subsection{Confinement may be an intrinsic property of quarks}

The experimental observations, if definitively confirmed under further
scrutiny, will indicate that the attribute of quark confinement is not
be the sole property of the QCD interaction alone and that quarks are
also confined in the QED interaction.  Consequently, an interesting
possibility is that the confinement attribute may be an intrinsic
property of quarks. This possibility is reinforced by the
observational absence of free quarks, which indicates further that in
the interaction of a quark and an antiquark in their many mutual
interactions, there may be a quark confinement principle which holds
that in the dynamics of quarks in different interactions, each
interaction always leads to the confinement of quarks.   That is, a
quark and its antiquark may be confined and bound as a neutral boson
in their many mutual interactions.  Specifically, a quark and its
antiquark certainly interact mutually in the weak-interaction and also
in the gravitational interaction.  There is no physical law that
forbids a quark and an antiquark to exchange a $Z^0$ boson or a
graviton to interact in the weak or the gravitational forces alone.
What is not forbidden is allowed, in accordance with Gell-Mann's
Totalitarian Principle \cite{Gel56}.  Consequently, a quark and its
antiquark may be confined and bound as a neutral boson in the weak
interaction and the gravitational interaction with the exchange of a
$Z^0$ boson or a graviton.
 
It is of interest to estimate the mass of such a confined neutral
boson with a weak strength of the coupling constant.  The quark and it
antiquark reside predominantly in (1+1)D.  If the work of Coleman
\cite{Col76} on confined fermions in QED in (1+1)D can be an
approximate analogous guide, such a confined boson falls into the
2D-weak-coupling regime.  In the 2D-weak-coupling regime the mass of
the composite boson would be approximately their rest masses, with
additional contributions from the weak coupling interaction that can
be calculated in the mass perturbation theory \cite{Col76}.  For
an interaction with a weak strength of coupling constant, we may expect the mass of
the confined and bound boson to be close to the rest mass of the quark
and the antiquark.  Such a particle however would decay by
quark-antiquark annihilations into two real photons, two virtual
photons, or an $e^+e^-$ pair.   The search for such a confined $q\bar
q$ pair may be signaled by a boson with a mass close to the sum of the
rest mass of the quark and its antiquark in the region of 4-10 MeV.
Experiments in high-energy hadron-hadron, nucleus-nucleus, and
$e^+e^-$ collisions in search of anomalous particles decaying into two
real or virtual photons or $e^+e^-$ in the region of a 4-10 MeV will
shed light on possible $q\bar q$ interactions with a weak strength of
coupling constants.  Rigorous theoretical work on the question of
quark confinement in the weak and gravitational interactions 
and the possibility of these weak-interaction-confined and gravitation-confined composite $q\bar q$ particles as dark matter 
material would be
of great interest.

The possibility of quarks confined in the QED interaction also implies
that the QED interaction between a quark and an antiquark differs from
the QED interaction between an electron and a positron.  It will be of
great interest to find out all the differences there can be and why
are they different.  For example, the QED interaction between an
electron and a positron may belong to the non-compact QED theory while
the QED interaction between a quark and an antiquark may belong to the
compact QED theory.  The possibilities of the compact and non-compact
QED bring with them the question whether the QED interaction is unique
or endowed with a multitude of experimentally testable possibilities
with different topological properties.  A related question is whether
the QED interaction between quarks in a nucleon may also contain the
linear QED confinement component that depends on the magnitudes and  signs of the
electric charges in addition to the standard Coulomb component  as in
Eq.\ (\ref{eq85}) below.

\subsection{New family of QED-confined particles and dark matter }

The success of the open-string description of the QCD and QED mesons
leads to the search for other neutral quark systems stabilized by the
QED interaction between the constituents in the color-singlet
subgroup, with the color-octet QCD gauge interaction as a spectator
field.  Of particular interest is the QED neutron with the $d$, $u$,
and $d$ quarks \cite{Won22,Won22a}.  They form a color product group
of ${\bb 3}$ $\otimes$ $ {\bb 3} $ $\otimes$ $ {\bb 3}$ = ${\bb 1}
\oplus {\bb 8} \oplus {\bb 8} \oplus {\bb {10}}$, which contains a
color singlet subgroup $\bb 1$ where the color-singlet currents and the
color-singlet QED gauge fields reside.  In the color-singlet
$d$-$u$-$d$ system with three different colors, the attractive QED
interaction between the $u$ quark and the two $d$ quarks overwhelms
the repulsion between the two $d$ quarks to stabilize the QED neutron
at an estimated mass of 44.5 MeV \cite{Won22}.  The analogous QED
proton has been found theoretically to be unstable because of the
stronger repulsion between the two $u$ quarks, and it does not provide
a bound state nor a continuum state for the QED neutron to decay onto
by way of the weak interaction.  Hence the QED neutron may be stable
against the weak interaction.  It may have a very long lifetime and
may be a good candidate for the dark matter.  Because QED mesons and
QED neutrons may arise from the coalescence of deconfined quarks
during the deconfinement-to-confinement phrase transition in different
environments such as in high-energy heavy-ion collisions, neutron-star
mergers \cite{Bau19,Wei20}, and neutron star cores \cite{Ann20}, the
search of the QED bound states in various environments will be of
great interest.

In a related matter, an assembly of the QED-confined $q\bar q$ mesons
can also be a good candidate for a part of the dark matter
\cite{Won20}.  Depending on its mass, the assembly can be an
$e^+ e^-$ emitter, a  $\gamma$ emitter, or the dark matter with no particle
emission if the mass exceeds their respective  threshold values \cite{Won20}.
 The dark matter models of the QED meson and the QED neutron utilize
 the material and the elements of the Standard Model, but 
 with a new confining combination.   Theoretical and experimental studies will be 
 needed to examine  the condense state of an
 assembly of QED mesons or QED neutrons.
 
In addition to new particles, the fact that the QED mesons are complex
objects brings with them additional degrees of freedom to lead to many
excited QED mesons states.  For example, there can be vibrational and
rotational states formed by these QED mesons.  We can get some idea on
the vibrational states from the spectrum of a stretched string as
shown in Fig.\ 7 of \cite{Won22}.  The possibility of adding quarks
with different flavors, angular momentum, and spin quantum numbers
will add other dimensions to the number of species of the QED-confined 
$q\bar q$ composite particles.

\subsection{Beyond the confining interaction of a quark and an antiquark in (3+1)D}

A quark and an antiquark reside predominately in (1+1)D, in which the
interaction between a quark and an antiquark is the linear confining
interaction for both QED and quasi-Abelian QCD, as discussed in
Section 3 and 4.  In the physical (3+1)D space-time, such a linear
interaction is only the dominant part of the full interaction between
the quark and the antiquark.  There will be additional residual
interactions arising from the presence of the transverse degrees of
freedom.  There are also contributions from the spin-spin, spin-orbit,
tensor, and other higher-order terms of the Breit interaction
\cite{Bar92,Won00,Won01,Cra09}.
 
For a confining string with a string tension $\sigma$, L\" uscher
\cite{Lus81,Lus80} considered the fluctuations in the transverse
direction of the flux tube as a massless bosonic field theory in two
dimensions with a classical action, for which the action can be
integrated out to lead to a potential between a static quark at ${\bf
  r}_1$ and an antiquark at $\bf r_2$ in the large string length limit
as
\begin{eqnarray}
V( {\bf r}_1 {\bf r}_2)=\sigma |{\bf r}_1 - {\bf r}_2| +c  -  \frac{\alpha}{|{\bf r}_1 - {\bf r}_2|} + O\left (\frac{1}{{|\bf r}_1 - {\bf r}_2|^2}\right ),
\label{eq83}
\end{eqnarray}
where $\alpha$ depends on the coupling constant, and $c$ is a
constant.  These are therefore long range residual interactions in
both the confined QCD and QED mesons.  They represent corrections that
arise from  expanding the potential between a quark and an antiquark 
 in powers of their  separation $ |{\bf r}_1 - {\bf r}_2| $.  A powerful tool
to study the non-perturbative behavior of the interquark potential
is the ``Effective String Theory'' in which the confining tube
contains the quark and the antiquark at the two ends
\cite{Nam70,Nam74,Got71,Lus81,Lus80,Pol91,Hel14,Aha13,Bon21}.  The
Nambu-Goto action can be integrated exactly in all geometries that
 are relevant for lattice gauge theories: the rectangle (Wilson
loop) in \cite{Bil13}, the cylinder (Polyakov loop correlators) in
\cite{Lus04,Lus05} and the torus (dual interfaces) in
\cite{Bil06}.

For quarks with color charge numbers $Q_1^\qcd$ and $Q_2^\qcd$
interacting in the QCD interaction, we can match the above equation
(\ref{eq83}) with the Cornell potential \cite{Eic75} and the
phenomenological quark-antiquark potentials in
\cite{Bar92,Won00,Won01,Won20} and \cite{Cra09}.  Upon neglecting the
spin-spin, spin-orbit, other higher order terms, and an unimportant
potential constant, we have the linear-plus-color-Coulomb interaction
of QCD
\begin{eqnarray}
V^\qcd( {\bf r}_1 {\bf r}_2)=  Q_1^\qcd Q_2^\qcd  \left [ -\sigma^\qcd |{\bf r}_1-  {\bf r}_2|  +  \frac{\alpha_s}{|{\bf r}_1 - {\bf r}_2|}    \right ] .
\end{eqnarray}
The quark and the antiquark also interact in the QED interaction.  We
can generalize the above to include both QCD ($\lambda=1$) and QED
($\lambda=0$) interactions to give
\begin{eqnarray}
V( {\bf r}_1 {\bf r}_2)&&=\sum_{\lambda=0}^1 Q_1^\lambda Q_2^\lambda  \left [ -\sigma^\lambda |{\bf r}_1-  {\bf r}_2|  +  \frac{\alpha_\lambda}{|{\bf r}_1 - {\bf r}_2|} 
\right ]\tau^\lambda, 
~ ~~ \lambda=\begin{cases} 0 & \text{QED} \cr 1  & \text{QCD} \cr \end{cases},
\label{eq85}
\end{eqnarray}
where $\tau^0$=$t^0$ is the generator of the U(1) gauge subgroup as
defined in Eqs.\ (\ref{eq14}), and $\tau^1$ is a fixed generator of the
SU(3) subgroup oriented randomly in the eight-dimensional
color-octet generator space in the quasi-Abelian approximation of the
non-Abelian QCD, as defined in (\ref{eq23b}) and discussed in Section
3.  The generators $\tau^0$ and $\tau^1$ satisfy $2{\rm
  tr}(\tau^\lambda \tau^{\lambda'}) = \delta^{\lambda \lambda'}$.

The above equation is for a single flavor.  In the case with many
flavors and flavor mixing, their effects can be taken into account by
replacing $Q_i^\lambda$ by the effective charge  $\tilde Q_i^\lambda$, where $\tilde
Q_i^\lambda$ is defined by Eq.\ (\ref{eq47a}) in subsections 4C and 4G.  It can be further generalized to the case when the
quark constituent and the antiquark constituent possess different
flavors.  For a composite $q_1 \bar q_2$ particle with 
many flavors 
and flavor mixing,
the above interaction between the quark $q_q$ and the antiquark $\bar q_2$
becomes
\begin{eqnarray}
V( {\bf r}_1 {\bf r}_2)&&=\sum_{\lambda=0}^1 \tilde Q_{q_1}^\lambda \tilde Q_{\bar q_2}^\lambda  \left [ -\sigma^\lambda |{\bf r}_1-  {\bf r}_2|  +  \frac{\alpha_\lambda}{|{\bf r}_1 - {\bf r}_2|} 
\right ]\tau^\lambda.
\label{eq86}
\end{eqnarray}
When there is no flavor mixing, as in the case of the charm and the
beauty quarks, the effective charge are just those of the standard
quark model, with $Q_{\{u,d,c,s,t,b\}}^\qcd$=1 and
$Q_{\{u,c,t\}}^\qed$=2/3, $Q_{\{d,s,b\}}^\qed$=$-1/3$, and $Q_{\bar
  q}^\lambda$=$ - Q_q^\lambda$.

In addition to the above linear-plus-Coulomb-type interaction, one can
include spin-spin, spin-orbit, tenor, and the full Breit interaction
to study the spectroscopy of hadrons, as carried out for a $q\bar q$
system for example in \cite{Bar92,Won00,Won01,Cra09}.  Such an
interaction will result in an open-string-type solution of the bound
meson states.
  
It is of interest to examine the meson-meson 
polarization in a many-meson system, assuming that the bound state
solutions for the quark and antiquark system within each meson have
already been obtained.  The task is to study the color-Coulomb and
electric-Coulomb interaction between constituents of different mesons
to see how such perturbative interactions may affect the dynamics of
the many-meson system\footnote{  The linear interaction between constituents
of different mesons are screened and can be included  by assuming a screening length as is done in \cite{Won04}.  We shall not included them  
in the present 
survey.  }.

When we work with pions as a possible molecular component, we should
keep in mind however that $\tilde Q_q^\qcd(\pi)$=$\tilde Q_{\bar q}^\qcd(\pi)$=0.  That is, 
 the effective color charge of a quark or an antiquark in a  pion is zero, as discussed in subsection 4G and in
Table \ref{tb2}.  This arises because of the cancellation of color
charges for two-flavor isovector QCD $q\bar q$ composite particles.
With the absence of the effective color charges, there is no
reactionary response to the confining interaction in $\pi^0$.  The
confinement of the constituents of $\pi^0$ comes only from the quark
condensate and quark masses  $m_q$ and $m_{\bar q}$ (see Section 4D).  There is
consequently no color-polarization of the pion under the color-Coulomb
interaction with neighboring color mesons.  So, in our subsequent
discussions on the color-polarization of QCD mesons, such a
 zero effective quark and antiquark color charges of a pion need to be kept in our mind.  On the other hand, for the discussions in electric charges and
electric-polarization in the QED interaction between mesons, pions
contain effective electric charges, and can still be polarized by QED
 forces between mesons.  So, pions can be included when we
examine QCD-QED molecular states arising from the electric-Coulomb
polarization between mesons.
 
\subsection{  Dipole-dipole interaction between neutral mesons}

In a single neutral meson (color-neutral with $\lambda=1$ or electric-neutral with $\lambda=0$, as the
case may be), the confining interaction between the quark and its
antiquark causes the constituents to execute a yo-yo motion, whose
classical trajectories for massless quarks are schematically depicted
in Fig.\ \ref{fig4a}.  There will be no net static dipole moment
 for  a composite neutral $q\bar q$ system.

We consider a system with $N$ neutral mesons and study the long-range color-Coulomb 
or electric-Coulomb interaction between the constituents from different mesons as
perturbations, starting with the $N$=2 system of two neutral mesons.
Owing to the two composite systems carrying color and electric
charges, the action of the long-range color-Coulomb and
electric-Coulomb interactions will  polarize the mesons \cite{Won04}.
As a consequence, they acquire dynamical dipole moments, leading to a
dipole-dipole interaction between the composite systems as discussed in
detail by Peskin and Bhanot \cite{Pes79,Bha79}.  Under appropriate
conditions, the polarization can lead to meson molecular states as
discussed previously in \cite{Won04}.  Among many other descriptions, meson molecular states have been
studied experimentally and theoretically since the earlier works of
\cite{Cho03,Won04,Tor04,Clo04,Pak04,Swa04,Vol04}, and in references cited in   the recent works of
\cite{Guo18,Yan20,Don21,Luo22}.

 A proper way to examine the meson polarization and meson molecular
 states in the $N$-meson system is to investigate each of the
 neighboring 2-meson pair as a four-body problem, with the interaction
 of the type in Eq.\ (\ref{eq86}), and reduce the problem to a
 two-body bound state problem as discussed in \cite{Won04}.  It will
 also be necessary to take into account the antisymmetry of the quarks
 if they are identical.  The problem can then be generalized from the
 2-meson system to the case of the $N$-meson system.
 
Following Wheeler's First Moral Principle\footnote{
Wheeler's First Moral Principle states  ```{\it Never make a calculation until you know the answer}'.   Make an estimate before every calculation, try a simple physical argument before every derivation.  Guess and answer to every puzzle.  A right guess reinforces the instinct.  A wrong answer brings the refreshment of surprises.'' \cite{Tay92}.
}
 \cite{Tay92}, it is worth
studying a simple massive dipole-dipole model before we carry out extensive
calculations.  We can gain new insight as to
 interesting geometrical configurations
of the $N$-meson phase space
where interesting physics may lie,
to merit further considerations.

\begin{figure} [H]
\centering
\includegraphics[scale=0.8]{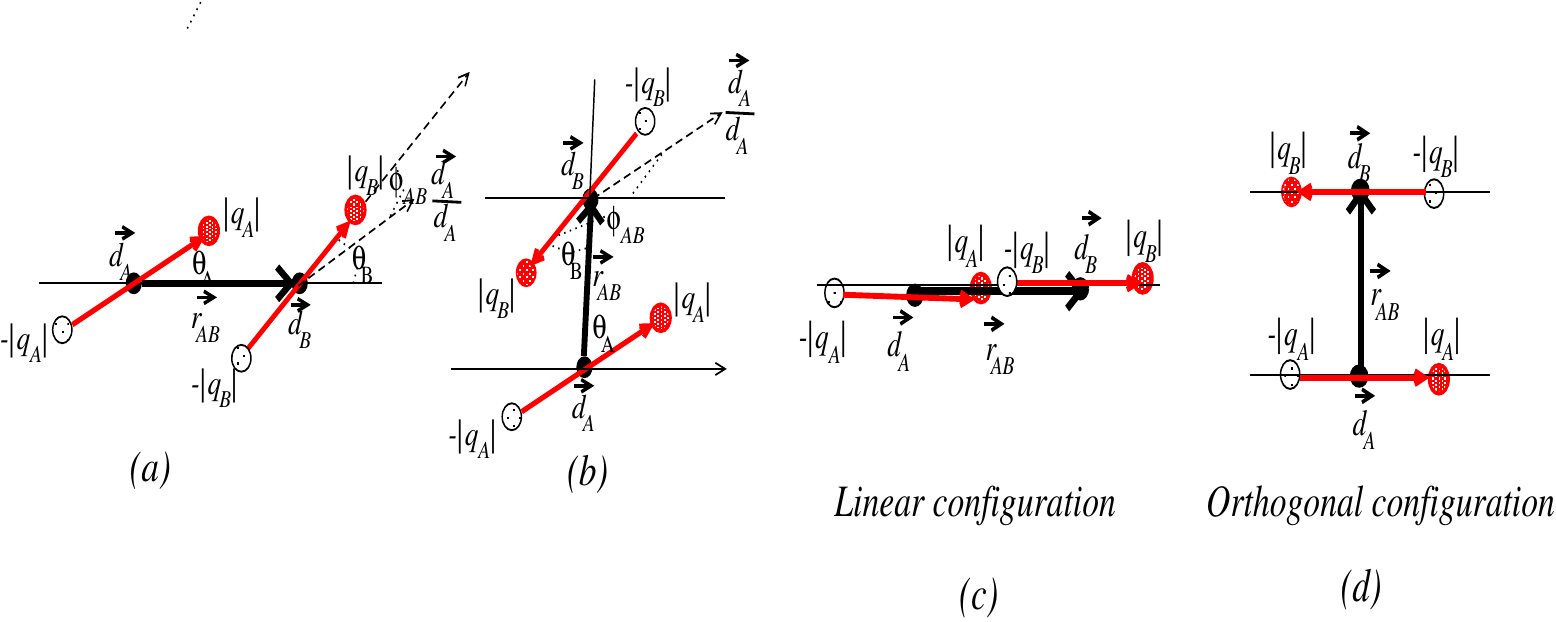}
\caption{The quark and antiquark configurations in the interaction
  between meson $A$ and $B$ with quark charges $\pm q_A$ and $\pm
  q_B$, respectively.  The two mesons are separated by the radius
  vector $\bb r_{AB}$ from meson $A$ to meson $B$. As a result of the
  long-range Coulomb interaction, mesons $A$ and $B$ acquire dynamical
  dipole moments $\bb d_A$ and $\bb d_B$, respectively.  In Figs. (\ref{fig12a}$a$)
  and \ref{fig12a}($b$), $\theta_A$ is the angle between  $\bb d_A$ and $\bb r_{AB}$,
  $\theta_B$ is  the angle between 
$\bb   d_B$ and $\bb r_{AB}$, and $\phi_{AB}$
  is the azimuthal angle between $\bb d_A$ and $\bb d_B$
  when we place $\bb d_A$ and $\bb d_B$  in the polar coordinates 
   with $\bb r_{AB}$ as the polar axis.
  Figs.\ \ref{fig12a}($c$) and \ref{fig12a}($d$) are the linear and orthogonal
  equilibrium configurations with respect to angular variations at a fixed $\bb r_{AB}$.}
\label{fig12a}
\end{figure}

We consider in Fig.\ \ref{fig12a} a meson $A$ with charges $ \pm |q_A|$,
and meson $B$ with charges $\pm|q_B|$, which can be color or electric,
static or dynamical, pure or mixed flavor,  as the case may be.  For each of the mesons $A$
and $B$, we presume that their linear-plus Coulomb interactions of
(\ref{eq86}) lead to their individual known bound states.  In the
neighborhood of each other, they are subject to the polarization
interaction from the long-range interaction of the constituents of 
other mesons as perturbations.  As a consequence, the mesons acquires
dynamical dipoles $\bb d_A$=$|q_A|\bb a_A$ and $\bb d_B$=$|q_B|\bb a_B$, respectively,
where the dynamical dipole $\bb d_i$ is directed from $-|q_i|$
to $+|q_i|$.  The potential
generated by the dynamical dipole $\bb d_A$ at a radius vector $\bb r_{AB}$ from meson $A$ to meson $B$ 
is given by

\begin{eqnarray}
V(\bb r_{AB}) &&=  -\alpha \bb d_A \cdot  \nabla (\frac{1}{r_{AB}}),
\end{eqnarray}
where $\alpha$=$(g^\lambda)^2/4\pi$.  The interaction energy between dipoles $\bb d_A$ and
$\bb d_B$ is
\begin{eqnarray}
W &&= (-|q_B|) V( \bb r_{AB}  -\frac{ \bb a_B}{2}) + |q_B| V( \bb r_{AB}  +\frac{ \bb a_B}{2})
\nonumber\\
&&=\alpha \left [\frac{\bb d_A \cdot \bb d_B }{r_{AB}^3} - 3 \frac {(\bb d _A\cdot \bb r_{AB}) (   \bb d_B \cdot \bb r_{AB}) } {r_{AB}^5} \right  ].
\end{eqnarray}
In terms of the angle $\theta_A$  between  $\bb d_A$ and $\bb r_{AB}$,  the angle  $\theta_B$ between $\bb d_A$ and  $\bb r_{AB}$,  and   the azimuthal angle
$\phi_{AB}$  between $\bb d_A$ and $\bb d_B$, when we place $\bb d_A$ and $\bb d_B$
 in the polar coordinate system with
 $\bb r_{AB}$ as the polar axis, (Fig.\ \ref{fig12a}($a$) and ($b$)), the
interaction energy $W$ between $A$ and $B$ is
\begin{eqnarray}
W (r_{AB},\theta_A,\theta_B,\phi_{AB})
&&=\alpha \frac{d_A d_B}{(r_{AB})^3} \bigl (
 \sin \theta_A \sin \theta_B 
\cos \phi_{AB}
 - 2 \cos \theta_A  \cos \theta_B \bigr ) .
\end{eqnarray}

For a fixed  $\bb r_{AB}$, we can get the equilibrium 
configurations 
with respect to angular variations 
by  taking the first and second derivatives with respect to $\theta_A$,
$\theta_B$ and $\phi_{AB}$.
Assuming that at the angular equilibrium, the
dynamical dipole moments $\bb d_A$ and $\bb d_B$ are weak functions of
the angles,  we find  that for a fixed $\bb r_{AB}$, there are two equilibrium configurations with respect to the angular variations:

\begin{enumerate}

\item
The linear equilibrium configuration in which the dynamical dipoles are linearly
aligned along the radius vector $\bb r_{AB}$ between $A$ and $B$ as in
a linear array:
\begin{eqnarray}
 \theta_A=0,~~ \theta_B=0, ~~\phi_{AB}={\rm ignorable}.
\end{eqnarray}
The dipole configurations shown in Fig.\ \ref{fig12a}($a$),  with a small $\theta_A$ and
$\theta_B$,  
approach
the linear equilibrium configuration  shown in
Fig.\ \ref{fig12a}($c$).

\item
The orthogonal equilibrium  configuration in which the dynamical dipoles are
orthogonal to the radius vector $\bb r_{AB}$ between mesons $A$ and
$B$, with opposing dynamical dipoles:
 
\begin{eqnarray}
 \theta_A=\frac{\pi}{2},~~ \theta_B=\frac{\pi}{2} , ~~\phi_{AB}=\pi.
\end{eqnarray}
The dipole configurations shown in Fig.\ \ref{fig12a}($b$), 
with opposing  dipoles   orthogonal to the radius vector $\bb r_{AB}$, and 
$\theta_A$ and $\theta_B$ close to  right
angles, approach  the
orthogonal equilibrium configuration shown in Fig.\ 13($d$).

 \end{enumerate}
 These equilibrium configurations for a fixed  $\bb r_{AB}$ can also be figured out by elementary physics arguments.
 
To obtain the molecular
state of mesons $A$ and $B$ for a given 
linear or an orthogonal configuration, it is necessary to solve the bound
state problem  involving $A$ and $B$ as a two-body problem 
in  the relative coordinate
$r_{AB}$,  
with the
reduced mass of the two mesons
 in the two body potential $W(r_{AB})$ 
\cite{Won04}.  The dipole-dipole interaction $W(r_{AB})$  at the equilibrium configurations    
are attractive at large  separations  $r_{AB}$,  at which the dipole-dipole
interactions are good approximations.  However, at small $r_{AB}$, it
is necessary to forgo the dipole approximation and study the
Coulomb-plus-linear interaction as a four-body problem, as is
done in \cite{Won04}, where  the two-body potential for 
mesons $A$ and $B$ is found  to be negative and contain a potential pocket.  
In the case of mesons containing constituents of identical particles,
it is necessary to take into account the antisymmetry of these
constituents.  The effects of antisymmetry is small when the mesons
are far apart and thus such effects are not important for large $r_{AB}$ 
 when the dipole approximation can be
approximately valid.  At small $r_{AB}$, the
antisymmetry gives rise to an effective repulsion between the mesons,
and consequently a 6-12 type potential of $(a/r^{12} - b/r^6)$ with
$a,b>0$ containing a potential pocket.  The potential obtained from these considerations will also be attractive at large distance and a 
potential pocket at short distances that will have bound meson molecular states in the limit of large
meson mass.  

Many QCD meson molecular states have been studied in 
\cite{Cho03,Won04,Tor04,Clo04,Pak04,Swa04,Vol04} and in references cited in \cite{Guo18,Yan20,Don21,Luo22}.    With QED
obeying similar dynamics as the QCD mesons, there should be similar linear and
orthogonal molecular states  for a system of two or many QED mesons.  

 From the Bohr-Summerfeld quantization rule, the
interaction potential such as the attractive $W(r_{AB})$ with a potential pocket would lead to a bound state if the action
integral along the radial $r_{AB}$ degree of freedom is equal to $2n\pi$.  The condition for molecular bound states depends crucially
on the reduced mass of the system, which depends in turn on the rest masses of
the quarks. 
Bound molecular states will begin to appear, when the reduced mass exceeds a threshold.  
Quantitative calculations will need to be carried out to find out
where the quark mass thresholds lie for the existence of the molecular
QCD meson and QED meson states. 

Given the large range of the quark masses of different flavors from a
few MeV to hundreds of GeV,
it is useful to consider 
a massive dipole-dipole model in the large mass limit 
 such that the masses of interacting dipoles are assumed 
 to exceed the threshold for  bound molecular states,
 for  systems with an attractive interaction potential 
 $W(r_{AB})$
 as a function of the dipole-dipole  separation $r_{AB}$,
 with the mesons oriented at  angles  of angular equilibrium (such as the linear or the orthogonal equilibrium of Figs. \ref{fig12a}($c$) and ($d$)).
  Our interest is to have an idea on the geometric configurations of the meson dipoles in the large mass limit so as to point to possible configurations that merit further analysis. 
  Such a massive dipole-dipole model is not the ultimate
definitive determination of stable configurations. 
On the other hand, the large range of quark masses ensure however that it is not just of academic interest. 
The model serves 
as a  guide to interesting meson geometrical configurations that may be possible when the
quark masses are large enough to exceed the mass threshold for molecular formation, and therefore worthy of further considerations with more rigorous analysis.

\subsection{ Interesting  molecular configurations  in the massive dipole-dipole model}

We can consider $N$ neutral mesons  in the massive dipole-dipole model to explore different molecular configurations of interest. 
The mesons will interact with each
other to lead to dynamical dipole moments, resulting in many different equilibrium configurations.  
The interactions are operative in the $\tau^0$ and
$\tau^1$ color space, as the color charge densities generates color
potentials while the electric charge densities generates electric
potentials. 
The  dipole-dipole interaction  energy $W$ of $N$ mesons is given by  
\begin{eqnarray}
W( \{{\bf r}_{i,j}^\lambda \}, \{\theta_i ^\lambda \},\{\phi_{i,j}^\lambda \} )
=\frac{1}{2}\sum_{\lambda=0}^1\sum _{i=1 }^{N} \sum _{j=1, j\ne i }^{N}
\left [ \frac{\alpha_\lambda d_i ^\lambda d_{j}^\lambda }{r_{ij}^3 }
(\sin \theta_i ^\lambda \sin \theta_{j}^\lambda \cos \phi_{i, j}^\lambda  - 2 \cos \theta_i^\lambda \theta_{j}^\lambda)
\right ]\tau^\lambda. 
\end{eqnarray}
 If we consider the approximation of only pairwise
dipole-dipole interaction between neighboring dipoles, then the
potential energy $W$ of $N$ mesons is given by
\begin{eqnarray}
W( \{{\bf r}_{i,i+1}^\lambda \}, \{\theta_i ^\lambda \},\{\phi_{i,i+1}^\lambda \} )=\sum_{\lambda=0}^1\sum _{i=1}^{N} 
\left [ \frac{\alpha_\lambda d_i ^\lambda d_{i+1}^\lambda }{r_{i,i+1}^3 }
(\sin \theta_i ^\lambda \sin \theta_{i+1}^\lambda \cos \phi_{i, i+1}^\lambda  - 2 \cos \theta_i^\lambda \theta_{i+1}^\lambda)
\right ]\tau^\lambda,
\end{eqnarray}
where in the case of a closed figure such as a polygon, the index $N+1$ reverts to be 1. 

With the linear and the orthogonal  equilibrium configurations, 
 there can be many 
different configurations for a system of many neutral mesons
(color-neutral or electric neutral, as the case may be).  Some
examples of the possible configurations of molecular states are
depicted in Fig.\ \ref{fig13}.
\begin{figure} [H]
\centering
\includegraphics[scale=0.80]{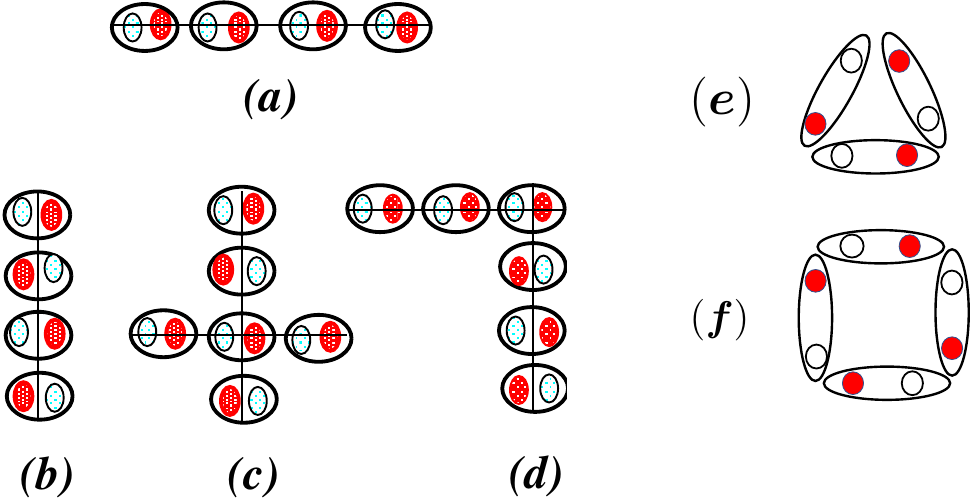}
\caption{ Interesting configurations of QCD and QED neutral mesons
 in the massive dipole-dipole
  model
 that may merit further
  considerations.  In these configurations, a meson is depicted with
  its positive charge constituent in a solid eclipse, and its negative
  charge constituent in an open eclipse.  Examples shown here include:
  a linear chain in Fig.\ \ref{fig13}($a$), an orthogonal chain as in
  Fig.\ \ref{fig13}($b$), the extension both longitudinally and
  transversely in Fig.\ \ref{fig13}($c$),  a linear chain turning
  in a right angle $\theta_1=\pi/2$ in Fig.\ \ref{fig13}($d$), a triangle in 
  Fig.\ \ref{fig13}($e$), and a square in  Fig.\ \ref{fig13}($f$).}
\label{fig13}
\end{figure}

One can build the linear chain as in Fig.\ \ref{fig13}($a$), the
orthogonal chain as in Fig.\ \ref{fig13}($b$), the extension both
longitudinally and transversely, as in Fig.\ \ref{fig13}($c$), or
turning in a right angle $\theta_1=\pi/2$ at a corner, as in
Fig.\ \ref{fig13}($d$).  By extending such chains and building up an orthogonal configuration into a third direction, it is possible
to construct molecular states in three dimensions as in organic chemistry.  

In addition to these linear and orthogonal configurations in two dimensions and perhaps also in three dimensions, there may be molecular  polygon meson systems.  
There may  be regular polygons if all the meson dipoles are the same, with $d_i$ = $d$.  
There may also be  irregular if some of the meson dipoles are different.   
We can consider the regular polygon as an simple illustration.  Such  regular polygon configurations arise in the massive 
  dipole-dipole model because  the interaction energy $W^{(N)}(r)$ as a function of the separation $r$ between the nearest neighboring dipoles  are always negative and attractive for 
   $N\ge 2$.  For example for $N=3$, the molecular state in the form of a regular triangle 
   as shown in Fig.\ \ref{fig13}($e$) has an interaction energy 
\begin{eqnarray}
W^{(3)}(r)=- \frac{3}{4} \frac { d^2 }{r^3}.
\end{eqnarray}
For $N=4$,  the molecular state in the form of a square 
as shown in Fig.\ \ref{fig13}($f$) has  
   the interaction energy 
\begin{eqnarray}
W^{(4)}(r) 
= (- 9 - \frac{1}{\sqrt{2}})\frac{ d^2 }{r^3}.
\end{eqnarray}
These interaction energies are  always negative and attractive, and  in the massive dipole-dipole model, a negative interaction energy with a potential pocket will lead to a bound state in the large mass limit.
For the  case of a polygon with $N$ dipoles, the interaction energy  from the contributions from the nearest neighboring dipoles is
\begin{eqnarray}
W^{(N)}({\rm all~ nearest~ neighboring ~pairs}) = \frac{N(N-1)}{2} \frac{ d^2}{r^3}[1   - 5 \cos^2( \frac{ \pi}{N})],
\end{eqnarray}
which is also always negative, and the contribution from next-to-next neighbors diminishes in strength.
In the massive dipole-dipole model, such a  negative interaction energy will lead to a stable  bound molecular state of  a polygon with $N$ mesons  in the  massive dipole-dipole model.  

It is interesting to take note the recent results on the stable $D^*D^*D^*$ triangular molecular state  from a  quantitative analysis in Ref.\  \cite{Luo22}.  The   $D^*D^*D^*$  triangular state for  $N=3$, together with the  molecular state X(3872) for $N=2$ , may 
 indicate  that the molecular state mass threshold in the large mass limit of the dipole-dipole model may  have been reached already with the charm quark mass.  
If this is the case, we should expect the massive dipole-dipole model to be approximately valid for other $N \ge$ 3  cases for  open charm mesons.  Open charm mesons in the form of 
square, pentagon, hexagon, and higher polygons may be possible.  Future experimental and theoretical research on these exotic  open charm meson polygons  will be of interest.

\subsection{Molecular states with both QCD and QED mesons}

The QCD and QED configurations in the massive dipole-dipole model considered in the last subsection
deal with color-neutral QCD mesons or electric-neutral QED mesons by
themselves in their respective $\tau^1$ and $\tau^0$ color subspace.
In addition to these  purely QCD and purely QED mesons,
there may be 
molecular states in the $\tau^0$ sector comprising of  both QCD and QED mesons
interacting  the QED interaction in the massive dipole-dipole model.

In such a mixture of QCD and QED mesons,
we can include both  electric-neutral and charged QCD mesons.
A charged QCD meson arises from  a quark and an  antiquark 
possessing electric
charges of the same sign.  For such a charged QCD meson,  
we assume that the forces leading to
quark confinement  in the  QCD meson  is dominated  by the  strong QCD
interaction.  Thus, even  though there occurs a
repulsive linear electric interaction between  electric constituent charges of
the same sign, such a QED  electric-linear repulsion will be overwhelmed by the
attractive QCD confining color-linear interaction  between the quark and the
antiquark and the confining force from the quark condensate.  The  QCD interaction is  expected to lead to  the confined
QCD  meson state.   As a  consequence, a charged QCD  meson $A$  with
constituents $q_A$ and $\bar q_A$ possesses a charge electric monopole
with a net electric charge $Q_A$
\begin{eqnarray}
Q_A= \frac{Q_{q_{_A}} + Q_{\bar q{_{_A}}}}{2}.
\end{eqnarray}
The charged meson also  possesses
electric dipole charges $\pm|Q_{q_{_A}} + Q_{{\bar q}_A}|/2$
and   a static  electric dipole 
$\bb d_A({\rm static})$,
\begin{eqnarray}
\bb d_A({\rm static})= 
\frac{|Q_{q_{_A}} + Q_{{\bar q}_A}| }{2} {\bb a_A},
\end{eqnarray}
where $\bb a_A$ is the open-string length vector of meson $A$ 
 from the lesser-charged quark to the greater-charged
quark of $A$. In the presence of neighboring mesons with
electric dipole moments, the charged meson will also be polarized to
acquire an additional dynamical dipole moment.  While the quantitative
magnitude of the total dipole moment needs to be worked out in detail,
it suffice to consider here that the QCD meson possesses a total
electric dipole vector sum $\bb d_A$ in its interaction with the neighboring
QED meson.  The effect of the electric monopole leads to an additional
electric monopole-dipole interaction energy with a neighboring dipole $\bb
d_B$, and the total electric interaction energy $W$ is given by
\begin{eqnarray}
W=\alpha_c  \left [ \frac{Q_A \bb d_B \cdot \bb r_{AB} }{r_{AB}^3} + \frac{\bb d_A \cdot \bb d_B }{r_{AB}^3} - 3 \frac {(\bb d _A\cdot \bb r_{AB}) (   \bb d_B \cdot \bb r_{AB}) } {r_{AB}^5} \right ] . 
\end{eqnarray}
With an additional monopole-dipole interaction,  the orientation angles at the linear equilibrium  configuration 
are unchanged.  That is,  they remain to be $\theta_A$=$\theta_B$=0, with ignorable $\phi_{AB}$.  So,
the linear   molecular state configurations for mixed QCD and QED
mesons are unchanged as in the case of neutral QCD and QED mesons.  
However, the orientation angles at the orthogonal equilibrium configuration are modified by the presence of the monopole-dipole interaction.  For this reason,  
we shall not consider the orthogonal
configuration in the present survey.

An interesting  (QED meson)-(QCD quarkonium)-(QED meson) combination as depicted in Fig. \ref{fig14}($a$).  In such a
configuration, the positive charge of the QCD quarkonium will
attract a QED meson dipole pointing away from the quarkonium while the
negative charge of the quarkonium will attract a QED dipole pointing
into the heavy quarkonium.  Such a chain of QED-QCD-QED mesons is a 
molecular state  in the massive dipole-dipole model at large mass limit. 
Whether such a structure with the X17 or E38 as the QED meson have reduced masses  exceeded  the mass thresholds for the occurrence  of molecular states 
will need to be worked out quantitatively in detail.  
\begin{figure} [H]
\centering
\includegraphics[scale=0.35]{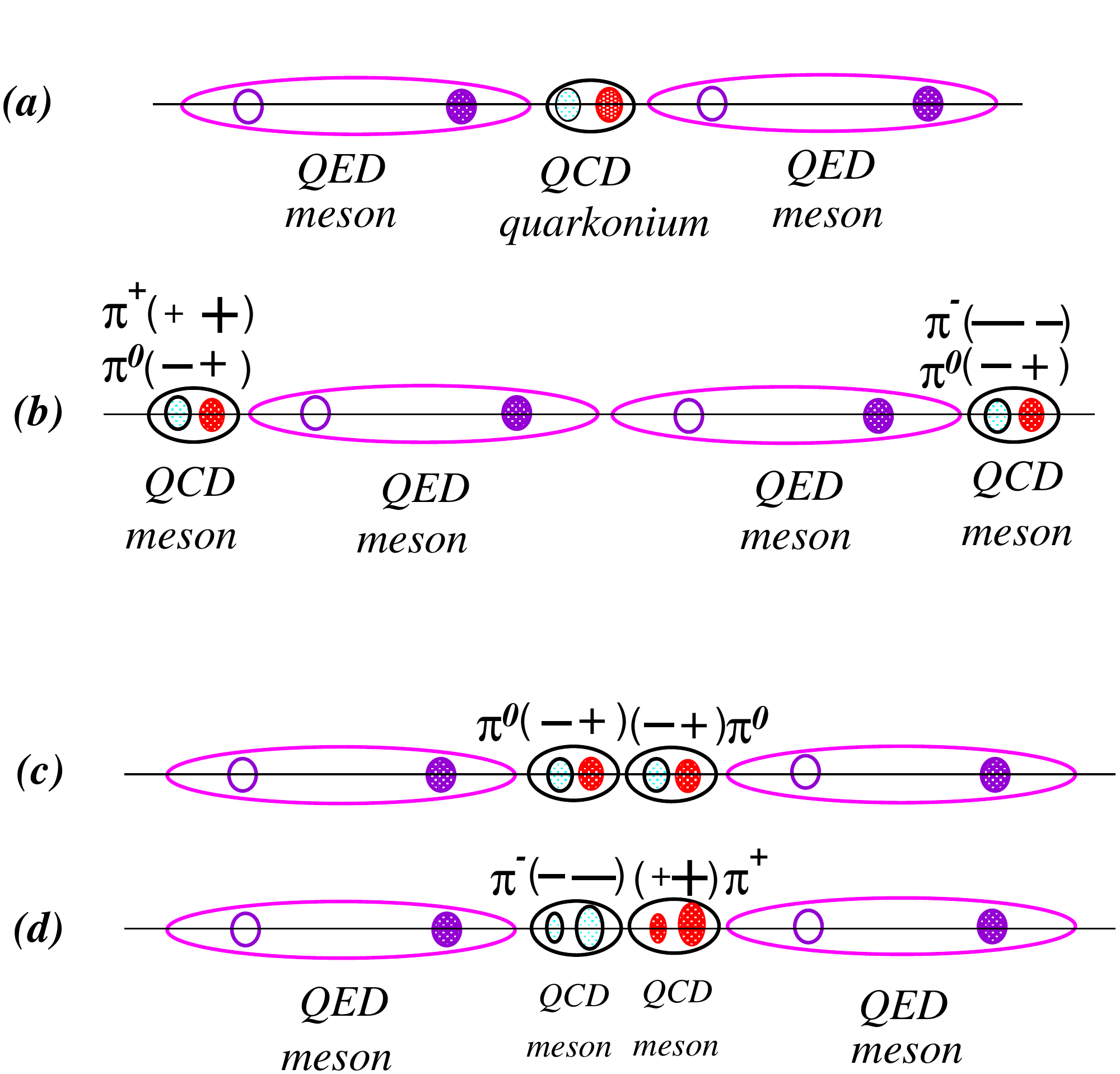}
\caption{ Examples of interesting linear molecular 
  configurations in the massive dipole-dipole model of QCD-QED mesons
  that may merit further considerations.  In these configurations, a
  a solid eclipse stands a positive charge and an open eclipse
 for a negative charge. Examples shown here include a linear chain for (QED
  meson)-(QCD quarkonium)-(QED meson) in Fig.\ \ref{fig14}($a$), 
  the (QCD meson)-(QED meson)-(QED meson)-(QCD meson) in
  Fig.\ \ref{fig14}($b$),  and the 
   (QED meson)-(QCD meson)-(QCD meson)-(QED meson)
  in 
  Fig.\ \ref{fig14}($c$) and \ref{fig14}($d$).  Pions as
  examples of the QCD mesons are indicated schematically.  
   }
\label{fig14}
\end{figure}

Other interesting combinations involve charged or neutral QCD and QED
mesons as depicted in
Fig.\ \ref{fig14} ($b$)-($d$).  In the meson chain in $(b$)-($d$), the QCD meson can be either neutral or charged with electric charge of the appropriate sign.  These configurations are the
equilibrium configurations in the massive dipole-dipole model.

Even though quarks in pions does not carry color charges because of
the cancellation of their color charges in the isovector state, they
carry electric charges. As a consequence, pion can be  electrically
polarized by other mesons and they in turn can polarize
other mesons.  The possible equilibrium configurations in
Fig.\ \ref{fig14}($b$)-($d$) in the massive dipole-dipole
model provides impetus to explore whether the $\pi$-(QED meson)-(QED
meson)-$\pi$  or the 
 (QED meson)-$\pi$-$\pi$-(QED meson)
molecules may be related to  the perplexing ABC anomaly
\cite{Aba60,Boo61,Ban73,Adl11,Bas17,Kom18} and the R360 anomaly \cite{Abr09,Abr19}.  If the molecular
states can resolve these anomalies, they will provide additional
support for the description of the anomalous X17 and E38 particles as
QED mesons.

In exploring the molecular description of the ABC and R360
resonances, we need to estimate the masses in these linear QCD-QED
meson molecules.  In a molecular state of the type in
Figs.\ \ref{fig14}, the molecular binding energy of the mesons would
be of order of a few MeV (see for example the molecular state in
Ref. \cite{Won04}).  So, the mass of the molecular states consisting
of
\begin{eqnarray}
(\text{mass of molecular state}) =\sum_i m_i - (\text{molecular binding energy}) 
\end{eqnarray}
where $m_i$ are the mass of the meson $i$ constituents of the
molecular state.  Within an error of the order of a few MeV, the mass
of a molecular state is approximately
\begin{eqnarray}
(\text{mass of a molecular state})  \approx \sum_i m_i.
\end{eqnarray}
We can use the above approximate molecular-state relation  to estimate the masses of the molecular states
in Figs.\ \ref{fig14}($b$)-($d$).
If the QED mesons in these figures are the X17 particles each with a mass of 17 MeV, then 
the masses in these linear QCD-QED meson molecules will be 
 \begin{eqnarray}
m_{\pi^+} +m({\rm X17)}+m({\rm X17}) + m_{\pi^-}&&=  314{\rm ~ MeV},
\nonumber\\ 
{\rm and} ~~~m_{\pi^0}+m({\rm X17)}+m({\rm X17}) + m_{\pi^0}&&=  304{\rm ~ MeV},
\end{eqnarray}
which fall within the energy of the ABC resonance at $M_{\rm
  ABC}$=310 MeV with a width of $\Gamma_{\rm ABC}$=10 MeV
\cite{Aba60}.  Other measurements gave the masses as $M_{\rm ABC}$=316
MeV$\pm$10 with a width $\Gamma_{\rm ABC}$=55$\pm $10 MeV in $d + p
\to ^3$He+ X collisions with a  deuteron beam at 2.83 MeV at a scattering
angle of 0.3$^o$ of the fused $^3$He \cite{Ban73}.  Another
measurement gives $M_{\rm ABC}$= 298$\pm 5$ MeV to 319$\pm 8$ MeV with
a width from $\Gamma_{\rm ABC}$=38$\pm 5 $ MeV to 44+$\pm 9$ MeV
\cite{Kom18}.
There is thus an approximate relation 
\begin{eqnarray}
m(X17) \approx \frac{1}{2} ( m_{\rm ABC} - 2 m_\pi  ).
\label{eq102}
\end{eqnarray}
 
If the QED mesons in Fig.\ \ref{fig14}($b$) are E38 particles each with a mass of 38 MeV, then the masses in these linear QCD-QED meson molecules will be 
 \begin{eqnarray}
m_{\pi^+} +m({\rm E38)}+m({\rm E38}) + m_{\pi^-}&=&  354
{\rm ~ MeV},
\nonumber\\ 
\text{and}~~~m_{\pi^0}+m({\rm E38)}+m({\rm E38}) + m_{\pi^0}&=&   344
{\rm ~ MeV},
\end{eqnarray}
which fall within the energy of the R360 resonance of
$M_{\rm R360}$=360$\pm 7 \pm 9$ MeV with a width of $\Gamma_{\rm
  R360}$=$63.7\pm 17.8$ MeV observed by Dubna \cite{Abr09}.  The
SACLAY group also observed similarly an anomalous resonance 
at $M_{X}$=365$\pm 23$ MeV with a
width of $\Gamma_{\rm X}$=$51\pm 10$ MeV
in 
the reaction of 
$d + p
\to ^3$He+ X  for  a deuteron beam at 3.82 MeV at a scattering
angle of 0.3$^o$ \cite{Ban73}.  There is thus an approximate relation 
\begin{eqnarray}
m(E38) \approx \frac{1}{2} ( m_{\rm R360} - 2 m_\pi  ).
\label{eq104}
\end{eqnarray}
 Equations (\ref{eq102}) and (\ref{eq104}) relating  the ABC and  R360 resonances with the X17 and E38 resonances suggests that the proposed description of QED-confined mesons have the prospect of linking  the five anomalies of (i) the anomalous soft photons, (ii) the X17 particle, (iii) the E38 particle, (iv) the ABC resonance, and (v) the R360 resonance  in as single consistent QED meson framework.

As the masses of the configurations  in Fig.\ \ref{fig15}($c$) and ($d$) falling within the measured ABC
and R360 resonances, it is thus worth exploring whether the molecular 
states may provide the appropriate descriptions for these two anomalous
resonances.  The ABC resonance is produced in nuclear collisions with
the following characteristics occurring near the threshold of two
pion production \cite{Aba60,Boo61,Ban73,Adl11,Bas17,Kom18}:
  
 \begin{enumerate}
  
 \item
 the ABC resonance occurs at energies near but slightly beyond the
 threshold for two pion production (of 270-280 MeV) at $M_{\rm ABC}\sim$ 310 MeV
 with a relatively narrow width of order 50-60 MeV

 \item
 most likely observed in the forward and backward directions

 \item
 isoscalar nature of the $\pi \pi$(X)  pair

\item
the occurrence of the ABC resonance is accompanied by the fusion of
the colliding nuclei into a fused nuclear system \cite{Aba60,Boo61,Ban73,Adl11,Bas17,Kom18} as in 
$d $+$ p $$\to$$ ^3$He$+$X$^0$, $d + p \to ^3$H+X$^+$
\cite{Aba60,Ban73}, and $d + d \to ^4$He+$X^0$ \cite{Adl11}

\item
no occurrence of the ABC resonance when there is no fusion of the
colliding nuclei, as in $pn \to pp \pi^0 \pi^-$ and $pn \to pn \pi^0
\pi^0$ even though the $d^*(2380)$ resonance appears in these
reactions.  Thus, the occurrence of the ABC resonance may be
independent of the $d^*({2380})$ resonance \cite{Bas17}.
  
  \end{enumerate}

\begin{figure} [H]
\centering
\includegraphics[scale=0.50]{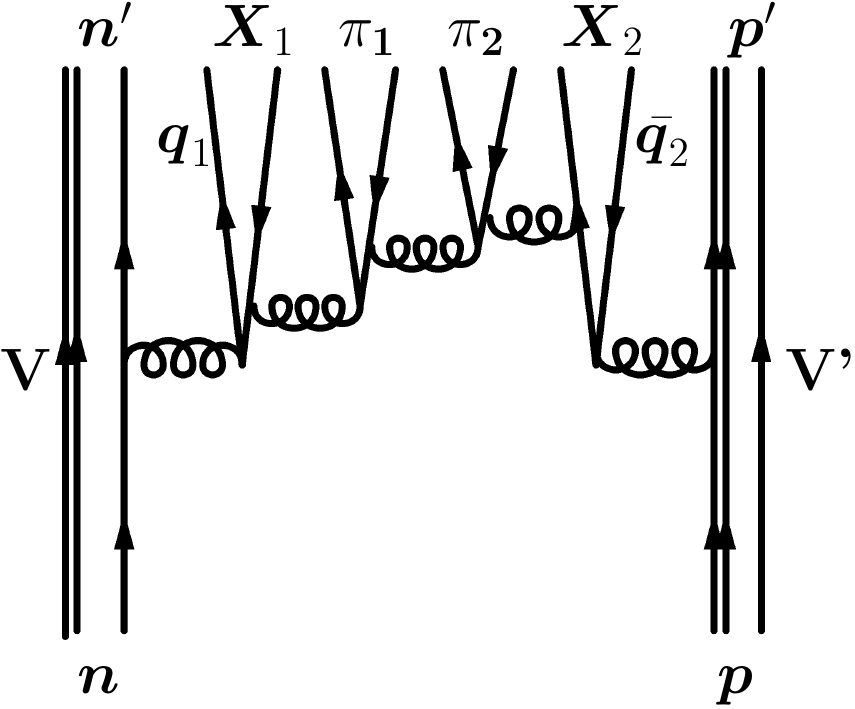}
\caption{ The possible mechanism of the production of the molecule state  $\pi$-(QED meson)-(QED
  meson)-$\pi$ in the collision of $n$ and $p$ near the two-pion
  threshold by the fusion of two virtual gluons, with the subsequent
  string fragmentation of the $q_1$-$\bar q_2$ open string into the four mesons,
  $X_1,\pi_1,\pi_2, X_2$, near the $\pi$$\pi$  threshold.  }
\label{fig15}
\end{figure}
 
 We can attempt to  interpret the molecular state description of the
 ABC and the R360 resonances  by describing the possible production mechanism in the following way.
 The  basic process may be  the collision of a projectile nucleon $p$ and a
 target nucleon $n$ as depicted in Fig.\ \ref{fig15}, entering as a
 sub-process in the more complicated nuclear collisions.  In such a
 $p+n$ collision, we may envisage the emission of a gluon from $n$ at the vertex $V$ and 
 and a gluon from $p$ at the vertex $V'$  as in Fig.\ \ref{fig15}.
 The fusion of the two gluons lead to a $q_1$ and $\bar q_2$ string which pull
 apart from each other.   The fragmentation of the receding  $q_1\bar q_2$ string 
 in the inside-outside cascade picture of Bjorken, Casher, Kogut, and Susskind
\cite{Bjo73,Cas74}  produces  the pions and other
 particles, $X_1,\pi_1,\pi_2,X_2$,
 as depicted in
 Fig.\ \ref{fig15}.      Because the produced particles arise by
 pulling the receding
 $q_1\bar q_2$ open string apart from the colliding nucleons, the
 fragmentation process will be facilitated in the forward and backward
 direction of the colliding baryon system.  Hence, the productions of
 the ABC and R360 resonances  may be  favored in the forward and backward directions.
 
 The isospin properties and the parity of the X17 as an isoscalar particle and
 E38 particles as an isovector particle require that they must be produced in pairs in order to have the
 quantum number of the vacuum.
 The produced $\pi^0\pi^0$  or $\pi^+\pi^-$  pair can also form a coupled isoscalar, even-parity  state in the QCD sector.
 The pair of X17 or E38  QED mesons can
 combine with two pions to form an isoscalar, even-parity  object,  depicted schematically in Fig.\ \ref{fig15}.
 The narrowness of the widths of the ABC and R360 resonances indicates that they may likely be    molecular resonances with a weak binding and long-range interactions, with the mesons  far separated.  They may not likely be  resonances 
from the exchange of the strong-interaction scalar QCD meson for which the width would be much greater.
 
 The requirement that the ABC resonance must necessarily accompany the
 fusion reaction may  perhaps be understood  in terms of the color flow of the
 projectile and target nucleons in Fig.\ \ref{fig15}.  The incident $p$
 and $n$ are initially in a colorless color-single state.  After $p$
 and $n$ each emits a virtual gluon at $V$ and $V'$, the scattered
 nucleons $p'$ and $n'$ will be colored objects.  With the produced
 $q\bar q$ open string materializing as the ABC resonance of the colorless
 molecular state of $\pi$-X17-X17-$\pi$, the process
 would not be observable if the scattered $p'$ and $n'$ remain as
 colored objects and cannot combine with other colored objects to form a
 colorless state.  The scattered colored $p'$ and $n'$ can bleach
 their colors by exchanging a virtual gluon and  fusing together into a
 colorless bound state of the fused nuclei.  Hence, the fusion reaction
 always accompany the production of the molecular  state of the
 ABC resonance.   Such a fusion of the baryons and the fusion of the gluons 
with the subsequent production of the ABC resonance
 can take place in conjunction because they occur in different parts of the three-dimensional space.
    
 It is interesting to note the similarity of the proposed mechanism  for the production of the ABC resonance
  in Fig.\ \ref{fig15} and the proposed mechanism  for the production of the X17
 particle in $p+^3$H$_{\rm ground ~state}$ in Fig.\ \ref{fig3}($a$), because
 the process of color bleaching leading to a
 fused nucleus  are the same, as discussed in Section
 2.
 
If the molecular states of $\pi$-X17-X17-$\pi$ and $\pi$-E38-E38-$\pi$
are proper descriptions of such resonances, we would expect that 
X17 particles may appear as decay products of the ABC resonance, and
E38 particles may appear as decay products of the R360 resonance.
A search for these particles accompanying the ABC resonance and the
R360 resonance in these reactions in the forward and backward directions will be of interest.  In this regard, it is
interesting to note the possible mutual evidential support that
reactions involved in the production of R360 are also the reactions
where the E38 particle has been observed \cite{Abr09,Abr19}.
Further research on these interesting suggestions with regard to the ABC and the R360
resonances will be of great interest.

\subsection{ Molecular states with baryons} 

The consideration of molecular mesons can be generalized to include
the discussions on baryons when we approximate a baryon as a
confined state of a quark and a diquark, with the diquark playing the
role of an antiquark.  In such a picture, a charged baryon possesses
an electric monopole in addition to an electric dipole moment.  The
color and electric charges in the presence of neighboring mesons or baryons
 will be polarized and they  will acquire additional dynamical dipole moments.  Likewise, we
need to consider both the dynamical color and electric polarization in
the molecular states involving baryons.  Because a
baryon that carry an electric charge  repel another
baryon with electric charges of the same sign, molecular states with 
baryons  would be favorable with a charged baryon with a neutral baryon or an antibaryon with a charge of opposite signs.  In this regard,  molecular states consisting of a baryon and an antibaryon  opens  a new degree of freedom to the construction of
hadron molecular states as discussed in \cite{Don21}.

An example of particular interest is the $d^*(2380)$ resonance
\cite{Adl14,Adl14a,Wor14,Wor16} that occur often with the ABC effect.
Although a 6-quark description is possible \cite{Bas13} and chiral quark model \cite{Gol89,Pin08} is used to 
describe the baryon-baryon resonance, it can also be depicted 
equivalently as a molecular state of
$\Delta^+[IJ=(1/2)(3/2)^+]$-$\Delta^0[IJ=(1/2)(3/2)^0]$ state with a
$D_{03}$($I=0$,$J^\pi$=$3^+$) dipole-dipole molecule. 
The molecular state  binds its two $\Delta$ resonances by a
binding energy of $E_{\rm binding}$=2$ \times 1232$- 2380=82 MeV.
Because of the molecular binding, the two $\Delta$ constituents lie
below their rest mass energies, and as a result, the decay life time is
lengthened with a narrower width of about 70 MeV, much smaller than
the expected width of twice the $\Delta$ decay width of 110 MeV.  They
are formed in the collision of a proton and a neutron so that they can
occur as a stretched configuration between the $\Delta^+$ and  $\Delta^0$
predominantly in the forward and backward directions.  It will be interesting to  examine the molecular properties of the $d^*(2380)$, in addition to the other descriptions.

\section{Conclusion and discussions} 

As advised by the late John Archibald Wheeler on many occasions, ``In the
exploration on the new frontiers of physics, we make progress by
walking on two legs, with one leg on firm grounds, and the other in a
venturing spirit''.  On the question of quark confinement in QED, on
firm grounds are (i) the non-isolation of quarks, (ii) a quark and its
antiquark interaction in the QED interaction, and (iii) a quark and its
antiquark are confined in the QED interaction in (1+1)D (Schwinger).  In venturing spirit is the
question whether they might be confined in the physical world of
(3+1)D, when they interact in the QED interaction alone.

A prerequisite for asking such a question is to inquire whether 
there are experimental circumstances in which a $q\bar q$ pair may be
produced and may interact in QED interaction alone.  We find that in
hadron-hadron, $AA$, $e^+e^-$, and $e^-A$ collisions, there can be
situations in which a quark and an antiquark may be produced with a
center-of-mass energy below the pion mass gap $m_\pi$ for 
collective QCD excitation, and the quark and its antiquark can interact
non-perturbatively in the QED interaction alone.

It is well known that in the Schwinger confinement mechanism, massless fermions
interacting in the Abelian QED gauge interactions in (1+1)D are
confined for all strengths of the gauge interaction, as in an open
string, leading to a confined and bound neutral boson with a mass
proportional to the magnitude of the coupling constant.

Light quarks have masses of only a few MeV and they can be
approximated as massless. A quark and an antiquark cannot be
isolated, so they reside predominantly in (1+1)D.  They can be produced and interact in the QED interaction alone.
The conditions under which the Schwinger confinement mechanism 
can be applicable are met when a light quark and a light antiquark are produced and 
interact  with the QED interaction alone.
Consequently, we can 
apply the Schwinger mechanism to quarks to conclude that a light quark and its antiquark are confined
in the QED interaction in (1+1)D.  The requirement of the massless
condition for quarks is actually not as restrictive as it may appear
to be because Coleman, Jackiw, and Susskind showed that
the Schwinger confinement persist even when quarks are
massive  in (1+1)D.  Therefore, a quark and its antiquark are confined in the QED
interaction for all flavors and gauge interaction strength in (1+1)D.
 
On questions of quark confinement and $q\bar q$ QCD bound states,  the
non-Abelian QCD interaction can also be approximated as a
quasi-Abelian interaction. As a consequence, the Schwinger confinement mechanism
can be applied to quarks interacting in both the QED interaction and
the QCD interaction, leading to confined $q\bar q$ pairs in QED and
QCD open string states in (1+1)D, with the composite boson masses
depending on the magnitudes of the QCD and QED coupling constants.
Such a viewpoint is consistent with the QCD string description of
hadrons in the Nambu\cite{Nam70} and Goto\cite{Nam70} string model,
the string fragmentation models of particle production of Bjorken,
Casher, Kogut, and Susskind \cite{Cas74,Art74,And83}, the Abelian
projection model \cite{tHo80}, and the Abelian dominance model
\cite{Sei07,Suz08}.

In a phenomenological analysis, we inquire whether an open string in 
the phenomenological  open string model of QCD and QED $q\bar q$ systems in (1+1)D  can be
the idealization of a flux tube in (3+1)D and can show up as a bound and confined boson.  In such a
phenomenological open string model, we need an important relationship
to ensure that the boson mass calculated in the lower (1+1)D can
properly represent the mass of a physical boson in (3+1)D.  The  open
string in (1+1)D can describe a physical meson in (3+1)D if the structure of the
flux tube is properly taken into account.  This can be achieved by
relating the coupling constant in (1+1)D with the coupling constant in
(3+1)D and the flux tube radius $R_T$ \cite{Won09,Won10,Kos12,Won22a}.
Using such a relationship, we find that that $\pi^0, \eta$, and
$\eta'$ can be adequately described as open string $q\bar q$ QCD
mesons.  By extrapolating into the $q\bar q$ QED sector in which a
quark and an antiquark interact with the QED interaction, we find an
open string isoscalar $I(J^\pi)$=$0(0^-)$ QED meson state at
17.9$\pm$1.5 MeV and an isovector $(I(J^\pi)$=$1(0^-), I_3$=0) QED
meson state at 36.4$\pm$3.8 MeV.

On the experimental front, it has been observed that anomalous soft
photons with a transverse momentum of many tens of MeV/c are 
proportionally produced when hadrons are produced and are not produced
when hadrons are not produced, indicating that the production of 
hadrons are always accompanied by 
the production of neutral
particles with masses in the region of many tens of MeV/c$^2$.
In search of axion with a mass of many tens of MeV, anomaly pointing
to the production of an X17 particle with a mass of about 17 MeV has
been observed in the decay of He$^4$, $^8$Be, and $^{12}$C excited
states.  There have been also observation of the E38 particle at Dubna
with a mass of 38 MeV.  The predicted masses of the isoscalar and
isovector QED mesons in the open-string model of the QCD and QED mesons
are close to the masses of the reported X17 and
E38 particles observed recently, making them good candidates for these
particles.  Experimental confirmation of the reported X17 and the E38
particles in the same experimental setup 
will shed light on the question of quark confinement for
quarks interacting in the  Abelian U(1) QED interaction.

On the theoretical front, there is a need for theoretical
clarification on the question of confinement with regard to lattice
gauge calculations.  The lattice gauge calculations indicate that a
static quark and a static antiquark interacting in the compact QED
interaction will not be confined in (3+1)D.  However, the deconfined
solution for static quark and static antiquark in compact QED in
(3+1)D in lattice gauge calculations contradicts the experimental
absence of fractional charges.  This indicates that the present
lattice gauge calculations for compact QED in (3+1)D may not be
complete and definitive, because the important Schwinger dynamical
quark effects associated with light quarks has not been included.
 
We have constructed a ``stretch (2+1)D'' flux tube model to investigate
the importance of the Schwinger mechanism on quark confinement in QED
in (3+1)D \cite{Won22} by utilizing the Polyakov's transverse
confinement in conjunction with Schwinger's longitudinal confinement,
We find that the stretch (2+1)D flux tube model leads to quark
confinement in compact QED in (3+1)D \cite{Won22a}.  Such a quark
confinement result of the stretch (2+1)D flux tube model is consistent
with the experimental absence of fractional charges.  Furthermore, it
gives predictions on the masses of neutral QED mesons and QED mesons
in agreement with the experimental QCD and QED meson spectra.  It is
therefore worthy of further considerations.  It is important to find
out whether future lattice gauge calculations with dynamical light
quarks, in a configuration such as the stretch (2+1)D flux tube
configuration, will lead to confined quarks in compact QED in (3+1)D.

The success of the open-string description of the QCD and QED mesons
leads to the search for other neutral quark systems stabilized by the
QED interaction between the constituents in the color-singlet
subgroup, with the color-octet QCD gauge interaction as a spectator
field.  Of particular interest is the QED neutron with the $d$, $u$,
and $d$ quarks \cite{Won22c}.  They form a color product group of
${\bb 3}$ $\otimes$ $ {\bb 3} $ $\otimes$ $ {\bb 3}$ = ${\bb 1} \oplus
{\bb 8} \oplus {\bb 8} \oplus {\bb {10}}$, which contains a color
singlet subgroup $\bb 1$ where the color-singlet current and the
color-singlet QED gauge field reside.  In the color-singlet
$d$-$u$-$d$ system with three different colors, the attractive QED
interaction between the $u$ quark and the two $d$ quarks overwhelms
the repulsion between the two $d$ quarks to stabilize the QED neutron
at an estimated mass of 44.5 MeV.  The analogous QED proton has been
found theoretically to be unstable, and it does not provide a bound state nor a
continuum state for the QED neutron to decay onto by way of the weak
interaction.  Hence the QED neutron may be stable against the weak
interaction.  It may have a very long lifetime and may be a good
candidate for the dark matter.  Because QED mesons and QED neutrons
may arise from the coalescence of deconfined quarks during the
deconfinement-to-confinement phrase transition in different
environments such as in high-energy heavy-ion collisions, neutron-star
mergers \cite{Bau19,Wei20}, and neutron star cores \cite{Ann20}, the
search of the QED bound states in various environments will be of
great interest.

 It is necessary to address the question why the QCD mesons and the QCD neutron have not been observed to decay to the lower energy QED counterparts, even though the QCD hadrons  lie higher in energy than those of the QED mesons  and the QED neutron.
The QCD meson and neutron and their analogous  QED counterparts lie at different local energy minima in the configuration space..
The QCD mesons  have a flux tube radius of about 0.4 fm and a longitudinal  length $L_\qcd$ of about a fermi, as one can infer from the lattice gauge calculations and phenomenological considerations. 
 For the QED meson, a flux tube radius about the same as  that of the QCD meson appears to be a reasonable concept, because such a flux tube radius  gives the QED meson masses in agreement with the observed X17 and E38 masses.  The longitudinal lengths $L_\qed$ of the QED mesons are an order of magnitude greater than those of the QCD mesons \cite{Won22c}.
The QCD meson and their analogous  QED meson lie at different local energy minima in the longitudinal length space $L$.
In between the two energy minima, there is a  barrier between the QCD meson and the analogous QED counterpart.
To make a transition from a QCD meson energy minimum  at  $L_{\rm QCD}$ so as  
to come to the region of the analogous QED meson in the local energy minimum at $L_{\rm QED}$, the QCD meson needs to tunnel under a barrier from $L_\qcd$ 
to $L_\qed$
with a barrier height of order 
$(L_\qcd - L_\qed)\times 1$ GeV/fm, where 1 GeV/fm is the magnitude of the QCD string tension coefficient.   This gives a barrier height of order  10 GeV.  Such a high barrier makes the transition   probability prohibitively  small.   Similarly, a QCD neutron has a spatial length scale of order 1 fm, whereas the QED neutron has a spatial length scale of order 20-30 fm \cite{Won22c}.  They lie in different local energy minima in the length scale space $L$.
 The high barrier between the QCD neutron energy minimum and the QED neutron energy minimum makes
the transition probability prohibitively small.

Experimentally, in the environment in which  a pair of  color-singlet quark and an antiquark  are produced, such as by the fusion of two virtual photons or two virtual gluons at the eigenenergy of a QED meson (e.g. 17 or 38 MeV, with the proper quantum numbers), as in Figs.\ 1, 2 and 3,   the color-singlet $q$ and $\bar q$ pair created at a local point 
will expand   to the full longitudinal extension to  become a   QED meson and will carry out a longitudinal yo-yo motion appropriate for the QED meson eigenstate,  as described in Fig. 12 and in  Ref. \cite{Won22c}.
Similarly, in the  surface region of a quark-gluon plasma
during the confinement-to-deconfinement phase transition in 
high-energy heavy-ion collisions,  deconfined $u$ and $d$ quarks with low energies may  search for
partners to form a low-energy $d$-$u$-$d$ bound state. A color-singlet combination of low-energy $u$ and $d$ quarks may form a  $d$-$u$-$d$ QED neutron eigenstate by their mutual  QED interactions, if these quarks posses the proper eigenenergy.  
A search for the QED neutron in high-energy heavy-ion collisions may be of great interest.

\acknowledgments{
The author is indebted to Prof. V. F. Perepelitsa whose talk at the
International Symposium on Multiparticle Dynamics, in 2009
introduced the author to the subject of anomalous soft photons which
raised author's interest  on the question of quark confinement in the QED interaction. The author
would like to thank Profs. Y. Jack Ng, A. Koshelkin, H. Sazdjian,
Soren Sorensen, D. Blaschke, Kh. U. Abraamyan, Gang Wang, Xi-Guang Cao,
G. Wilk, Y. Sharon, L. Zamick, and I-Yang Lee for
helpful communications. The research was supported
in part by the Division of Nuclear Physics, U.S. Department of Energy
under Contract DE-AC05-00OR22725 with UT-Battelle, LLC.
}

\end{document}